\documentclass[11 pt, a4paper]{article}
\usepackage[a4paper, hmargin=1.25in,vmargin=1.25in]{geometry}
\usepackage{amsmath,amssymb,amsthm,mathrsfs,amsfonts,mathtools,calculator,dsfont,bm} 
\usepackage[dvipsnames,svgnames, x11names]{xcolor}
\usepackage{natbib,bibentry,bibunits,lmodern}
\usepackage{grffile} 
\usepackage[bottom,hang,flushmargin]{footmisc} 
\usepackage{enumitem} 
\newlist{problems}{enumerate}{1}
\setlist[problems]{label={\arabic*.}, ref={\arabic{part}.\thechapter.\arabic*}}
\usepackage{pdfpages,selectp} 
\usepackage[hyphens]{url}
\usepackage[labelsep=period]{caption} 
\usepackage{subcaption}
\usepackage{fancyvrb} 
\usepackage{environ,trimspaces,xspace}
\usepackage[titletoc]{appendix}
\usepackage{fancyhdr,setspace,titlesec,titling} 
\usepackage{siunitx,booktabs,longtable,threeparttable,multirow,makecell,float,rotating} 
\usepackage{tikz,pgfplots} 
\pgfplotsset{compat=1.11}
\usetikzlibrary{arrows,intersections,plotmarks}
\usetikzlibrary{backgrounds,positioning,fit,petri}
\usetikzlibrary{mindmap,trees,calendar,external,shapes}
\usetikzlibrary{decorations.fractals,decorations.pathmorphing,shadows,shadings,patterns}
\usetikzlibrary{spy,calc} 
\usepgfplotslibrary{groupplots}
\usepackage[bookmarksnumbered,bookmarksopen,psdextra,unicode,hyperindex,breaklinks,hypertexnames=false]{hyperref}
\usepackage[open,openlevel=2,numbered,atend]{bookmark} 
\usepackage{tabularx}
\usepackage{cleveref}
\renewcommand{\thetable}{\Roman{table}}
\usepackage{pdflscape}

\definecolor{Ocean}{rgb}{0,0,0.75}
\makeatletter
\renewcommand\paragraph{\@startsection{paragraph}{4}{\z@}%
            {-2.5ex\@plus -1ex \@minus -.25ex}%
            {1.25ex \@plus .25ex}%
            {\normalfont\normalsize\bfseries}}
\makeatother
\hypersetup{colorlinks, linkcolor = {NavyBlue}, citecolor = {NavyBlue}, urlcolor ={NavyBlue}}

\newtheorem{thm}{Theorem}[section]

\newtheorem{lem}{Lemma}[section]
\newtheorem{pro}{Proposition}[section]
\newtheorem{ass}{Assumption}[section]
\theoremstyle{definition}

\newtheorem{rem}{Remark}[section]

\theoremstyle{definition}

\newtheorem*{ex*}{Example}

\newtheoremstyle{exctd}
{\topsep} {\topsep}%
{\upshape}
{}
{\bfseries\scshape}
{.}
{1em}
{\thmname{#1} \thmnumber{ #2}\thmnote{#3} (cont.)}
\theoremstyle{exctd}

\newcommand{\amin}{\operatornamewithlimits{arg\,min}}
\newcommand{\amax}{\operatornamewithlimits{arg\,max}}

\usepackage{dcolumn}
\newcolumntype{L}{D{.}{.}{2,7}}

\pgfmathdeclarefunction{npdf}{3}{%
\pgfmathparse{1/(#3*sqrt(2*pi))*exp(-((#1-#2)^2)/(2*#3^2))}%
}
\tikzset{
    declare function={
        ncdf(\x,\m,\s)=1/(1 + exp(-0.07056*((\x-\m)/\s)^3 - 1.5976*(\x-\m)/\s));
    }
}


\defaultbibliographystyle{C:/Dropbox/Research_Related/BibMaker/ecta}
\defaultbibliography{C:/Dropbox/Research_Related/BibMaker/Mybibliography_Infinity}

\begin{document}
\begin{bibunit}
\pdfbookmark[1]{Title}{title}
\title{Robust Estimation of Conditional Factor Models}
\author{Qihui Chen\\ School of Management and Economics\\ The Chinese University of Hong Kong, Shenzhen\\ qihuichen@cuhk.edu.cn}
\date{\today}
\maketitle

\begin{abstract}
This paper develops estimation and inference methods for conditional quantile factor models. We first introduce a simple sieve estimation, and establish asymptotic properties of the estimators under large $N$. We then provide a bootstrap procedure for estimating the distributions of the estimators. We also provide two consistent estimators for the number of factors. The methods allow us not only to estimate conditional factor structures of distributions of asset returns utilizing characteristics, but also to conduct robust inference in conditional factor models, which enables us to analyze the cross section of asset returns with heavy tails. We apply the methods to analyze the cross section of individual US stock returns.
\end{abstract}

\begin{center}
\end{center}
\newpage

\section{Introduction}\label{Sec: 1}
Classical factor models \citep{ChamberlainRothschild_FactorStuctures_1982} may not be satisfactory for empirical asset pricing, because they ignore assets' characteristics information that can be informative \citep{RosenbergMcKibben_ThePrediction_1973,FamaFrench_Commonrisk_1993,DanielTitman_Characteristics_1997}. Recently, \citet*{Connoretal_EfficientFFFactor_2012}, \citet*{Kellyetal_Characteristics_2019} and \citet*{Chenetal_SeimiparametricFactor_2021} introduce conditional factor models by incorporating assets' characteristics information into pricing errors and risk exposures. The conditional factor models provide a parsimonious and meaningful way of utilizing characteristics to analyze the cross-sectional differences of average asset returns. However, these works adopt least squares methods and require at least finite fourth moment, which is restrictive for individual stock returns that may cross-sectionally exhibit heavy tails and large observations even for monthly data \citep*{Guetal_EmpiricalAsset_2020}.

In this paper,we study a conditional quantile factor model,
which allows us to estimate conditional factor structures of average asset returns utilizing characteristics in the presence of large return observations, as well as distributions of asset returns. Specifically, we develop estimation and inference methods for the following conditional quantile factor model: for $\tau\in(0,1)$,
\begin{align}\label{Eqn: Model}
Q_{y_{it}|z_{it},f_t(\tau)}(\tau, z_{it},f_t(\tau)) = \alpha(\tau,z_{it}) +\beta(\tau,z_{it})^{\prime}f_t(\tau), i=1,\ldots,N,t=1,\ldots,T,
\end{align}
where $y_{it}$ is the excess return of asset $i$ in time period $t$, $z_{it}$ is an $M\times 1$ vector of pre-specified characteristics known at the beginning of time period $t$, $f_t(\tau)$ is a $K(\tau)\times 1$ vector of unobserved latent factors, $Q_{y_{it}|z_{it},f_t(\tau)}(\tau, z_{it},f_t(\tau))$ denotes the $\tau$th quantile of the conditional distribution of $y_{it}$ given $z_{it}$ and $f_t(\tau)$, $\beta(\tau,\cdot)$ is a $K(\tau)\times 1$ vector of unknown factor loading functions, $\alpha(\tau,\cdot)$ is an unknown intercept function. Here, the intercepts and the factor loadings are functions of characteristics. The factors, the intercept function, the factor loading functions, and the number of factors are allowed to be quantile-dependent. In the case of symmetric distributions, \eqref{Eqn: Model} at $\tau=0.5$ describes a conditional factor structure of average returns, which allows us to distinguish between the ``risk'' (i.e., $\beta(0.5,z_{it})$) and ``mispricing'' (i.e., $\alpha(0.5,z_{it})$) explanations of the role of characteristics in explaining average returns. Assuming that $Q_{y_{it}|z_{it},f_t(\tau)}(\tau, z_{it},f_t(\tau))$ is strictly increasing in the first argument, we may rewrite \eqref{Eqn: Model} as
\begin{align}\label{Eqn: Modelalt}
y_{it} = \alpha(\tau,z_{it}) +\beta(\tau,z_{it})^{\prime}f_t(\tau) + \varepsilon_{it}(\tau) \text{ with }P(\varepsilon_{it}(\tau)\leq 0|z_{it},f_t(\tau))=\tau.
\end{align}

As the first contribution of the paper, we introduce a simple sieve estimation for the parameters in the model. First, we approximate the nonparametric functions $\alpha(\tau,\cdot)$ and $\beta(\tau,\cdot)$ by sieve methods, and develop an easy-to-compute estimator for $\alpha(\tau,\cdot)$, $\beta(\tau,\cdot)$ and $f_{t}(\tau)$. The estimators can be easily obtained by conducting principle component analysis (PCA) on an estimated coefficient matrix, which is obtained by running the cross-sectional quantile regression of $y_{it}$ on sieves of $z_{it}$ for each time period $t$. Throughout the paper, we refer to the method as \textit{the QR-PCA}. Second, we establish asymptotic properties of the estimators under large $N$, including rate of convergence and asymptotic normality. In particular, we establish a strong Gaussian approximation for the distributions of the estimators of the large dimensional coefficient matrices in the sieve approximation of $\alpha(\tau,\cdot)$ and $\beta(\tau,\cdot)$ as well as the estimator of $f_{t}(\tau)$. Third, we provide two methods for determining the number of factors $K(\tau)$. Thus, we do not need to specify the number of factors \textit{a priori} in conducting the QR-PCA.

As the second contribution, we develop a bootstrap procedure to estimate the aforementioned asymptotic distributions for the purpose of inference. The main challenge lies in that the asymptotic distribution of the estimator of the coefficient matrix in the sieve approximation of $\beta(\tau,\cdot)$ involves a rotational transformation matrix, which depends on the estimator of $f_{t}(\tau)$. Consequently, the rotational transformation matrix in the bootstrap world could be different, rendering the failure of the bootstrap. To dispense with this issue, we enforce the same factor estimator in the bootstrap world as in the real world when estimating the large dimensional coefficient matrix in the sieve approximation of $\beta(\tau,\cdot)$. The bootstrap procedure allows us to conduct inference on $\alpha(\tau,\cdot)$, $\beta(\tau,\cdot)$ and $f_{t}(\tau)$. All the asymptotic results are shown to hold uniformly over $\tau$ in the set of quantile indices of interest. In addition, the results are attractive in several aspects: (i) they are robust to heavy tailed errors; (ii) they do not require large $T$; (iii) they allow for temporal dependence and $z_{it}$ being nonstationary; (iv) they are applicable for unbalanced panels. Our simulation studies show that our estimators and bootstrap procedure perform very well in finite samples, even when $T$ is small.

Finally, to demonstrate the empirical relevance of our methods, we study the cross section of individual US stock returns. We use the same data set of monthly observations and the same functional form specifications of the intercept function and the factor loading functions as \citet{Chenetal_SeimiparametricFactor_2021}, and implement the QR-PCA at various quantiles. The main interesting findings are as follows. First, the intercept function is statistically different from zero at $1\%$ level for all quantiles in all specifications, regardless of the number of estimated factors (up to 10). Second, the quantile factors extracted by the QR-PCA vary across quantiles in all specifications. Third, the median factors are very different from \citet{Chenetal_SeimiparametricFactor_2021}'s mean factors in all specifications, though their estimated dimensions are identical. Fourth, the median factors can help improve the mean factors’ ability to explain the cross section of portfolio returns. We also conduct analysis using daily observations and have similar findings.

There are several studies on qauntile factor models, including \citet{AndoBai_QuantilePanel_2018}, \citet*{ChenDoladoGonzalo_QuantileFactor_2015} and \citet*{Maetal_SemiparametricQuantile_2020}. The first two study the quantile version of the classical factor model, and the third one studies the quantile version of the \citet{Connoretal_EfficientFFFactor_2012} model, which is most related to our study. Our paper is distinct in several aspects. First, we allow for time-varying characteristics, nonzero intercepts and multivariate factor loading functions, which are important in empirical asset pricing. To the best of our knowledge, our paper is the first one studying time-varying quantile factor models. Second, our asymptotic results hold uniformly over $\tau$ and are valid for small $T$, while all of their asymptotic results hold only for each $\tau$ and require $T\to\infty$.  The small $T$ property allows us to conduct rolling small sub-sample analyses to accommodate more dynamics, which appears appealing in empirical asset pricing. Third, our results are applicable for unbalanced panels, which is useful since individual stocks have varying life spans. Fourth, we provide inference methods for both the intercepts and factor loadings, and estimators for the number of factors. This allows us to test the significance of characteristics, and pin down the number of factors. For the use of empirical process techniques, we need to impose cross-sectional independence in the proofs. However, our simulation studies show that our estimators continue to work well in the presence of cross-sectional dependence.

Conditional factor models have attracted much attention in empirical asset pricing. \citet{Connoretal_EfficientFFFactor_2012} propose a kernel estimation by assuming time-invariant characteristics, zero pricing errors and univariate risk exposure functions. \citet{Kellyetal_Characteristics_2019} employ alternating least squares under linear specifications of the pricing error function and the risk exposure functions. \citet{Chenetal_SeimiparametricFactor_2021} introduce a simple and tractable sieve estimation to allow for nonlinearity. In contrast to the characteristic-based approach \citep*{RosenbergMcKibben_ThePrediction_1973,Lewellen_Crosssection_2014,Greenetal_Characteristics_2017} and the risk-based approach \citep{FamaFrench_Commonrisk_1993,FamaFrench_FiveFactor_2015}, conditional factor models allow us to disentangle the characteristics' role in capturing the risk exposures from the pricing errors without the need to pre-specify factors. We complement the literature by providing robust estimation and inference methods for conditional factor models, which allow us to analyze the cross section of asset returns with heavy tails.

In addition to providing robust inference, quantile analyses of asset returns may provide more information about return distributions, which may be relevant in applications. For example, left tail measures such as expected shortfall and value-at-risk are widely used in financial risk management; right tail information affects investors' behavior in financial markets; a prospect investor is affected by the whole distribution of return process \citep{BarberisHuang_StockAsLottery_2008}. \citet*{Gowllandetal_BeyondCentral_2009} provide examples of how quantile regressions can be utilized to analyze the effectiveness of individual factors for quantitative investment purposes. The QR-PCA enables us to estimate conditional factor structures of distributions of asset returns utilizing characteristics.

In particular, our study is related to the literature on the ``idiosyncratic volatility pricing puzzle'' \citep*{Angetal_CSVolatility_2006}. An important concern of the literature is about how to correctly extract the common factors in idiosyncratic volatilities. Most existing works need to pre-specify the factors for means of asset returns, and thus may fail to extract the idiosyncratic volatility factors due to omitted factors; see, for example, \citet*{Duarteetal_SystematicRisk_2014} and \citet*{Herskovicetal_CommonFactorIVariance_2016}, among others. \citet*{Renaultetal_APTIVariance_2019} propose to extract the idiosyncratic volatility factors from the squared returns, which nevertheless may fail in the presence of heavy tails. The QR-PCA provides a simple and robust way of extracting the common factors in idiosyncratic volatilities utilizing characteristics without the need to pre-specify the factors for the means of asset returns.

The structure of the paper is given as follows. Section \ref{Sec: 2} introduces the estimation method\textemdash the QR-PCA. Section \ref{Sec: 3} establishes asymptotic properties of the estimators, including rate of convergence and asymptotic distribution. Section \ref{Sec: 4} introduces a bootstrap procedure for inference purposes. Section \ref{Sec: 5} provides two methods for determining the number of factors. Section \ref{Sec: 6} presents simulation studies.  Section \ref{Sec: 7} gives an empirical analysis. Section \ref{Sec: 8} briefly concludes. Proofs are collected in the appendices.

For convenience of the reader, we collect standard pieces of notation here, which will be used throughout the paper. We denote the Euclidian norm of a column vector $x$ by $\|x\|$. We denote the operator norm of a matrix $A$ by $\|A\|_2$, the Frobenius norm by $\|A\|_{F}$, and the vectorization by $\mathrm{vec}(A)$. We denote the $k$th largest eigenvalue of a symmetric matrix $A$ by $\lambda_{k}(A)$, and the smallest and largest eigenvalues by $\lambda_{\min}(A)$ and $\lambda_{\max}(A)$.

\section{The QR-PCA}\label{Sec: 2}
To illustrate the idea of the {QR-PCA}, we first assume that $\alpha(\tau,\cdot) = 0$ and $\beta(\tau,z_{it}) = \Gamma(\tau)^{\prime}z_{it}$ for some $M\times K(\tau)$ matrix $\Gamma(\tau)$. Thus, \eqref{Eqn: Modelalt} reduces to
\begin{align}\label{Eqn: Model: Illustration}
y_{it} = z_{it}^{\prime}\Gamma(\tau)f_t(\tau) + \varepsilon_{it}(\tau) \text{ with }P(\varepsilon_{it}(\tau)\leq 0|z_{it},f_t(\tau))=\tau.
\end{align}
We observe that $\eqref{Eqn: Model: Illustration}$ can be alternatively viewed as a panel data quantile regression model with time-varying slope coefficients $\Gamma(\tau)f_t(\tau)$ with a reduced rank. Based on this, we may estimate $\Gamma(\tau)$ and $f_t(\tau)$ in two steps: first estimating $\Gamma(\tau)f_t(\tau)$ by running the cross-sectional quantile regression of $y_{it}$ on $z_{it}$ for each time period $t$ and then applying the standard PCA to the estimated coefficient matrix to extract estimators for $\Gamma(\tau)$ and $f_t(\tau)$. The two-step procedure is referred to as the QR-PCA.

We allow for nonzero $\alpha(\tau,\cdot)$ and nonparametric specifications of $\alpha(\tau,\cdot)$ and $\beta(\tau,\cdot) = (\beta_{1}(\tau,\cdot),\ldots,\beta_{K(\tau)}(\tau,\cdot))^{\prime}$. To dispense with the curse of dimensionality in the presence of multivariate $z_{it}$, we assume that $\alpha(\tau,\cdot)$ and $\beta_k(\tau,\cdot)$'s are separable. Specifically, we assume that there are $\{\alpha_{m}(\tau,\cdot)\}_{m\leq M}$ and $\{\beta_{km}(\tau,\cdot)\}_{m\leq M}$ such that
\begin{align}\label{Eqn: Sep}
\alpha(\tau,z_{it}) = \sum_{m=1}^{M}\alpha_{m}(\tau,z_{it,m}) \text{ and } \beta_{k}(\tau,z_{it}) = \sum_{m=1}^{M}\beta_{km}(\tau,z_{it,m}),
\end{align}
where $(z_{it,1},\ldots,z_{it,M})^{\prime}\equiv z_{it}$ or $z_{it,m}$ denotes the $m$th element of $z_{it}$. We use the sieve method for nonparametric estimation of $\alpha_{m}(\tau,\cdot)$ and $\beta_{km}(\tau,\cdot)$. Let $\{\phi_{j}(\cdot)\}_{j\geq 1}$ be a sequence of basis functions (e.g., polynomials, B-spline), which can be used to approximate $\alpha_{m}(\tau,\cdot)$ and $\beta_{km}(\tau,\cdot)$ well. We assume that
\begin{align}\label{Eqn: Sieve}
\alpha_{m}(\tau,z_{it,m}) &= \sum_{j=1}^{J}a_{m,j}(\tau) \phi_{j}(z_{it,m}) + r_{m,J}(\tau,z_{it,m}),\\
\beta_{km}(\tau,z_{it,m}) &= \sum_{j=1}^{J}b_{km,j}(\tau) \phi_{j}(z_{it,m}) + \delta_{km,J}(\tau,z_{it,m}).
\end{align}
Here, $\{a_{m,j}(\tau)\}_{j\leq J}$ and $\{b_{km,j}(\tau)\}_{j\leq J}$ are the sieve coefficients; $r_{m,J}(\tau,\cdot)$ and $\delta_{km,J}(\tau,\cdot)$ are ``remaining functions'' representing the approximation errors; $J$ is the sieve size. The sieve method requires that $\sup_{\tau,z}|r_{m,J}(\tau,z)|\to 0$ and $\sup_{\tau,z}|\delta_{km,J}(\tau,z)|\to 0$ as $J\to\infty$. We next write the model using matrices. To the end, we let $\bar{\phi}(z_{it,m}) \equiv (\phi_{1}(z_{it,m}),\ldots,\phi_{J}(z_{it,m}))^{\prime}$ and ${\phi}(z_{it}) \equiv (\bar{\phi}(z_{it,1})^{\prime},\ldots, \bar{\phi}(z_{it,M})^{\prime})^{\prime}$. Denote $a(\tau)\equiv (a_{1,1}(\tau),$ $\ldots,a_{1,J}(\tau),\ldots, a_{M,1}(\tau),\ldots,a_{M,J}(\tau))^{\prime}$ and $B(\tau)\equiv(b_1(\tau),\cdots, b_{K(\tau)}(\tau))$, where $b_{k}(\tau)= (b_{k1,1}(\tau),\ldots,b_{k1,J}(\tau),\ldots, b_{kM,1}(\tau),\ldots,$ $b_{kM,J}(\tau))^{\prime}$. Here, $a(\tau)$ is a $JM\times 1$ vector of the sieve coefficients for $\alpha(\tau,\cdot)$, and $B(\tau)$ is a $JM\times K(\tau)$ matrix of the sieve coefficients for $\beta(\tau,\cdot)$. Set $r(\tau,z_{it}) \equiv  \sum_{m=1}^{M}r_{m,J}(\tau,z_{it,m})$ and $\delta(\tau,z_{it}) \equiv (\sum_{m=1}^{M}\delta_{1m,J}(\tau,z_{it,m}),\ldots, $ $\sum_{m=1}^{M}\delta_{K(\tau)m,J}(\tau,z_{it,m}))^{\prime}$, which are the sieve approximation errors for $\alpha(\tau,z_{it})$ and $\beta(\tau,z_{it})$, respectively. Then
\begin{align}\label{Eqn: Sieve: together}
\alpha(\tau,z_{it}) = a(\tau)^{\prime}\phi(z_{it}) + r(\tau,z_{it}) \text{ and } \beta(\tau,z_{it}) = B(\tau)^{\prime}\phi(z_{it}) + \delta(\tau,z_{it}).
\end{align}
Thus, $\alpha(\tau,z_{it})$ and $\beta(\tau,z_{it})$ can be well approximated by $a(\tau)^{\prime}\phi(z_{it})$ and $B(\tau)^{\prime}\phi(z_{it})$, and estimation of $\alpha(\tau,\cdot)$ and $\beta(\tau,\cdot)$ boils down to estimation of $a(\tau)$ and $B(\tau)$.

We now introduce the estimation of $a(\tau)$, $B(\tau)$ and $f_t(\tau)$ based on the above sieve approximation in \eqref{Eqn: Sieve: together} by employing the QR-PCA. Under the basic sieve assumption, plugging \eqref{Eqn: Sieve: together} into \eqref{Eqn: Modelalt} implies that
\begin{align}\label{Eqn: Model: SieveApprox}
y_{it}\approx \phi(z_{it})^{\prime}[a(\tau) +B(\tau)f_t(\tau)]+ \varepsilon_{it}(\tau) \text{ with }P(\varepsilon_{it}(\tau)\leq 0|z_{it},f_t(\tau))=\tau.
\end{align}
Thus, we obtain an approximate model with linear intercept and loading functions with possibly large dimensions, which are due to possibly large $J$. Despite of the possibly large dimensions, we may continue to employ the QR-PCA to estimate $a(\tau)$, $B(\tau)$ and $f_t(\tau)$. As different from \eqref{Eqn: Model: Illustration}, we are interested in point identification and consistent estimation of $a(\tau)$ and $\alpha(\tau,\cdot)$. To the end, we impose $a(\tau)^{\prime}B(\tau)=0$. To describe the estimation procedure, let $\tilde{Y}_t(\tau)$ be the $JM\times 1$ vector of estimated coefficients in the cross-sectional $\tau$th quantile regression of $y_{it}$ on $\phi(z_{it})$ for each time period $t$:
\begin{align}\label{Eqn: Ytildet}
\tilde{Y}_t(\tau)=\amin_{b}\sum_{i=1}^{N}\rho_{\tau}(y_{it} - \phi(z_{it})^{\prime}b),
\end{align}
where $\rho_{\tau}(u) = (\tau-1\{u\leq 0\})u$ is the check function, and $\bar{\tilde{Y}}(\tau) \equiv \sum_{t=1}^{T}\tilde{Y}_{t}(\tau)/T$. We may estimate $a(\tau)$, $B(\tau)$ and $f_t(\tau)$ as follows. First, since $\tilde{Y}_t(\tau)\approx a(\tau)+B(\tau){f}_t(\tau)$, we may remove $a(\tau)$ by subtracting $\bar{\tilde{Y}}(\tau)$ from $\tilde{Y}_{t}(\tau)$ and estimate $B(\tau)$ by applying the standard PCA to $\{\tilde{Y}_{t}(\tau)-\bar{\tilde{Y}}(\tau)\}_{t\leq T}$. Second, since $\bar{\tilde{Y}}(\tau)\approx a(\tau)+B(\tau)\bar{f}(\tau)$ and $a(\tau)^{\prime}B(\tau)=0$ where $\bar{f}(\tau)=\sum_{t=1}^{T}f_t(\tau)/T$, we may estimate $a(\tau)$ according to $a(\tau)\approx[I_{JM}-B(\tau)[B(\tau)'B(\tau)]^{-1}B(\tau)^{\prime}]\bar{\tilde{Y}}(\tau)$. Third, since $\tilde{Y}_t(\tau)\approx a(\tau)+B(\tau){f}_t(\tau)$ and $a(\tau)^{\prime}B(\tau)=0$, we may estimate $f_t(\tau)$ according to $f_t(\tau)\approx[B(\tau)'B(\tau)]^{-1}B(\tau)^{\prime}\tilde{Y}_t(\tau)$.

We formally define the estimators as follows. Denote the estimators of $a(\tau)$, $B(\tau)$, $\alpha(\tau,\cdot)$, $\beta(\tau,\cdot)$ and $F(\tau)=(f_1(\tau),\ldots,f_{T}(\tau))^{\prime}$ by $\hat{a}(\tau)$, $\hat{B}(\tau)$, $\hat{\alpha}(\tau,\cdot)$, $\hat{\beta}(\tau,\cdot)$ and $\hat{F}(\tau)$. Let $M_T\equiv I_{T} - 1_T1_T^{\prime}/T$, where $1_{T}$ denotes a $T\times 1$ vector of ones. To conduct PCA, we use the following normalization: $B(\tau)^{\prime}B(\tau)=I_{K(\tau)}$ and $F(\tau)^{\prime}M_T F(\tau)/T$ is diagonal with diagonal entries in descending order. Let $\tilde{Y}(\tau) \equiv (\tilde{Y}_1(\tau),\ldots, \tilde{Y}_T(\tau))$. Then the eigenvectors corresponding to the first $K(\tau)$ largest eigenvalues of the $JM\times JM$ matrix $\tilde{Y}(\tau)M_T\tilde{Y}(\tau)^{\prime}/T$ give the columns of $\hat{B}(\tau)$, $\hat{a}(\tau) = [I_{JM}-\hat{B}(\tau)\hat{B}(\tau)^{\prime}]\bar{\tilde{Y}}(\tau)$, and
\begin{align}\label{Eqn: Estimators}
\hat{\alpha}(\tau,z)= \hat{a}(\tau)^{\prime}\phi(z), \hat{\beta}(\tau,z) = \hat{B}(\tau)^{\prime}\phi(z),\hat{F}(\tau) = (\hat{f}_1(\tau),\ldots,\hat{f}_T(\tau))^{\prime} = \tilde{Y}(\tau)^{\prime}\hat{B}(\tau).
\end{align}
We assume that $K(\tau)$ is known here, and conduct asymptotic analysis and develop bootstrap inference in Sections \ref{Sec: 3} and \ref{Sec: 4}. In Section \ref{Sec: 5}, we develop two consistent estimators of $K(\tau)$, so all the asymptotic results carry over to the unknown $K(\tau)$ case using a conditioning argument.

\begin{rem}\label{Rem: Unbalanced}
The QR-PCA is applicable for unbalanced panels. Notice that the crucial step is to obtain $\tilde{Y}_t(\tau)$. When unbalanced panels are present, we may obtain $\tilde{Y}_t(\tau)$ by using available observations in time period $t$. The asymptotic results established in Sections \ref{Sec: 3}-\ref{Sec: 5} continue to hold when $\min_{t\leq T} N_t \to\infty$, where $N_t$ is the number of available observations in time period $t$.
\end{rem}


\begin{rem}\label{Rem: VolatilityFactor}
The QR-PCA can be used to extract idiosyncratic volatility factors in addition to mean factors. To illustrate, we consider the following model: $y_{it} = \alpha(z_{it}) + \beta(z_{it})^{\prime}f_{t} +[\delta(z_{it}) + \gamma(z_{it})^{\prime}g_{t}]e_{it}$, where $e_{it}$ is an idiosyncratic error, $g_t$ are the idiosyncratic volatility factors, and $f_{t}$ are the mean factors. Assume that the distribution of $\varepsilon_{it}$ is symmetric and $\delta(z_{it}) + \gamma(z_{it})^{\prime}g_{t}>0$. It is easy to see that $(f_{t}^{\prime},g_{t}^{\prime})^{\prime}$ are the $\tau$th quantile factors for $\tau\neq 0.5$, thus can be extracted by the QR-PCA.
\end{rem}

In view of \eqref{Eqn: Model: SieveApprox}, a natural alternative estimation method is given by solving the following minimization problem
\begin{align}\label{Eqn: LeastSE}
\min_{a,B,\{f_t\}_{t\leq T}}\sum_{i=1}^{N}\sum_{t=1}^{T}\rho_{\tau}(y_{it} - \phi(z_{it})^{\prime}a -\phi(z_{it})^{\prime}Bf_t).
\end{align}
The main challenges for this method are that the objective function is nonconvex in $(a,B,\{f_t\}_{t\leq T})$, and the estimators do not have an analytical closed form. These make it difficult not only to find a computational algorithm to obtain the estimators, but also to derive their asymptotic properties. One may suggest the following alternating quantile regression procedure to find the estimators: (i) specifying an initial estimator for $f_t$; (ii) searching $a$ and $B$ via quantile regression for a given $f_t$; (iii) searching $f_t$ via quantile regression for a given $a$ and $B$; (iv) iterating (ii) and (iii) until convergence. However, its convergence to the estimators defined in \eqref{Eqn: LeastSE} is not guaranteed due to the nonconvexity. Indeed, when the quantile regressions are replaced with least square regressions, the procedure reduces to the alternating least squares studied in \citet*{Parketal_FactorDynamics_2009}. They point out that the alternating least squares estimators may not converge to the least squares estimators, defined in \eqref{Eqn: LeastSE} by replacing the check function with the quadratic loss function. In fact, the same problem occurs even in the alternating least squares for classical static factor models; its convergence requires certain conditions on the initial value, see section 7.3 of \citet{GolubVanLoan_MatrixComputation_2013}. In contrast to the alternating quantile regression procedure, our QR-PCA is computationally simple and has good asymptotic properties; see Sections \ref{Sec: 3} and \ref{Sec: 4}.

\section{Asymptotic Analysis}\label{Sec: 3}
In this section, we establish asymptotic properties of our estimators. Specifically, we establish rate of convergence and asymptotic distribution.

\subsection{Rate of Convergence}
Let $\mathcal{T}$ be the set of quantile indices of interest and $\bar{\mathcal{T}}$ denote the convex hull of $\mathcal{T}$. We first establish a rate of convergence for the estimators $\hat{a}(\tau)$, $\hat{B}(\tau)$, $\hat{F}(\tau)$, $\hat{\alpha}(\tau,\cdot)$ and $\hat{\beta}(\tau,\cdot)$. To the end, we impose the following assumptions.

\begin{ass}[Data generating process]\label{Ass: DGP}
For each $t\leq T$, the following conditions hold: (i) $\{(y_{it},z_{it})\}_{i=1}^{n}$ is an independently and identically distributed (i.i.d.) random sample from the distribution of the pair $(y_{t},z_{t})$; (ii) the conditional density $f_{y_t|z_{t}}(y|z)$ is bounded from above uniformly over $y\in\mathcal{Y}_{z,t}$ and $z\in\mathcal{Z}_t$, where $\mathcal{Y}_{z,t}$ denotes the support of $y_t$ given $z_{t}=z$ and $\mathcal{Z}_t$ denotes the support of $z_t$; (iii) $f_{y_t|z_{t}}(Q_{y_{t}|z_{t}}(\tau,z)|z)$ is bounded away from zero uniformly over $\tau\in\bar{\mathcal{T}}$ and $z\in\mathcal{Z}_t$, where $Q_{y_t|z_t}(\cdot,z)$ is the conditional quantile function of $y_t$ given $z_t=z$; (iv) the derivative of $y\mapsto f_{y_t|z_{t}}(y|z)$ is continuous and bounded in absolute value from above uniformly over $y\in\mathcal{Y}_{z,t}$ and $z\in\mathcal{Z}_t$.
\end{ass}

Assumption \ref{Ass: DGP}(i) requires cross-sectional independence of the data that allows us to use empirical process techniques, but it can be extended to allow for cross-sectional dependence at the expense of more technicalities. No temporal dependence restriction is imposed, so temporal dependence is allowed. In particular, $z_{it}$ is allowed to be nonstationary across $t$. Assumptions \ref{Ass: DGP}(ii)-(iv) are standard in the literature; see, for example, \citet{Koenker_Quantile_2005} and \citet*{Bellonietal_ConditionalQuantile_2019}. In particular, they do not require conditional moment of $y_t$ given $z_t$ to exist.

\begin{ass}[Basis functions]\label{Ass: Basis}
For each $t\leq T$, the following conditions hold: (i) the eigenvalues of $E[\phi(z_{it})\phi(z_{it})^{\prime}]$ are bounded from above and away from zero; (ii) $\|f_{t}(\tau)\|$ is bounded uniformly over $\tau\in\mathcal{T}$; (iii) $\max_{m\leq M}\sup_{\tau\in\mathcal{T},z\in\mathcal{Z}_{t,m}}|r_{m,J}(\tau,z)|=O(J^{-\kappa})$ and $\max_{k\leq K(\tau),m\leq M}\sup_{\tau\in\mathcal{T},z\in\mathcal{Z}_{t,m}}|\delta_{km,J}(\tau,z)|=O(J^{-\kappa})$ for some constant $\kappa>0$, where $\mathcal{Z}_{t,m}$ denotes the support of $z_{t,m}$; (iv) $\sup_{\tau\in\mathcal{T}}\|a(\tau)+B(\tau)f_t(\tau) - \beta_{t}(\tau)\| = O(J^{-\kappa}\xi_{J}^{-1})$, where $\beta_{t}(\tau) =
\amin_{b} E[\rho_{\tau}(y_{t}-\phi(z_{t})^{\prime}{b})]$ and $\xi_{J}=\sup_{z\in\mathcal{Z}}\|\bar{\phi}(z)\|$ with $\mathcal{Z} = \cup_{m\leq M, t\leq T}\mathcal{Z}_{t,m}$.
\end{ass}

Assumptions \ref{Ass: Basis}(i) and (iii) are standard in the literature; see, for example, \citet{Newey_SeriesEstimator_1997} and \citet{Chen_Handbook_2007}. In particular, Assumption \ref{Ass: Basis}(iii) can be easily satisfied by using B-slpine or polynomial basis functions under certain smoothness of $\alpha(\tau,\cdot)$ and $\beta(\tau,\cdot)$. For simplicity of proof, we assume $f_{t}(\tau)$ to be fixed parameters in Assumption \ref{Ass: Basis}(ii). All the results continue to hold, when $\{f_{t}(\tau)\}_{t\leq T}$ are assumed to be random variables that are independent of $\{z_{it}\}_{i\leq N,t\leq T}$. Assumption \ref{Ass: Basis}(iv) is not new in the literature. A similar condition has been imposed by \citet{Bellonietal_ConditionalQuantile_2019} in sieve estimation of conditional quantile process; see their lemma 1 for preliminary sufficient conditions and comment 1 for related discussions.

\begin{ass}[Intercept and loading functions and factors]\label{Ass: InterceptLoadingsFactors}
(i) $a(\tau)^{\prime}B(\tau)=0$ for each $\tau\in\mathcal{T}$; (ii) $\|a(\tau)\|$ is bounded uniformly over $\tau\in\mathcal{T}$; (iii) the eigenvalues of $B(\tau)'B(\tau)$ are bounded from above and away from zero uniformly over $\tau\in\mathcal{T}$; (iv) the eigenvalues of $F(\tau)^{\prime}M_TF(\tau)/T$ are bounded away from zero uniformly over $\tau\in\mathcal{T}$.
\end{ass}

Assumptions \ref{Ass: InterceptLoadingsFactors}(i) and (ii) are imposed for the identification of $a(\tau)$ and $\alpha(\tau,\cdot)$. Similar conditions are also imposed in existing studies of semiparametric factor models for conditional means; see, for example, \citet{Connoretal_EfficientFFFactor_2012} and \citet{Chenetal_SeimiparametricFactor_2021}. Assumptions \ref{Ass: InterceptLoadingsFactors}(iii) and (iv) are similar to the \textit{pervasive} condition in \citet{StockWatson_PCA_2002}; a similar condition is also imposed in \citet{ChenDoladoGonzalo_QuantileFactor_2015} and \citet{Maetal_SemiparametricQuantile_2020}. Since the dimension of $B(\tau)$ is $JM\times K(\tau)$, Assumption \ref{Ass: InterceptLoadingsFactors}(iii) requires $JM\geq K(\tau)$. Since the rank of $M_T$ is $T-1$, Assumption \ref{Ass: InterceptLoadingsFactors}(iv) requires $T\geq K(\tau)+1$, which further requires $T\geq 2$. These restrictions on $J$ and $T$ are mild, since we assume $K(\tau)$ to be fixed throughout the paper.

To proceed, let $H(\tau)\equiv [F(\tau)^{\prime}M_T\hat{F}(\tau)][\hat{F}(\tau)^{\prime}M_T\hat{F}(\tau)]^{-1}$, which is a rotational transformation matrix that defines the convergence limits of $\hat{B}(\tau)$ and $\hat{F}(\tau)$. The rate of convergence result is established as follows.

\begin{thm}\label{Thm: RateConv}
Suppose Assumptions \ref{Ass: DGP}-\ref{Ass: InterceptLoadingsFactors} hold. Let $\hat{a}(\tau)$, $\hat{B}(\tau),\hat{F}(\tau)$, $\hat{\alpha}(\tau,\cdot)$ and $\hat{\beta}(\tau,\cdot)$ be given in \eqref{Eqn: Estimators}. Assume (i) $N\to\infty$; (ii) $T\geq K(\tau)+1$ is finite; (iii) $J\to\infty$ with $J\xi^{2}_{J}\log^{2} N=o(N)$ and $J^{-\kappa}\log N=o(1)$. Then
\begin{align*}
\sup_{\tau\in\mathcal{T}}\|\hat{a}(\tau) - a(\tau)\|^{2}&=O_{p}\left(\frac{{J}}{{N}}+\frac{1}{J^{2\kappa}\xi_{J}^{2}}\right),\notag\\
\sup_{\tau\in\mathcal{T}}\|\hat{B}(\tau) - B(\tau) H(\tau)\|^{2}_{F}&=O_{p}\left(\frac{{J}}{{N}}+\frac{1}{J^{2\kappa}\xi_{J}^{2}}\right),\notag\\
\sup_{\tau\in\mathcal{T}}\frac{1}{T}\|\hat{F}(\tau)-F(\tau)[H(\tau)^{\prime}]^{-1}\|_{F}^{2}&=O_{p}\left(\frac{{J}}{{N}}+\frac{1}{J^{2\kappa}\xi_{J}^{2}}\right),\notag\\
\sup_{t\leq T}\sup_{\tau\in\mathcal{T}}\sup_{z\in\mathcal{Z}_t}|\hat{\alpha}(\tau,z)-\alpha(\tau,z)|^{2}&=O_{p}\left(\frac{{J\xi_{J}^{2}}}{{N}}+\frac{1}{J^{2\kappa}}\right),\notag\\
\sup_{t\leq T}\sup_{\tau\in\mathcal{T}}\sup_{z\in\mathcal{Z}_t}\|\hat{\beta}(\tau,z)-H(\tau)^{\prime}\beta(\tau,z)\|^{2}&=O_{p}\left(\frac{{J\xi_{J}^{2}}}{{N}}+\frac{1}{J^{2\kappa}}\right).
\end{align*}
If condition (iii) is replaced with $(iii^{\prime})$ $J\to\infty$ with $J^3\xi^{2}_{J}\log^{2} N=o(N)$ and $J^{-\kappa+1}\log N$ $=o(1)$, then for each $\tau\in\mathcal{T}$,
\begin{align*}
\frac{1}{T}\|\hat{F}(\tau)-F(\tau)[H(\tau)^{\prime}]^{-1}\|_{F}^{2}&=O_{p}\left(\frac{{1}}{{N}}+\frac{1}{J^{2\kappa}\xi_{J}^{2}}\right).
\end{align*}
\end{thm}

Theorem \ref{Thm: RateConv} establishes a rate of convergence of the estimators $\hat{a}(\tau)$, $\hat{B}(\tau)$ and $\hat{F}(\tau)$ that holds uniformly over $\tau\in\mathcal{T}$, a rate of convergence of the estimators $\hat{\alpha}(\tau,z)$ and $\hat{\beta}(\tau,z)$ that holds uniformly over $\tau\in\mathcal{T}$ and $z\in\mathcal{Z}_t$, and a faster rate of convergence of the estimator $\hat{F}(\tau)$ that holds for each $\tau\in\mathcal{T}$. In particular, $\hat{a}(\tau)$ and $\hat{B}(\tau)$ have a standard nonparametric rate, while $\hat{F}(\tau)$ can attain the optimal rate $1/N$, which is the fastest rate that one would attain if $\alpha(\tau,\cdot)$ and $\beta(\tau,\cdot)$ were known. To see the latter, let us assume $NJ^{-2\kappa}\xi_{J}^{-2}=O(1)$, which is possible in the case of sufficiently large $\kappa$ under the restriction $J^3\xi^{2}_{J}\log^{2} N=o(N)$. The last result of the theorem implies that the rate of $\hat{F}(\tau)$ is $1/N$ for each $\tau\in\mathcal{T}$. It is noted that the uniform rate of $\hat{F}(\tau)$ is hardly further improved without additional assumption; see Lemma \ref{Lem: TechA4} for an optimal uniform rate under additional conditions. The optimal rate finding implies that the nonparametric specification of $\alpha(\tau,\cdot)$ and $\beta(\tau,\cdot)$ does not deteriorate the optimal rate of estimating $F$ as long as $\alpha(\tau,\cdot)$ and $\beta(\tau,\cdot)$ are sufficiently smooth. Several additional interesting findings are summarized as follows. First, due to the lack of identification, ${B}(\tau)$, ${\beta}(\tau,\cdot)$ and ${F}(\tau)$ can only be consistently estimated up to a rotational transformation, as usually occurred in high-dimensional factor analyses. Second, as distinct from high-dimensional factor analyses \citep{StockWatson_PCA_2002,BaiNg_NumberofFactors_2002}, our result does not require $T\to\infty$, rendering the estimators appealing to empirical studies in finance. Third, the consistency of $\hat{F}(\tau)$ requires $J\to\infty$. This implies that misspecifying $\alpha(\tau,\cdot)$ or $\beta(\tau,\cdot)$ as a parametric form may lead to inconsistent estimation of $F(\tau)$. Fourth, as implied by Assumption \ref{Ass: DGP}, the result is robust to heavy tailed errors, which may be prevalent in financial data, and allows for temporal dependence and nonstationarity of $z_{it}$.

\subsection{Asymptotic Distribution}
We next establish the asymptotic distribution of $\hat{a}(\tau)$, $\hat{B}(\tau)$ and $\hat{F}(\tau)$. To the end, we impose the following assumption.

\begin{ass}[Asymptotic distribution]\label{Ass: AsymDis}
The following conditions hold: (i) the eigenvalues of $[F(\tau)^{\prime}M_TF(\tau)/T]B(\tau)'B(\tau)$ are distinct uniformly over $\tau\in\mathcal{T}$; (ii) there is a zero-mean Gaussian process $\mathbb{N}(\cdot)= (\mathbb{N}_1(\cdot),\ldots,\mathbb{N}_T(\cdot))$ on $\mathcal{T}$ with a.s. continuous path such that the covariance function of $\mathbb{N}(\cdot)$ coincides with that of $\mathbb{U}(\cdot) = (\mathbb{U}_1(\cdot),\ldots,\mathbb{U}_T(\cdot))$ and $\sup_{\tau\in\mathcal{T}}\|\mathbb{U}(\tau)-\mathbb{N}(\tau)\|_{F}$ $= O_{p}(\delta_{N})$ for some $\delta_{N}\downarrow 0$ as $N\to\infty$, where $\mathbb{U}_t(\tau) = \sum_{i=1}^{N}\phi(z_{it})(\tau-1\{u_{it}\leq \tau\})/\sqrt{N}$ with $u_{it}=F_{y_{t}|z_{t}}(y_{it}|z_{it})$ and $F_{y_t|z_{t}}(y|z)$ the conditional distribution function of $y_t$ given $z_t=z$.
\end{ass}

Distinct eigenvalue condition has been commonly imposed in the literature to obtain asymptotic normality; see, for example, \citet{Bai_Inferential_2003}. Assumption \ref{Ass: AsymDis}(ii) requires that $\mathbb{U}(\cdot)$ be coupled to a Gaussian process, where $\mathbb{U}(\cdot)$ determines the asymptotic distribution of $\hat{a}(\tau)$, $\hat{B}(\tau)$ and $\hat{F}(\tau)$. The condition may be verified by resorting to the Yurinskii's coupling \citep{Pollard_Probability_2002}; see Proposition \ref{Pro: Strong}. Here, temporal dependence is allowed.

Let $J_{t}(\tau) \equiv E[f_{y_t|z_t}(Q_{y_t|z_t}(\tau,z_{t})|z_t)\phi(z_{t})\phi(z_{t})^{\prime}]$ for each $t$, which plays a crucial role in the analysis. The asymptotic distribution result is established as follows.

\begin{thm}\label{Thm: AsympDis}
Suppose Assumptions \ref{Ass: DGP}-\ref{Ass: AsymDis} hold. Let $\hat{a}(\tau)$, $\hat{B}(\tau)$ and $\hat{F}(\tau)$ be defined in \eqref{Eqn: Estimators}. Assume (i) $N\to\infty$; (ii) $T\geq K(\tau)+1$ is finite; (iii) $J\to\infty$ with $J^3\xi^{2}_{J}\log^{2} N=o(N)$ and $J^{-\kappa+1}\log N=o(1)$. Let $r_{N}\equiv ({J^{3/4}\xi_{J}^{1/2}\log^{1/2}N})/{N^{1/4}}+J^{-(\kappa-1)/2}\log^{1/2}N + \sqrt{N}J^{-k}\xi_{J}^{-1}$. Then
\begin{align*}
\sup_{\tau\in\mathcal{T}}\|\sqrt{N}[\hat{a}(\tau) - a(\tau)]-\mathbb{G}_{a}(\tau)\|&=O_{p}\left(r_{N}+\delta_{N}\right),\notag\\
\sup_{\tau\in\mathcal{T}}\|\sqrt{N}[\hat{B}(\tau) - B(\tau) H(\tau)]-\mathbb{G}_{B}(\tau)\|_{F}&=O_{p}\left(r_{N}+\delta_{N}\right),\notag\\
\sup_{\tau\in\mathcal{T}}\frac{1}{T}\|\sqrt{N}\{\hat{F}(\tau)-F(\tau)[H(\tau)^{\prime}]^{-1}\}-\mathbb{G}_{F}(\tau)\|_{F}&=O_{p}\left(r_{N}+\delta_{N}\right).
\end{align*}
where $\mathbb{G}_{a}(\tau) = -B(\tau)\mathcal{H}(\tau)\mathbb{G}_{B}(\tau)^{\prime} a(\tau)-[I_{JM}-B(\tau)\mathcal{H}(\tau)\mathcal{H}(\tau)^{\prime}B(\tau)^{\prime}][\mathbb{G}_{B}(\tau)\mathcal{H}(\tau)^{-1}\bar{f}(\tau)$ $-\mathbb{D}(\tau)1_{T}/T]$, $\mathbb{G}_{B}(\tau) = \mathbb{D}(\tau)M_TF(\tau)B(\tau)^{\prime}B(\tau)\mathcal{M}(\tau)/T$, $\mathbb{G}_{F}(\tau) = 1_{T}a(\tau)^{\prime}\mathbb{G}_{B}(\tau)-F(\tau)$ $\times[\mathcal{H}(\tau)^{\prime}]^{-1}\mathbb{G}_{B}(\tau)^{\prime}B(\tau)\mathcal{H}(\tau)+\mathbb{D}(\tau)B(\tau)\mathcal{H}(\tau)$, $\mathbb{D}(\tau) = (J_{1}^{-1}(\tau)\mathbb{N}_1(\tau),\ldots,J_{T}^{-1}(\tau)\mathbb{N}_{T}(\tau))$, and $\mathcal{H}(\tau)$ and $\mathcal{M}(\tau)$ are nonrandom matrices given in Lemma \ref{Lem: TechB1}.
\end{thm}

Theorem \ref{Thm: AsympDis} establishes a strong approximation of $\sqrt{N}[\hat{a}(\tau) - a(\tau)]$, $\sqrt{N}[\hat{B}(\tau) - B(\tau) H(\tau)]$ and $\sqrt{N}\{\hat{F}(\tau)-F(\tau)[H(\tau)^{\prime}]^{-1}\}$ by zero-mean Gaussian processes $\mathbb{G}_{a}(\tau)$, $\mathbb{G}_{B}(\tau)$ and $\mathbb{G}_{F}(\tau)$, where the approximation holds uniformly over $\tau\in\mathcal{T}$. Therefore, $\sqrt{N}[\hat{a}(\tau) - a(\tau)]$, $\sqrt{N}[\hat{B}(\tau) - B(\tau) H(\tau)]$ and $\sqrt{N}\{\hat{F}(\tau)-F(\tau)[H(\tau)^{\prime}]^{-1}\}$ behave like Gaussian random vectors (matrices) for each $\tau\in\mathcal{T}$. We also note that the result is robust to heavy tailed errors, allows for temporal dependence and nonstationarity of $z_{it}$, and does not require $T\to\infty$. For the purpose of inference, we need to estimate the distribution of $\mathbb{G}_{a}(\tau)$, $\mathbb{G}_{B}(\tau)$ and $\mathbb{G}_{F}(\tau)$. We next develop a bootstrap procedure.

\section{Bootstrap Inference}\label{Sec: 4}
In this section, we develop a bootstrap procedure to estimate the distribution of $\mathbb{G}_{a}(\tau)$, $\mathbb{G}_{B}(\tau)$ and $\mathbb{G}_{F}(\tau)$ for the purpose of inference.

We adopt the weighted bootstrap method. To describe the method, consider a set of weights $\{w_i\}_{i\leq N}$ that are i.i.d. draws from the standard exponential distribution independent of the data. For each time period $t$, define the weighted bootstrap estimator $\tilde{Y}^{\ast}_t(\tau)$ as the solution to the weighted quantile regression problem
\begin{align}\label{Eqn: Ytildetstar}
\tilde{Y}^{\ast}_t(\tau)=\amin_{b}\sum_{i=1}^{N}w_{i}\rho_{\tau}(y_{it} - \phi(z_{it})^{\prime}b).
\end{align}
We use the same weights $\{w_{i}\}_{i\leq N}$ for all $t$'s to maintain the temporal dependence of the data. To define the bootstrap estimators of $a(\tau)$, $B(\tau)$ and $F(\tau)$, let $\tilde{Y}^{\ast}(\tau)\hspace{-0.09cm}\equiv\hspace{-0.09cm}(\tilde{Y}^{\ast}_1(\tau),\ldots, \tilde{Y}^{\ast}_T(\tau))$ and $\bar{\tilde{Y}}^{\ast}(\tau)\equiv \sum_{t=1}^{T}\tilde{Y}^{\ast}_t(\tau)/T$. The bootstrap estimators are
\begin{align}\label{Eqn: BootEstimators}
\hat{B}^{\ast}(\tau) &= \tilde{Y}^{\ast}(\tau)M_T\hat{F}(\tau)[\hat{F}(\tau)^{\prime}M_T\hat{F}(\tau)]^{-1},\notag\\
\hat{a}^{\ast}(\tau) &= (I_{JM}-\hat{B}^{\ast}(\tau)[\hat{B}^{\ast}(\tau)^{\prime}\hat{B}^{\ast}(\tau)]^{-1}\hat{B}^{\ast}(\tau)^{\prime})\bar{\tilde{Y}}^{\ast}(\tau),\\
\hat{F}^{\ast}(\tau) & =\tilde{Y}^{\ast}(\tau)^{\prime}\hat{B}^{\ast}(\tau)[\hat{B}^{\ast}(\tau)^{\prime}\hat{B}^{\ast}(\tau)]^{-1}.\notag
\end{align}
An alternative bootstrap estimator is given by $\hat{B}^{\ast\ast}(\tau)$, whose columns are given by the eigenvectors associated with the first $K(\tau)$ largest eigenvalues of $\tilde{Y}^{\ast}(\tau)M_T\tilde{Y}^{\ast}(\tau)^{\prime}/T$. However, the distribution of $\sqrt{N}[\hat{B}^{\ast\ast}(\tau) - \hat{B}(\tau)]$ conditional on the data may not be able to estimate the distribution of $\mathbb{G}_{B}(\tau)$. To see this, we notice that
\begin{align}\label{Eqn: AltBootstrap}
\hat{B}^{\ast\ast}(\tau) = \tilde{Y}^{\ast}(\tau)M_T\hat{F}^{\ast\ast}(\tau)[\hat{F}^{\ast\ast}(\tau)^{\prime}M_T\hat{F}^{\ast\ast}(\tau)]^{-1},
\end{align}
where $\hat{F}^{\ast\ast}(\tau) = \tilde{Y}^{\ast}(\tau)^{\prime}\hat{B}^{\ast\ast}(\tau)$. Thus, we may obtain an asymptotic expansion of $\sqrt{N}[\hat{B}^{\ast\ast}(\tau)-B(\tau)H^{\ast}(\tau)]$ similar to that of $\sqrt{N}[\hat{B}(\tau)-B(\tau)H(\tau)]$, where $H^{\ast}(\tau)= [F(\tau)^{\prime}M_T\hat{F}^{\ast\ast}(\tau)][\hat{F}^{\ast\ast}(\tau)^{\prime}M_T\hat{F}^{\ast\ast}(\tau)]^{-1}$. However, the difference between $H^{\ast}(\tau)$ and $H(\tau)$ may not be negligible, rendering the failure of the bootstrap estimator $\hat{B}^{\ast\ast}(\tau)$. To fix the problem, we use $\hat{F}(\tau)$ to replace $\hat{F}^{\ast\ast}(\tau)$ in \eqref{Eqn: AltBootstrap} to obtain $\hat{B}^{\ast}(\tau)$. In other words, we do not need to re-estimate $F(\tau)$ in estimating $B(\tau)$ in the bootstrap. After obtaining $\hat{B}^{\ast}(\tau)$, we just follow the plug-in rule to obtain $\hat{a}^{\ast}(\tau)$ and $\hat{F}^{\ast}(\tau)$, since $\hat{a}(\tau) = [I_{JM}-\hat{B}(\tau)\hat{B}(\tau)^{\prime}]\bar{\tilde{Y}}(\tau)=[I_{JM}-\hat{B}(\tau)[\hat{B}(\tau)^{\prime}\hat{B}(\tau)]^{-1}\hat{B}(\tau)^{\prime}]\bar{\tilde{Y}}(\tau)$ and $\hat{F}(\tau)= \tilde{Y}(\tau)^{\prime}\hat{B}(\tau)= \tilde{Y}(\tau)^{\prime}\hat{B}(\tau)[\hat{B}(\tau)^{\prime}\hat{B}(\tau)]^{-1}$.

We propose to estimate the distribution of $\mathbb{G}_{a}(\tau)$, $\mathbb{G}_{B}(\tau)$ and $\mathbb{G}_{F}(\tau)$ by the distribution of $\sqrt{N}[\hat{a}^{\ast}(\tau) - \hat{a}(\tau)]$, $\sqrt{N}[\hat{B}^{\ast}(\tau) - \hat{B}(\tau)]$ and $\sqrt{N}[\hat{F}^{\ast}(\tau)-\hat{F}(\tau)]$ conditional on the data. To establish the validity, we impose the following assumption.

\begin{ass}[Bootstrap weight]\label{Ass: BootWeight}
(i) $\{w_{i}\}_{i\leq N}$ is an i.i.d. sequence of random variables with the standard exponential distribution, and is independent of $\{y_{it},z_{it}\}_{i\leq N,t\leq T}$; (ii) there is a zero-mean Gaussian process $\mathbb{N}^{\ast}(\cdot)= (\mathbb{N}^{\ast}_1(\cdot),\ldots,\mathbb{N}^{\ast}_T(\cdot))$ on $\mathcal{T}$ conditional on $\{y_{it},z_{it}\}_{i\leq N,t\leq T}$ with a.s. continuous path such that the covariance function of $\mathbb{N}^{\ast}(\cdot)$ coincides with that of $\mathbb{U}(\cdot)$ and $\sup_{\tau\in\mathcal{T}}\|\mathbb{U}^{\ast}(\tau)-\mathbb{N}^{\ast}(\tau)\|_{F} = O_{p}(\delta_{N})$, where $\mathbb{U}^{\ast}(\cdot) = (\mathbb{U}^{\ast}_1(\cdot),\ldots, \mathbb{U}^{\ast}_T(\cdot))$ and $\mathbb{U}^{\ast}_t(\tau) = \sum_{i=1}^{N}(w_i-1)\phi(z_{it})(\tau-1\{u_{it}\leq \tau\})/\sqrt{N}$.
\end{ass}

Assumption \ref{Ass: BootWeight}(ii) requires that $\mathbb{U}^{\ast}(\cdot)$ be coupled to a Gaussian process conditional on the data. The condition can also be verified by resorting to the Yurinskii's coupling; see Proposition \ref{Pro: StrongBoot}. Here, we use the same weights $\{w_{i}\}_{i\leq N}$ for all $t$'s to maintain the temporal dependence.

The validity of the bootstrap is established as follows.

\begin{thm}\label{Thm: BootDis}
Suppose Assumptions \ref{Ass: DGP}-\ref{Ass: AsymDis} and \ref{Ass: BootWeight} hold. Let $\hat{a}(\tau)$, $\hat{B}(\tau)$, $\hat{F}(\tau)$, $\hat{a}^{\ast}(\tau)$, $\hat{B}^{\ast}(\tau)$ and $\hat{F}^{\ast}(\tau)$ be given in \eqref{Eqn: Estimators} and \eqref{Eqn: BootEstimators}. Assume (i) $N\to\infty$; (ii) $T\geq K(\tau)+1$ is finite; (iii) $J\to\infty$ with $J^3\xi^{2}_{J}\log^{2} N=o(N)$ and $J^{-\kappa+1}\log N=o(1)$. Then
\begin{align*}
\sup_{\tau\in\mathcal{T}}\|\sqrt{N}[\hat{a}^{\ast}(\tau) - \hat{a}(\tau)]-\mathbb{G}^{\ast}_{a}(\tau)\|&=O_{p}\left(r_{N}+\delta_{N}\right),\notag\\
\sup_{\tau\in\mathcal{T}}\|\sqrt{N}[\hat{B}^{\ast}(\tau) - \hat{B}(\tau)]-\mathbb{G}^{\ast}_{B}(\tau)\|_{F}&=O_{p}\left(r_{N}+\delta_{N}\right),\notag\\
\sup_{\tau\in\mathcal{T}}\frac{1}{T}\|\sqrt{N}[\hat{F}^{\ast}(\tau)-\hat{F}(\tau)]-\mathbb{G}^{\ast}_{F}(\tau)\|_{F}&=O_{p}\left(r_{N}+\delta_{N}\right),
\end{align*}
where $\mathbb{G}^{\ast}_{a}(\tau) = -B(\tau)\mathcal{H}(\tau)\mathbb{G}^{\ast}_{B}(\tau)^{\prime} a(\tau)-[I_{JM}-B(\tau)\mathcal{H}(\tau)\mathcal{H}(\tau)^{\prime}B(\tau)^{\prime}][\mathbb{G}^{\ast}_{B}(\tau)\mathcal{H}(\tau)^{-1}\bar{f}(\tau)$ $-\mathbb{D}^{\ast}(\tau)1_{T}/T]$, $\mathbb{G}^{\ast}_{B}(\tau) = \mathbb{D}^{\ast}(\tau)M_TF(\tau)B(\tau)^{\prime}B(\tau)\mathcal{M}(\tau)/T$, $\mathbb{G}^{\ast}_{F}(\tau) = 1_{T}a(\tau)^{\prime}\mathbb{G}^{\ast}_{B}(\tau)-F(\tau)[\mathcal{H}(\tau)^{\prime}]^{-1}\mathbb{G}^{\ast}_{B}(\tau)^{\prime}B(\tau)\mathcal{H}(\tau)+\mathbb{D}^{\ast}(\tau)B(\tau)\mathcal{H}(\tau)$, $\mathbb{D}^{\ast}(\tau)=(J_{1}^{-1}(\tau)\mathbb{N}^{\ast}_1(\tau),\ldots,J_{T}^{-1}(\tau)\times $ $\mathbb{N}^{\ast}_{T}(\tau))$, and $\mathcal{H}(\tau)$ and $\mathcal{M}(\tau)$ are nonrandom matrices given in Lemma \ref{Lem: TechB1}. In particular, $\mathbb{G}^{\ast}_{a}(\tau)$, $\mathbb{G}^{\ast}_{B}(\tau)$ and $\mathbb{G}^{\ast}_{F}(\tau)$ are zero-mean Gaussian processes conditional on $\{y_{it},z_{it}\}_{i\leq N,t\leq T}$, and their covariance functions coincide with those of $\mathbb{G}_{a}(\tau)$, $\mathbb{G}_{B}(\tau)$ and $\mathbb{G}_{F}(\tau)$, respectively.
\end{thm}

Theorem \ref{Thm: BootDis} implies that the distribution of $\sqrt{N}[\hat{a}^{\ast}(\tau) - \hat{a}(\tau)]$, $\sqrt{N}[\hat{B}^{\ast}(\tau) - \hat{B}(\tau)]$ and $\sqrt{N}[\hat{F}^{\ast}(\tau)-\hat{F}(\tau)]$ conditional on the data can well approximate the distribution of $\mathbb{G}_{a}(\tau)$, $\mathbb{G}_{B}(\tau)$ and $\mathbb{G}_{F}(\tau)$, which is equal to the distribution of $\mathbb{G}^{\ast}_{a}(\tau)$, $\mathbb{G}^{\ast}_{B}(\tau)$ and $\mathbb{G}^{\ast}_{F}(\tau)$, where the approximation holds uniformly over $\tau\in\mathcal{T}$. In practice, the latter can be simulated. We stress that the result is robust to heavy tailed errors, allows for temporal dependence and nonstationarity of $z_{it}$, and does not require $T\to\infty$.

\textbf{Inference on  $\alpha(\tau,\cdot)$ and $\beta(\tau,\cdot)$}. We can immediately use the results in Theorems \ref{Thm: AsympDis} and \ref{Thm: BootDis} to conduct inference on the coefficients of $\phi_{j}(z_{it,m})$'s in $\alpha(\tau,z_{it})$ and $\beta(\tau,z_{it})$. First, we can test whether $\phi_{j}(z_{it,m})$'s are individually significant in $\alpha(\tau,z_{it})$. Second, we can test whether $\alpha(\tau,\cdot) = 0$ by comparing $N\hat{a}(\tau)^{\prime}\hat{a}(\tau)$ with the $1-\alpha$ quantile of $N[\hat{a}^{\ast}(\tau)-\hat{a}(\tau)]^{\prime}[\hat{a}^{\ast}(\tau)-\hat{a}(\tau)]$ conditional on the data for $0<\alpha<1$. Third, we can test joint significance of $\phi_{j}(z_{it,m})$'s in $\beta(\tau,z_{it})$ for some given $j$'s and $m$'s, which is equivalent to whether several rows of $B(\tau)H(\tau)$ are jointly equal to zero.
However, we are not able to conduct individual significance test for each entry of $\beta(z_{it})$ due to the lack of identification, and we cannot use the results in Theorems \ref{Thm: AsympDis} and \ref{Thm: BootDis} to test whether  $\beta(\tau,\cdot) = 0$ due to the nonsingularity requirement in Assumption \ref{Ass: InterceptLoadingsFactors}(iii).

\begin{rem}\label{Rem: UnbalancedBoot}
The bootstrap procedure is also applicable for unbalanced panels. Notice that the crucial step is to obtain $\tilde{Y}^{\ast}_t(\tau)$. In the case of unbalanced panels, we may obtain $\tilde{Y}^{\ast}_t(\tau)$ by using available observations in time period $t$, where we need to generate the  bootstrap weights $\{w_{i}\}_{i\leq N_{\max}}$ and $N_{\max}$ is the total number of observation units. Theorem \ref{Thm: BootDis} continues to hold when $\min_{t\leq T} N_t \to\infty$, where $N_t$ is the number of available observations in time period $t$.
\end{rem}

\section{Estimating the Number of Factors}\label{Sec: 5}
In this section, we consider two estimators for the number of factors $K(\tau)$. The first one is obtained by maximizing the ratio of two adjacent eigenvalues; the second one is obtained by counting the number of ``large'' eigenvalues. Let $\lambda_{k}(\tilde{Y}(\tau)M_T\tilde{Y}(\tau)^{\prime}/T)$ be the $k$th largest eigenvalue of the $JM\times JM$ matrix $\tilde{Y}(\tau)M_T\tilde{Y}(\tau)^{\prime}/T$. We define the first estimator of $K(\tau)$ as
\begin{align}\label{Eqn: Kestiamtor}
\hat{K}(\tau) = \amax_{1\leq k\leq K_{\max}}\frac{\lambda_{k}(\tilde{Y}(\tau)M_T\tilde{Y}(\tau)^{\prime}/T)}{\lambda_{k+1}(\tilde{Y}(\tau)M_T\tilde{Y}(\tau)^{\prime}/T)},
\end{align}
where $K_{\max}$ is a pre-specified integer. Our second estimator of $K(\tau)$ is given by
\begin{align}\label{Eqn: Kestiamtoralt}
\tilde{K}(\tau)= \#\{1\leq k\leq JM: \lambda_{k}(\tilde{Y}(\tau)M_T\tilde{Y}(\tau)^{\prime}/T)\geq \lambda_{N}\},
\end{align}
where $\# A$ denotes the number of elements in set $A$ and $0<\lambda_{N}\to 0$ is a pre-specified tuning parameter.

It is straightforward to establish the consistency of $\tilde{K}(\tau)$. However, to establish the consistency of $\hat{K}(\tau)$, we impose the following assumption.

\begin{ass}[Determination of $K(\tau)$]\label{Ass: NumFactors}
$\sup_{\tau\in\mathcal{T}}J^{2}/[N\lambda_{\min}(\mathbb{E}(\tau)\mathbb{E}(\tau)^{\prime})]=o_{p}(1)$, where $\mathbb{E}(\tau)= (J_{1}^{-1}(\tau)\mathbb{U}_1,\ldots, J_{T}^{-1}(\tau)\mathbb{U}_T)$.
\end{ass}

Assumption \ref{Ass: NumFactors} restricts the dependence of the data across $t$. The consistency of the estimators is established as follows.

\begin{thm}\label{Thm: NumFactors}
(A) Suppose Assumptions \ref{Ass: DGP}-\ref{Ass: InterceptLoadingsFactors} and \ref{Ass: NumFactors} hold. Let $\hat{K}(\tau)$ be given in \eqref{Eqn: Kestiamtor}. Assume (i) $N\to\infty$; (ii) $T\geq 2[K(\tau)+1]$ is finite; (iii) $J\to\infty$ with $r_{N}\sqrt{J} = o(1)$. Assume that $K(\tau)\leq K_{\max}\leq T-K(\tau)-1$. Then
\[P(\hat{K}(\tau)={K}(\tau) \text{ for all } \tau\in\mathcal{T})\to 1.\]
(B) Suppose Assumptions \ref{Ass: DGP}-\ref{Ass: InterceptLoadingsFactors} hold. Let $\tilde{K}(\tau)$ be given in \eqref{Eqn: Kestiamtoralt}. Assume (i) $N\to\infty$; (ii) $T\geq K(\tau)+1$ is finite; (iii) $J\to\infty$ with $J\xi^{2}_{J}\log^{2} N=o(N)$ and $J^{-\kappa}\log N=o(1)$; (iv) $0<\lambda_N\to 0$ and $\lambda_N\min\{N/J,J^{2\kappa}\xi_{J}^{2}\}\to\infty$. Then
\[P(\tilde{K}(\tau)={K}(\tau) \text{ for all } \tau\in\mathcal{T})\to 1.\]
\end{thm}

Theorem \ref{Thm: NumFactors} demonstrates that both $\hat{K}(\tau)$ and $\tilde{K}(\tau)$ are consistent uniformly over $\tau\in\mathcal{T}$. Again, the result does not require $T\to\infty$. In practice, one needs to choose $K_{\max}$ and $\lambda_{N}$. In our simulation studies below, we let $K_{\max}$ be the largest integer no larger than $JM/2$ and $\lambda_{N}=1/\log N$, and $\hat{K}(\tau)$ and $\tilde{K}(\tau)$ have satisfactory performance.

\section{Monte Carlo Simulations}\label{Sec: 6}
In this section, we conduct several Monte Carlo simulations to investigate the finite sample performance of our estimators and bootstrap test. In particular, we focus on three relevant issues: (i) how well the QR-PCA performs in relation to \citet{Chenetal_SeimiparametricFactor_2021}'s regressed-PCA when the distribution of the idiosyncratic errors exhibits heavy tails; (ii) how well the QR-PCA performs in estimating the extra quantile factors that cannot be captured by the regressed-PCA; (iii) how well our bootstrap procedure performs in testing whether $\alpha(\tau,\cdot)=0$. In addition, we examine whether increasing $T$ helps improve the accuracy of our estimators and bootstrap test, and check how restrictive is the cross-sectional independence, imposed in Assumption \ref{Ass: DGP}(i).

\subsection{Robustness to Heavy Tailed Errors}\label{Sec: 61}
To examine (i), we consider the following data generating process:
\begin{align}\label{Eqn: SimulationDGP1}
y_{it} = \alpha(z_{it}) + \beta(z_{it})^{\prime}f_{t} + e_{it},i=1,\ldots,N, t=1,\ldots,T,
\end{align}
where $f_{t} = 0.3 f_{t-1} + \eta_t$ with $\eta_t$'s being independent draws from $N(0,I_2)$ and $f_0\sim N(0,I_2/0.91)$, $e_{it}$'s are independent draws from $t_{\nu}$ for $\nu=1,2,3$ with $t_{\nu}$ denoting the Student's $t$-distribution with $\nu$ degrees of freedom, and $z_{it}$'s, $\alpha(\cdot)$ and $\beta(\cdot)$ are specified as follows. Let $z_{it,1}=\sigma_{t}\cdot u_{it,1}, z_{it,2} = 0.3z_{i(t-1),2} + u_{it,2} \text{ and } z_{it,3}=u_{it,3}$, where $u_{it} = (u_{it,1},u_{it,2},u_{it,3})^{\prime}$'s are independent draws from $N(0,I_3)$, $\sigma_{t}$'s are independent draws from $U(1,2)$, and $z_{i0,2}$'s are independent draws from $N(0,1)$. Thus, the three entries of $z_{it}$ are varying over $t$ in three different ways. Here, $f_0$, $\eta_t$'s, $e_{it}$'s, $u_{it}$'s, $\sigma_{t}$'s, and $z_{i0,2}$'s are mutually independent. We assume
\begin{align}\label{Eqn: alphabeta1}
\alpha(z_{it}) = z_{it,1} + 0.5 z_{it,1}^{2} \text{ and } \beta(z_{it}) = (z_{it,2} + 0.5 z_{it,2}^{2}, 2z_{it,3} + z_{it,3}^{2})^{\prime}.
\end{align}
Thus, $M=3$, $K(\tau) = 2$, $\alpha(\tau,z_{it}) = \alpha(z_{it}) + Q_{t_\nu}(\tau)$, $\beta(\tau,z_{it})=\beta(z_{it})$ and $f_{t}(\tau) = f_{t}$ for $\tau\in(0,1)$, where $Q_{t_\nu}(\tau)$ is the $\tau$th quantile of $t_{\nu}$. Therefore, we may use the QR-PCA at $\tau = 0.5$ to estimate $\alpha(\cdot)$, $\beta(\cdot)$ and $f_{t}$, since $\alpha(0.5,z_{it}) = \alpha(z_{it})$ and the QR-PCA does not require the moments of $e_{it}$ to exist. On the other hand, the regressed-PCA may not work due to the possible heavy tail of the distribution of $e_{it}$.

To implement the QR-PCA and the regressed-PCA, we let $\phi(z_{it}) = (z_{it,1},z_{it,1}^{2},$ $z_{it,2},z_{it,2}^{2},z_{it,3},z_{it,3}^{2})^{\prime}$, leading to a zero sieve approximation error. Denote the regressed-PCA estimators by $\hat{a}$, $\hat{B}$, $\hat{F}$, $\hat{K}$ and $\tilde{K}$ (two estimators for the number of factors). We let $K_{\max}$ be the largest integer no larger than $JM/2$ in the implementation of $\hat{K}(0.5)$, and choose $\lambda_{NT} = 1/\log(N)$ in the implementation of $\tilde{K}(0.5)$ and $\tilde{K}$. We investigate the performance of $\hat{a}(0.5)$, $\hat{B}(0.5)$, $\hat{F}(0.5)$, $\hat{K}(0.5)$ and $\tilde{K}(0.5)$ under different $(N,T)$'s, and their comparison with $\hat{a}$, $\hat{B}$, $\hat{F}$, $\hat{K}$ and $\tilde{K}$. We report the correct rates of $\hat{K}(0.5)$, $\tilde{K}(0.5)$, $\hat{K}$ and $\tilde{K}$ in Table \ref{Tab: CorrectRate1}, and the mean square errors of $\hat{a}(0.5)$, $\hat{B}(0.5)$, $\hat{F}(0.5)$, $\hat{a}$, $\hat{B}$ and $\hat{F}$ in Table \ref{Tab: MSE1}. We let the number of simulation replications equal to $1,000$.

The main findings are as follows. First, as shown in Table \ref{Tab: CorrectRate1}, our estimators can select the correct number of factors with high accuracy in all cases, while the estimators by the regressed-PCA fail when $e_{it}$'s are $t_1$. Second, as shown in Table \ref{Tab: MSE1}, the QR-PCA performs very well in all cases, while the regressed-PCA fails when $e_{it}$'s are $t_1$. Specifically, the mean square errors of the QR-PCA estimators are quickly getting closer to zero as $N$ increases even when $T=10$. The mean square errors of the regressed-PCA estimators are persistently large in all combinations of $(N, T)$ when $e_{it}$'s are $t_1$, though the large means square errors vanish when the distribution of $e_{it}$'s changes from $t_1$ to $t_3$. Third, as shown in Tables \ref{Tab: CorrectRate1} and \ref{Tab: MSE1}, increasing $T$ helps improve the accuracy of our estimators. To sum up, our estimators have satisfactory finite sample performance in the presence of heavy tailed errors, while the regressed-PCA may break down.

\setlength{\tabcolsep}{3.5pt}
\begin{table}[!htbp]
\footnotesize
\centering
\begin{threeparttable}
\renewcommand{\arraystretch}{1.5}
\caption{Correct rates of $\hat{K}(0.5)$, $\tilde{K}(0.5)$, $\hat{K}$ and $\tilde{K}$}\label{Tab: CorrectRate1}
\begin{tabular}{cccccccccc}
\hline\hline
&Value of $\nu$&(N,T)&&$\hat{K}(0.5)$&$\tilde{K}(0.5)$&&$\hat{K}$&$\tilde{K}$&\\
\cline{2-9}
&\multirow{9}{*}{$\nu=1$}&$(50,10)$ &&0.931&0.976&&0.005&0.000&\\
&&$(100,10)$                        &&0.997&1.000&&0.002&0.000&\\
&&$(200,10)$                        &&1.000&1.000&&0.002&0.000&\\
&&$(500,10)$                        &&1.000&1.000&&0.000&0.000&\\
&&$(50,50)$                         &&1.000&0.993&&0.000&0.000&\\
&&$(100,50)$                        &&1.000&1.000&&0.000&0.000&\\
&&$(200,50)$                        &&1.000&1.000&&0.000&0.000&\\
&&$(500,50)$                        &&1.000&1.000&&0.000&0.000&\\
&&$(50,100)$                        &&0.999&0.998&&0.000&0.000&\\
&&$(100,100)$                       &&1.000&1.000&&0.000&0.000&\\
&&$(200,100)$                       &&1.000&1.000&&0.000&0.000&\\
&&$(500,100)$                       &&1.000&1.000&&0.000&0.000&\\
\cline{2-9}
&\multirow{9}{*}{$\nu=2$}&$(50,10)$ &&0.990&1.000&&0.811&0.851&\\
&&$(100,10)$                        &&1.000&1.000&&0.909&0.940&\\
&&$(200,10)$                        &&1.000&1.000&&0.944&0.966&\\
&&$(500,10)$                        &&1.000&1.000&&0.967&0.985&\\
&&$(50,50)$                         &&1.000&1.000&&0.894&0.796&\\
&&$(100,50)$                        &&1.000&1.000&&0.932&0.903&\\
&&$(200,50)$                        &&1.000&1.000&&0.950&0.946&\\
&&$(500,50)$                        &&1.000&1.000&&0.967&0.976&\\
&&$(50,100)$                        &&1.000&1.000&&0.900&0.769&\\
&&$(100,100)$                       &&1.000&1.000&&0.936&0.899&\\
&&$(200,100)$                       &&1.000&1.000&&0.967&0.960&\\
&&$(500,100)$                       &&1.000&1.000&&0.976&0.979&\\
\cline{2-9}
&\multirow{9}{*}{$\nu=3$}&$(50,10)$ &&0.994&1.000&&0.965&0.991&\\
&&$(100,10)$                        &&1.000&1.000&&0.994&0.999&\\
&&$(200,10)$                        &&1.000&1.000&&0.999&1.000&\\
&&$(500,10)$                        &&1.000&1.000&&1.000&1.000&\\
&&$(50,50)$                         &&1.000&1.000&&0.993&0.993&\\
&&$(100,50)$                        &&1.000&1.000&&0.998&0.999&\\
&&$(200,50)$                        &&1.000&1.000&&0.999&1.000&\\
&&$(500,50)$                        &&1.000&1.000&&1.000&1.000&\\
&&$(50,100)$                        &&1.000&1.000&&0.999&0.999&\\
&&$(100,100)$                       &&1.000&1.000&&0.999&1.000&\\
&&$(200,100)$                       &&1.000&1.000&&1.000&1.000&\\
&&$(500,100)$                       &&1.000&1.000&&1.000&1.000&\\
\hline\hline
\end{tabular}
\end{threeparttable}
\end{table}%

\setlength{\tabcolsep}{3.5pt}
\begin{table}[!htbp]
\footnotesize
\centering
\begin{threeparttable}
\renewcommand{\arraystretch}{1.38}
\caption{Mean square errors of  $\hat{a}(0.5)$ , $\hat{B}(0.5)$, $\hat{F}(0.5)$, $\hat{a}$, $\hat{B}$ and $\hat{F}$\tnote{\dag}}\label{Tab: MSE1}
\begin{tabular}{cccccccccccc}
\hline\hline
&Value of $\nu$&(N,T)&&$\hat{a}(0.5)$&$\hat{B}(0.5)$&$\hat{F}(0.5)$&&$\hat{a}$&$\hat{B}$&$\hat{F}$&\\
\cline{2-11}
&\multirow{9}{*}{$\nu=1$}&$(50,10)$ &&0.0433&0.0544&0.2097&&1.7501&1.5155&$7\times10^6$      &\\
&&$(100,10)$                        &&0.0139&0.0164&0.0614&&1.5241&1.4862&$9\times 10^7$     &\\
&&$(200,10)$                        &&0.0052&0.0065&0.0238&&1.4180&1.4495&$2\times10^6$      &\\
&&$(500,10)$                        &&0.0019&0.0023&0.0082&&1.8262&1.4358&$1\times10^7$      &\\
&&$(50,50)$                         &&0.0084&0.0133&0.2080&&2.3530&1.7954&$8\times 10^7$     &\\
&&$(100,50)$                        &&0.0026&0.0031&0.0646&&1.9289&1.7711&$2\times 10^8$     &\\
&&$(200,50)$                        &&0.0009&0.0011&0.0258&&3.2888&1.7679&$7\times 10^8$     &\\
&&$(500,50)$                        &&0.0003&0.0004&0.0090&&1.9535&1.7619&$2\times 10^8$     &\\
&&$(50,100)$                        &&0.0028&0.0071&0.2058&&1.2344&1.8606&\ $1\times 10^{10}$&\\
&&$(100,100)$                       &&0.0009&0.0015&0.0645&&1.0370&1.8420&$6\times 10^8$     &\\
&&$(200,100)$                       &&0.0003&0.0005&0.0258&&0.9350&1.8270&\ $1\times 10^{10}$&\\
&&$(500,100)$                       &&0.0001&0.0002&0.0090&&1.0671&1.8149&$9\times 10^8$     &\\
\cline{2-11}
&\multirow{9}{*}{$\nu=2$}&$(50,10)$ &&0.0232&0.0292&0.1056&&0.0755&0.1225&23.698&\\
&&$(100,10)$                        &&0.0090&0.0107&0.0405&&0.0422&0.0606&267.95&\\
&&$(200,10)$                        &&0.0039&0.0047&0.0173&&0.0220&0.0389&6.3480&\\
&&$(500,10)$                        &&0.0014&0.0018&0.0065&&0.0106&0.0174&1.0875&\\
&&$(50,50)$                         &&0.0044&0.0059&0.1123&&0.0274&0.0724&1840.6&\\
&&$(100,50)$                        &&0.0017&0.0019&0.0432&&0.0116&0.0349&14.634&\\
&&$(200,50)$                        &&0.0007&0.0008&0.0188&&0.0084&0.0186&2.9146&\\
&&$(500,50)$                        &&0.0003&0.0003&0.0069&&0.0034&0.0075&0.1675&\\
&&$(50,100)$                        &&0.0015&0.0032&0.1121&&0.0125&0.0639&30.337&\\
&&$(100,100)$                       &&0.0006&0.0009&0.0433&&0.0073&0.0275&4.6039&\\
&&$(200,100)$                       &&0.0002&0.0004&0.0189&&0.0024&0.0192&11.748&\\
&&$(500,100)$                       &&0.0001&0.0001&0.0070&&0.0007&0.0025&0.0636&\\
\cline{2-11}
&\multirow{9}{*}{$\nu=3$}&$(50,10)$ &&0.0197&0.0233&0.0868&&0.0289&0.0341&0.1335&\\
&&$(100,10)$                        &&0.0080&0.0097&0.0349&&0.0118&0.0141&0.0528&\\
&&$(200,10)$                        &&0.0034&0.0043&0.0159&&0.0053&0.0066&0.0254&\\
&&$(500,10)$                        &&0.0012&0.0015&0.0059&&0.0020&0.0024&0.0096&\\
&&$(50,50)$                         &&0.0036&0.0049&0.0929&&0.0051&0.0093&0.1467&\\
&&$(100,50)$                        &&0.0014&0.0017&0.0378&&0.0021&0.0028&0.0565&\\
&&$(200,50)$                        &&0.0007&0.0007&0.0171&&0.0010&0.0012&0.0261&\\
&&$(500,50)$                        &&0.0002&0.0003&0.0063&&0.0004&0.0004&0.0101&\\
&&$(50,100)$                        &&0.0012&0.0023&0.0935&&0.0017&0.0040&0.1301&\\
&&$(100,100)$                       &&0.0005&0.0008&0.0374&&0.0007&0.0013&0.0559&\\
&&$(200,100)$                       &&0.0002&0.0003&0.0170&&0.0003&0.0005&0.0265&\\
&&$(500,100)$                       &&0.0001&0.0001&0.0064&&0.0001&0.0002&0.0102&\\
\hline\hline
\end{tabular}
\begin{tablenotes}
      \item[\dag] {\footnotesize{The mean square errors of $\hat{a}$, $\hat{B}$ and $\hat{F}$ are given by $\sum_{\ell=1}^{1000}\|\hat{a}^{(\ell)}-a(0.5)\|^2/1000$, $\sum_{\ell=1}^{1000}\|\hat{B}^{(\ell)}-B(0.5)H(0.5)^{(\ell)}\|_{F}^2/1000$ and $\sum_{\ell=1}^{1000}\|\hat{F}^{(\ell)}- F(0.5)(H(0.5)^{{(\ell)}\prime})^{-1}$ $\|_{F}^2/{1000T}$, where $\hat{a}^{(\ell)}$, $\hat{B}^{(\ell)}$ and $\hat{F}^{(\ell)}$ denote estimates in the $\ell$th simulation replication, and $H(0.5)^{(\ell)}\equiv (F(0.5)^{\prime}M_T\hat{F}^{(\ell)})(\hat{F}^{{(\ell)}\prime}M_T\hat{F}^{(\ell)})^{-1}$ is a rotational transformation matrix. The mean square errors of $\hat{a}(0.5)$, $\hat{B}(0.5)$ and $\hat{F}(0.5)$ are defined similarly.}}
    \end{tablenotes}
\end{threeparttable}
\end{table}%

\subsection{Extracting Extra Quantile Factors}\label{Sec: 62}
To examine (ii), we consider the following data generating process:
\begin{align}\label{Eqn: SimulationDGP2}
y_{it} = \alpha(z_{it}) + \beta(z_{it})^{\prime}f_{t} + 3|g_t|\cdot e_{it},t=1,\ldots,T, i=1,\ldots,N,
\end{align}
where $z_{it}$'s, $\alpha(\cdot)$, $\beta(\cdot)$ and $f_{t}$ are the same as in Section \ref{Sec: 61}, $g_{t}$'s are independent draws from $U(0,1)$,
and $e_{it}$'s are specified as follows. To allow for both cross-sectional and serial correlations, by following \citet{BaiNg_NumberofFactors_2002} we assume
\begin{align}\label{Eqn: error2}
e_{it} = \rho \cdot e_{i,t-1} + v_{it} + \omega \cdot \sum_{j=i-L,j\neq i}^{i+L} v_{jt},
\end{align}
where $v_{it}$'s are independent draws from $N(0,1)$ or $t_{3}$. The autoregressive coefficient $\rho$ captures the serial correlation of $e_{it}$, while the parameters $\omega$ and $L$ capture the cross-sectional correlations of $e_{it}$. Thus, $M=3$; $\alpha(\tau,z_{it}) = \alpha(z_{it})$ for $\tau\in(0,1)$; $K(\tau) =3$, $\beta(\tau,z_{it})=(\beta(z_{it})\prime, 3Q_{e}(\tau))^{\prime}$ and $f_{t}(\tau) = (f_{t}^{\prime},|g_{t}|)^{\prime}$ for $\tau\neq 0.5$, where $Q_{e}(\tau)$ is the $\tau$th quantile of the distribution of $e_{it}$; $K(\tau) =2$, $\beta(\tau,z_{it})=\beta(z_{it})$ and $f_{t}(\tau) = f_{t}$ for $\tau= 0.5$. In particular, we may use the QR-PCA at $\tau\neq 0.5$ to extract the idiosyncratic volatility factor $|g_t|$, which cannot be extracted by the regressed-PCA.

We implement the QR-PCA at $\tau = 0.25, 0.5, 0.75$ for three models embedded in the previous specification: (M1) independent errors: $\rho =0$, $\omega=0$ and $v_{it}$'s are independent draws from $N(0,1)$; (M2): independent errors with heavy tails: $\rho =0$, $\omega=0$ and $v_{it}$'s are independent draws from $t_{3}$; (M3): serially and cross-sectionally correlated errors: $\rho =0.2$, $\omega=0.2$, $L=3$ and $v_{it}$'s are independent draws from $N(0,1)$. In particular, (M3) allows us to check how restrictive the cross-sectional independence is. To implement the QR-PCA, we choose $\phi(z_{it}) = (1,z_{it,1},z_{it,1}^{2},$ $z_{it,2},z_{it,2}^{2},z_{it,3},z_{it,3}^{2})^{\prime}$, leading to a zero sieve approximation error. We let $K_{\max}$ be the largest integer no larger than $JM/2$ in the implementation of $\hat{K}(\tau)$, and choose $\lambda_{NT} = 1/\log(N)$ in the implementation of $\tilde{K}(\tau)$. We investigate the performance of $\hat{a}(\tau)$, $\hat{B}(\tau)$, $\hat{F}(\tau)$, $\hat{K}(\tau)$ and $\tilde{K}(\tau)$ under different $(N,T)$'s. We report the mean square errors of $\hat{a}(\tau)$, $\hat{B}(\tau)$ and $\hat{F}(\tau)$ in Table \ref{Tab: MSE2}, and the correct rates of $\hat{K}(\tau)$ and $\tilde{K}(\tau)$ in Table \ref{Tab: CorrectRate2}. The number of simulation replications is set to $1,000$.

The main findings are as follows. First, as shown in Table \ref{Tab: CorrectRate2}, our estimators can select the correct number of factors with high accuracy in most cases. In particular, the estimators at $\tau=0.25$ and $\tau=0.75$ can extract the variance factor $|g_t|$ in addition to the mean factors $f_{t}$ with high accuracy. It should be noted that the accuracy at $\tau=0.5$ may be low when $N=50$ in the case of model M3. However, such low accuracy vanishes as $N$ increases. Second, as shown in Table \ref{Tab: MSE2}, the mean square errors of the QR-PCA estimators are getting closer to zero as $N$ increases even when $T=10$ in all three models, regardless of the value of $\tau$. It should be noted that the mean square errors of the factor estimators at $\tau=0.25$ and $0.75$ may be large when $N=50$. However, the mean square errors quickly get closer to zero as $N$ increases. Third, as shown in Tables \ref{Tab: CorrectRate2} and \ref{Tab: MSE2}, increasing $T$ helps improve the accuracy of our estimators. Fourth, as shown in third panels of Tables \ref{Tab: CorrectRate2} and \ref{Tab: MSE2}, the performance of our estimators is encouraging in the presence of cross-sectional dependence, although all of our theories impose cross-sectional independence. To sum up, our estimators have satisfactory finite sample performance in choosing the number of quantile-dependent factors and in estimating them.

\setlength{\tabcolsep}{3.5pt}
\begin{table}[!htbp]
\footnotesize
\centering
\begin{threeparttable}
\renewcommand{\arraystretch}{1.5}
\caption{Correct rates of $\hat{K}(\tau)$ and $\tilde{K}(\tau)$}\label{Tab: CorrectRate2}
\begin{tabular}{ccccccccccccc}
\hline\hline
&\multirow{2}{*}{Model}&\multirow{2}{*}{(N,T)}&&\multicolumn{2}{c}{$\tau = 0.25$}&&\multicolumn{2}{c}{$\tau = 0.5$}&&\multicolumn{2}{c}{$\tau = 0.75$}&\\
\cline{5-6}\cline{8-9}\cline{11-12}
&&&&$\hat{K}(\tau)$&$\tilde{K}(\tau)$&&$\hat{K}(\tau)$&$\tilde{K}(\tau)$&&$\hat{K}(\tau)$&$\tilde{K}(\tau)$&\\
\cline{2-12}
&\multirow{9}{*}{M1}&$(50,10)$    &&0.580&0.751&&0.652&0.763&&0.583&0.735&\\
&&$(100,10)$                      &&0.700&0.643&&0.875&0.973&&0.701&0.639&\\
&&$(200,10)$                      &&0.877&0.640&&0.960&1.000&&0.866&0.622&\\
&&$(500,10)$                      &&0.990&0.765&&0.997&1.000&&0.995&0.762&\\
&&$(50,50)$                       &&0.904&0.999&&0.724&0.308&&0.906&1.000&\\
&&$(100,50)$                      &&0.967&0.997&&0.965&0.972&&0.973&0.997&\\
&&$(200,50)$                      &&0.998&0.999&&0.999&1.000&&0.997&0.996&\\
&&$(500,50)$                      &&1.000&1.000&&1.000&1.000&&1.000&1.000&\\
&&$(50,100)$                      &&0.979&1.000&&0.729&0.103&&0.982&1.000&\\
&&$(100,100)$                     &&1.000&1.000&&0.998&0.990&&1.000&1.000&\\
&&$(200,100)$                     &&1.000&1.000&&1.000&1.000&&1.000&1.000&\\
&&$(500,100)$                     &&1.000&1.000&&1.000&1.000&&1.000&1.000&\\
\cline{2-12}
&\multirow{9}{*}{M2}&$(50,10)$    &&0.527&0.846&&0.566&0.560&&0.529&0.843&\\
&&$(100,10)$                      &&0.681&0.835&&0.850&0.941&&0.674&0.832&\\
&&$(200,10)$                      &&0.857&0.839&&0.959&1.000&&0.860&0.841&\\
&&$(500,10)$                      &&0.986&0.936&&0.990&1.000&&0.986&0.952&\\
&&$(50,50)$                       &&0.841&0.839&&0.596&0.080&&0.791&0.836&\\
&&$(100,50)$                      &&0.971&0.998&&0.932&0.826&&0.966&0.998&\\
&&$(200,50)$                      &&0.998&1.000&&0.997&1.000&&0.999&1.000&\\
&&$(500,50)$                      &&1.000&1.000&&1.000&1.000&&1.000&1.000&\\
&&$(50,100)$                      &&0.882&0.865&&0.514&0.004&&0.875&0.836&\\
&&$(100,100)$                     &&0.998&1.000&&0.990&0.811&&0.998&1.000&\\
&&$(200,100)$                     &&1.000&1.000&&1.000&1.000&&1.000&1.000&\\
&&$(500,100)$                     &&1.000&1.000&&1.000&1.000&&1.000&1.000&\\
\cline{2-12}
&\multirow{9}{*}{M3}&$(50,10)$    &&0.682&0.889&&0.382&0.372&&0.688&0.897&\\
&&$(100,10)$                      &&0.789&0.854&&0.634&0.802&&0.795&0.851&\\
&&$(200,10)$                      &&0.926&0.863&&0.802&0.971&&0.916&0.860&\\
&&$(500,10)$                      &&0.992&0.933&&0.952&1.000&&0.996&0.932&\\
&&$(50,50)$                       &&0.988&0.998&&0.101&0.001&&0.986&0.995&\\
&&$(100,50)$                      &&1.000&1.000&&0.444&0.232&&0.995&1.000&\\
&&$(200,50)$                      &&1.000&1.000&&0.825&0.925&&0.999&1.000&\\
&&$(500,50)$                      &&1.000&1.000&&0.998&1.000&&1.000&1.000&\\
&&$(50,100)$                      &&0.998&0.999&&0.033&0.000&&0.998&1.000&\\
&&$(100,100)$                     &&1.000&1.000&&0.339&0.044&&1.000&1.000&\\
&&$(200,100)$                     &&1.000&1.000&&0.893&0.931&&1.000&1.000&\\
&&$(500,100)$                     &&1.000&1.000&&1.000&1.000&&1.000&1.000&\\
\hline\hline
\end{tabular}
\end{threeparttable}
\end{table}%

\setlength{\tabcolsep}{3.5pt}
\begin{table}[!htbp]
\footnotesize
\centering
\begin{threeparttable}
\renewcommand{\arraystretch}{1.39}
\caption{Mean square errors of  $\hat{a}(\tau)$ , $\hat{B}(\tau)$ and $\hat{F}(\tau)$\tnote{\dag}}\label{Tab: MSE2}
\begin{tabular}{cccccccccccccccc}
\hline\hline
&\multirow{2}{*}{Model}&\multirow{2}{*}{(N,T)}&&\multicolumn{3}{c}{$\tau = 0.25$}&&\multicolumn{3}{c}{$\tau = 0.5$}&&\multicolumn{3}{c}{$\tau = 0.75$}&\\
\cline{5-7}\cline{9-11}\cline{13-15}
&&&&$\hat{a}(\tau)$&$\hat{B}(\tau)$&$\hat{F}(\tau)$&&$\hat{a}(\tau)$&$\hat{B}(\tau)$&$\hat{F}(\tau)$&&$\hat{a}(\tau)$&$\hat{B}(\tau)$&$\hat{F}(\tau)$&\\
\cline{2-15}
&\multirow{9}{*}{M1}&$(50,10)$    &&0.1424&0.4958&247.44&&0.0663&0.0651&0.1848&&0.1509&0.5173&111.72&\\
&&$(100,10)$                      &&0.0601&0.2343&5.3343&&0.0277&0.0270&0.0701&&0.0598&0.2331&0.9900&\\
&&$(200,10)$                      &&0.0257&0.1062&0.1902&&0.0129&0.0127&0.0319&&0.0263&0.1057&0.2004&\\
&&$(500,10)$                      &&0.0092&0.0384&0.0482&&0.0051&0.0046&0.0118&&0.0095&0.0400&0.0540&\\
&&$(50,50)$                       &&0.0523&0.4195&3.6213&&0.0134&0.0255&0.2292&&0.0503&0.4289&4.5562&\\
&&$(100,50)$                      &&0.0230&0.1952&0.8081&&0.0056&0.0089&0.0944&&0.0216&0.1915&0.8574&\\
&&$(200,50)$                      &&0.0092&0.0764&0.2401&&0.0028&0.0037&0.0439&&0.0088&0.0739&0.2551&\\
&&$(500,50)$                      &&0.0027&0.0196&0.0653&&0.0010&0.0013&0.0166&&0.0028&0.0196&0.0682&\\
&&$(50,100)$                      &&0.0424&0.3408&2.9921&&0.0064&0.0122&0.2395&&0.0411&0.3440&3.2297&\\
&&$(100,100)$                     &&0.0166&0.1420&0.7257&&0.0028&0.0039&0.1007&&0.0158&0.1451&0.7993&\\
&&$(200,100)$                     &&0.0057&0.0504&0.2301&&0.0013&0.0016&0.0467&&0.0055&0.0495&0.2432&\\
&&$(500,100)$                     &&0.0014&0.0114&0.0662&&0.0005&0.0005&0.0177&&0.0014&0.0117&0.0687&\\
\cline{2-15}
&\multirow{9}{*}{M2}&$(50,10)$    &&0.2444&0.6501&17043 &&0.0886&0.1022&0.2782&&0.2568&0.6714&$4\times 10^6$&\\
&&$(100,10)$                      &&0.1071&0.3133&25.521&&0.0369&0.0365&0.0930&&0.1017&0.3073&8.5110&\\
&&$(200,10)$                      &&0.0397&0.1301&0.3036&&0.0171&0.0161&0.0403&&0.0414&0.1341&0.3650&\\
&&$(500,10)$                      &&0.0137&0.0509&0.0808&&0.0063&0.0056&0.0143&&0.0140&0.0498&0.0833&\\
&&$(50,50)$                       &&0.1026&0.5262&13.345&&0.0185&0.0396&0.3427&&0.0937&0.5291&8.9553&\\
&&$(100,50)$                      &&0.0392&0.2491&1.5175&&0.0071&0.0119&0.1254&&0.0386&0.2473&1.6097&\\
&&$(200,50)$                      &&0.0148&0.0982&0.4178&&0.0033&0.0047&0.0549&&0.0146&0.0971&0.4414&\\
&&$(500,50)$                      &&0.0044&0.0254&0.1059&&0.0012&0.0016&0.0199&&0.0042&0.0252&0.1101&\\
&&$(50,100)$                      &&0.0883&0.4343&6.4971&&0.0086&0.0206&0.3479&&0.0810&0.4320&6.5577&\\
&&$(100,100)$                     &&0.0309&0.1873&1.3704&&0.0036&0.0057&0.1322&&0.0297&0.1883&1.4694&\\
&&$(200,100)$                     &&0.0101&0.0668&0.3959&&0.0016&0.0020&0.0580&&0.0098&0.0671&0.4190&\\
&&$(500,100)$                     &&0.0024&0.0148&0.1055&&0.0006&0.0007&0.0215&&0.0023&0.0149&0.1096&\\
\cline{2-15}
&\multirow{9}{*}{M3}&$(50,10)$    &&0.1441&0.5425&5561.7&&0.1076&0.1318&0.9181&&0.1457&0.5432&313.10&\\
&&$(100,10)$                      &&0.0685&0.2671&84.557&&0.0488&0.0480&0.0933&&0.0628&0.2542&3.5306&\\
&&$(200,10)$                      &&0.0286&0.1142&0.4940&&0.0243&0.0218&0.0409&&0.0300&0.1188&0.4960&\\
&&$(500,10)$                      &&0.0115&0.0494&0.1031&&0.0102&0.0081&0.0156&&0.0115&0.0495&0.1023&\\
&&$(50,50)$                       &&0.0425&0.4637&10.763&&0.0246&0.0535&0.3212&&0.0386&0.4715&12.431&\\
&&$(100,50)$                      &&0.0210&0.2283&2.2191&&0.0112&0.0164&0.1250&&0.0200&0.2266&2.2596&\\
&&$(200,50)$                      &&0.0096&0.0799&0.6083&&0.0050&0.0067&0.0567&&0.0095&0.0778&0.6231&\\
&&$(500,50)$                      &&0.0034&0.0138&0.1445&&0.0021&0.0024&0.0215&&0.0033&0.0140&0.1502&\\
&&$(50,100)$                      &&0.0329&0.3819&7.6268&&0.0115&0.0252&0.3105&&0.0292&0.3794&7.8000&\\
&&$(100,100)$                     &&0.0148&0.1639&1.9105&&0.0055&0.0077&0.1309&&0.0143&0.1648&2.0042&\\
&&$(200,100)$                     &&0.0061&0.0477&0.5640&&0.0025&0.0029&0.0603&&0.0059&0.0461&0.5851&\\
&&$(500,100)$                     &&0.0016&0.0050&0.1434&&0.0010&0.0010&0.0231&&0.0016&0.0050&0.1472&\\
\hline\hline
\end{tabular}
\begin{tablenotes}
      \item[\dag] {\footnotesize{The mean square errors of $\hat{a}(\tau)$, $\hat{B}(\tau)$ and $\hat{F}(\tau)$ are given by $\sum_{\ell=1}^{1000}\|\hat{a}(\tau)^{(\ell)}-a(\tau)\|^2/1000$, $\sum_{\ell=1}^{1000}\|\hat{B}(\tau)^{(\ell)}-B(\tau)H(\tau)^{(\ell)}\|_{F}^2/1000$ and $\sum_{\ell=1}^{1000}\|\hat{F}(\tau)^{(\ell)}- F(\tau)(H(\tau)^{{(\ell)}\prime})^{-1}\|_{F}^2/{1000T}$, where $\hat{a}(\tau)^{(\ell)}$, $\hat{B}(\tau)^{(\ell)}$ and $\hat{F}(\tau)^{(\ell)}$ denote estimates in the $\ell$th simulation replication and $H(\tau)^{(\ell)}\equiv (F(\tau)^{\prime}M_T\hat{F}(\tau)^{(\ell)})(\hat{F}(\tau)^{{(\ell)}\prime}M_T\hat{F}(\tau)^{(\ell)})^{-1}$ is a rotational transformation matrix.}}
    \end{tablenotes}
\end{threeparttable}
\end{table}%

\subsection{Bootstrap Testing for $\alpha(\tau,\cdot)=0$}\label{Sec: 63}
To examine (iii), we consider the following data generating process:
\begin{align}\label{Eqn: SimulationDGP3}
y_{it} = \beta(z_{it})^{\prime}f_{t} + e_{it}, i=1,\ldots,N, t=1,\ldots,T,
\end{align}
where $z_{it}$'s, $\beta(\cdot)$ and $f_{t}$ are the same as in Section \ref{Sec: 61}, $e_{it}$'s are specified as in \eqref{Eqn: error2} with $\rho=0.3$, $\omega=0$ and $v_{it}$'s being independent draws from $N(0,1)$.
Thus, $M=3$, $K(\tau) = 2$, $\alpha(\tau,z_{it}) =  Q_{t_\nu}(\tau)$, $\beta(\tau,z_{it})=\beta(z_{it})$ and $f_{t}(\tau) = f_{t}$ for $\tau\in(0,1)$. We implement the QR-PCA, and conduct the bootstrap test for $\mathrm{H}_{0}: \alpha(\tau,\cdot)=0$ versus $\mathrm{H}_{1}: \alpha(\tau,\cdot)\neq 0$ at different $\tau$'s, defined in Section \ref{Sec: 4}. Here, $\mathrm{H}_{0}$ holds if and only if $\tau = 0.5$.

To implement the QR-PCA, we choose $\phi(z_{it}) = (1,z_{it,1},z_{it,1}^{2},$ $z_{it,2},z_{it,2}^{2},z_{it,3},z_{it,3}^{2})^{\prime}$, leading to a zero sieve approximation error. We set the number of bootstrap draws to be $499$ in the implementation of the weighted bootstrap. We report the rejection rates of the bootstrap test at $\tau = 0.5,0.501,0.502,\ldots,0.51$ in Table \ref{Tab: RejectionRateAlpha}. The number of simulation replications is set to $1,000$.

The main findings are as follows. The bootstrap test may be undersized when $N$ is small. However, the rejection rate is improved by increasing $N$ to $500$. In addition, increasing $T$ helps improve the size performance of the test. When $N=500$ and $T=100$, the rejection rate is close to the significance level. In all cases, the rejection rate rises as $\tau$ increases, and approaches to one for large $\tau$ when $N=500$. Increasing $T$ also helps improve the power performance of the test. To sum up, our bootstrap test performs well in testing whether $\alpha(\tau,\cdot)=0$.

\setlength{\tabcolsep}{3.5pt}
\begin{table}[!htbp]
\footnotesize
\centering
\begin{threeparttable}
\renewcommand{\arraystretch}{1.5}
\caption{Rejection rates of testing $\alpha(\tau,\cdot) = 0 $\tnote{\dag}}\label{Tab: RejectionRateAlpha}
\begin{tabular}{ccccccccccccccc}
\hline\hline
&\multirow{2}{*}{(N,T)}&&\multicolumn{11}{c}{$\tau$}&\\
\cline{4-14}
&&&$0.5$&$0.501$&$0.502$&$0.503$&$0.504$&$0.505$&$0.506$&$0.507$&$0.508$&$0.509$&$0.51$&\\
\cline{2-14}
&$(50,10)$   &&0.008&0.014&0.016&0.022&0.032&0.054&0.075&0.106&0.159&0.201&0.255&\\
&$(100,10)$  &&0.019&0.024&0.042&0.076&0.117&0.165&0.244&0.334&0.436&0.546&0.651&\\
&$(200,10)$  &&0.021&0.032&0.064&0.134&0.244&0.393&0.565&0.709&0.826&0.909&0.957&\\
&$(500,10)$  &&0.029&0.061&0.164&0.378&0.643&0.843&0.960&0.995&1.000&1.000&1.000&\\
\cline{2-14}
&$(50,50)$   &&0.020&0.020&0.041&0.102&0.204&0.362&0.526&0.669&0.810&0.889&0.948&\\
&$(100,50)$  &&0.021&0.035&0.108&0.312&0.534&0.766&0.903&0.974&0.996&1.000&1.000&\\
&$(200,50)$  &&0.019&0.070&0.273&0.636&0.894&0.985&0.999&1.000&1.000&1.000&1.000&\\
&$(500,50)$  &&0.039&0.192&0.743&0.978&1.000&1.000&1.000&1.000&1.000&1.000&1.000&\\
\cline{2-14}
&$(50,100)$  &&0.018&0.042&0.153&0.398&0.689&0.875&0.972&0.994&0.999&1.000&1.000&\\
&$(100,100)$ &&0.018&0.090&0.429&0.811&0.974&0.999&1.000&1.000&1.000&1.000&1.000&\\
&$(200,100)$ &&0.032&0.225&0.791&0.993&1.000&1.000&1.000&1.000&1.000&1.000&1.000&\\
&$(500,100)$ &&0.049&0.595&0.994&1.000&1.000&1.000&1.000&1.000&1.000&1.000&1.000&\\
\hline\hline
\end{tabular}
\begin{tablenotes}
      \small
      \item[\dag] The significance level $\alpha = 5\%$.
    \end{tablenotes}
\end{threeparttable}
\end{table}%

\section{Individual U.S. Stock Returns}\label{Sec: 7}
\pgfplotstableread{
X Y Z
-0.87054775	0.00002506	0
-0.85168064	0.00009354	0
-0.83281352	0.00027637	0
-0.81394641	0.00065326	0
-0.79507930	0.00126346	0
-0.77621218	0.00208055	0
-0.75734507	0.00305224	0
-0.73847795	0.00407563	0
-0.71961084	0.00497074	0
-0.70074372	0.00564278	0
-0.68187661	0.00622580	0
-0.66300949	0.00682636	0
-0.64414238	0.00726172	0
-0.62527527	0.00735183	0
-0.60640815	0.00737632	0
-0.58754104	0.00789334	0
-0.56867392	0.00904608	0
-0.54980681	0.01039055	0
-0.53093969	0.01167605	0
-0.51207258	0.01359561	0
-0.49320546	0.01734015	0
-0.47433835	0.02328050	0
-0.45547123	0.03030319	0
-0.43660412	0.03661363	0
-0.41773701	0.04123292	0
-0.39886989	0.04483597	0
-0.38000278	0.04927897	0
-0.36113566	0.05638635	0
-0.34226855	0.06731796	0
-0.32340143	0.08297494	0
-0.30453432	0.10425160	0
-0.28566720	0.13147985	0
-0.26680009	0.16464700	0
-0.24793298	0.20498852	0
-0.22906586	0.25517493	0
-0.21019875	0.31729354	0
-0.19133163	0.39298875	0
-0.17246452	0.48665542	0
-0.15359740	0.60503578	0
-0.13473029	0.75104395	0
-0.11586317	0.91978815	0
-0.09699606	1.10259232	0
-0.07812894	1.29471761	0
-0.05926183	1.49904538	0
-0.04039472	1.72073482	0
-0.02152760	1.95447784	0
-0.00266049	2.17569728	0
0.01620663	2.34700174	0
0.03507374	2.43478130	0
0.05394086	2.42127545	0
0.07280797	2.31006747	0
0.09167509	2.12788799	0
0.11054220	1.91661197	0
0.12940931	1.71436796	0
0.14827643	1.53980801	0
0.16714354	1.39188602	0
0.18601066	1.26140326	0
0.20487777	1.14206516	0
0.22374489	1.03356905	0
0.24261200	0.93813212	0
0.26147912	0.85664730	0
0.28034623	0.78833377	0
0.29921335	0.73210026	0
0.31808046	0.68651869	0
0.33694757	0.64896236	0
0.35581469	0.61611961	0
0.37468180	0.58560936	0
0.39354892	0.55667212	0
0.41241603	0.52937031	0
0.43128315	0.50379416	0
0.45015026	0.47980860	0
0.46901738	0.45598890	0
0.48788449	0.42934706	0
0.50675160	0.39888253	0
0.52561872	0.36831937	0
0.54448583	0.34196108	0
0.56335295	0.31979502	0
0.58222006	0.29957705	0
0.60108718	0.28080167	0
0.61995429	0.26423641	0
0.63882141	0.24965448	0
0.65768852	0.23531137	0
0.67655564	0.21967510	0
0.69542275	0.20385203	0
0.71428986	0.19130798	0
0.73315698	0.18427987	0
0.75202409	0.18093611	0
0.77089121	0.17646994	0
0.78975832	0.16675784	0
0.80862544	0.15146175	0
0.82749255	0.13455414	0
0.84635967	0.12167405	0
0.86522678	0.11586038	0
0.88409389	0.11526152	0
0.90296101	0.11516048	0
0.92182812	0.11238542	0
0.94069524	0.10769745	0
0.95956235	0.10384463	0
0.97842947	0.10161687	0
0.99729658	0.09878633	0
1.01616370	0.09339185	0
1.03503081	0.08664115	0
1.05389793	0.08120017	0
1.07276504	0.07784524	0
1.09163215	0.07523733	0
1.11049927	0.07187144	0
1.12936638	0.06718337	0
1.14823350	0.06174455	0
1.16710061	0.05701693	0
1.18596773	0.05410991	0
1.20483484	0.05249643	0
1.22370196	0.05070937	0
1.24256907	0.04838466	0
1.26143618	0.04705356	0
1.28030330	0.04855302	0
1.29917041	0.05266504	0
1.31803753	0.05662814	0
1.33690464	0.05730035	0
1.35577176	0.05397749	0
1.37463887	0.04852506	0
1.39350599	0.04311567	0
1.41237310	0.03878795	0
1.43124021	0.03582153	0
1.45010733	0.03427663	0
1.46897444	0.03375183	0
1.48784156	0.03318081	0
1.50670867	0.03164788	0
1.52557579	0.02947167	0
1.54444290	0.02793014	0
1.56331002	0.02788966	0
1.58217713	0.02916204	0
1.60104425	0.03109716	0
1.61991136	0.03318322	0
1.63877847	0.03481087	0
1.65764559	0.03489088	0
1.67651270	0.03246557	0
1.69537982	0.02798466	0
1.71424693	0.02343610	0
1.73311405	0.02072392	0
1.75198116	0.02014147	0
1.77084828	0.02056526	0
1.78971539	0.02069364	0
1.80858250	0.01989778	0
1.82744962	0.01832320	0
1.84631673	0.01664323	0
1.86518385	0.01562247	0
1.88405096	0.01547317	0
1.90291808	0.01554678	0
1.92178519	0.01500474	0
1.94065231	0.01394356	0
1.95951942	0.01346410	0
1.97838654	0.01408911	0
1.99725365	0.01485161	0
2.01612076	0.01441012	0
2.03498788	0.01261228	0
2.05385499	0.01034114	0
2.07272211	0.00838076	0
2.09158922	0.00706724	0
2.11045634	0.00671500	0
2.12932345	0.00774304	0
2.14819057	0.01012517	0
2.16705768	0.01291321	0
2.18592479	0.01468122	0
2.20479191	0.01460198	0
2.22365902	0.01302387	0
2.24252614	0.01101917	0
2.26139325	0.00950617	0
2.28026037	0.00876169	0
2.29912748	0.00849327	0
2.31799460	0.00819688	0
2.33686171	0.00755025	0
2.35572883	0.00661103	0
2.37459594	0.00568059	0
2.39346305	0.00504473	0
2.41233017	0.00485960	0
2.43119728	0.00510527	0
2.45006440	0.00553191	0
2.46893151	0.00577023	0
2.48779863	0.00557819	0
2.50666574	0.00500089	0
2.52553286	0.00436807	0
2.54439997	0.00404256	0
2.56326708	0.00405325	0
2.58213420	0.00416130	0
2.60100131	0.00436297	0
2.61986843	0.00504491	0
2.63873554	0.00634142	0
2.65760266	0.00766572	0
2.67646977	0.00811587	0
2.69533689	0.00741360	0
2.71420400	0.00624889	0
2.73307112	0.00559881	0
2.75193823	0.00582731	0
2.77080534	0.00649283	0
2.78967246	0.00683224	0
2.80853957	0.00632669	0
2.82740669	0.00501948	0
2.84627380	0.00348152	0
2.86514092	0.00240747	0
2.88400803	0.00213031	0
2.90287515	0.00248127	0
2.92174226	0.00308316	0
2.94060937	0.00371272	0
2.95947649	0.00431935	0
2.97834360	0.00476752	0
2.99721072	0.00480546	0
3.01607783	0.00442832	0
3.03494495	0.00407614	0
3.05381206	0.00420288	0
3.07267918	0.00469799	0
3.09154629	0.00495305	0
3.11041340	0.00449404	0
3.12928052	0.00339416	0
3.14814763	0.00210782	0
3.16701475	0.00107573	0
3.18588186	0.00050030	0
3.20474898	0.00036869	0
3.22361609	0.00057688	0
3.24248321	0.00097039	0
3.26135032	0.00131102	0
3.28021744	0.00135693	0
3.29908455	0.00108115	0
3.31795166	0.00073390	0
3.33681878	0.00061288	0
3.35568589	0.00087491	0
3.37455301	0.00146840	0
3.39342012	0.00217409	0
3.41228724	0.00275432	0
3.43115435	0.00311679	0
3.45002147	0.00331254	0
3.46888858	0.00336058	0
3.48775569	0.00315231	0
3.50662281	0.00261663	0
3.52548992	0.00192373	0
3.54435704	0.00140443	0
3.56322415	0.00124975	0
3.58209127	0.00133765	0
3.60095838	0.00137125	0
3.61982550	0.00116742	0
3.63869261	0.00079481	0
3.65755973	0.00046335	0
3.67642684	0.00034505	0
3.69529395	0.00049734	0
3.71416107	0.00085404	0
3.73302818	0.00122135	0
3.75189530	0.00135729	0
3.77076241	0.00115842	0
3.78962953	0.00075556	0
3.80849664	0.00037507	0
3.82736376	0.00014123	0
3.84623087	0.00004035	0
3.86509798	0.00000977	0
3.88396510	0.00000828	0
3.90283221	0.00003269	0
3.92169933	0.00011938	0
3.94056644	0.00033783	0
3.95943356	0.00072911	0
3.97830067	0.00117254	0
3.99716779	0.00140322	0
4.01603490	0.00125982	0
4.03490202	0.00085689	0
4.05376913	0.00044616	0
4.07263624	0.00017968	0
4.09150336	0.00005826	0
4.11037047	0.00002658	0
4.12923759	0.00005856	0
4.14810470	0.00018176	0
4.16697182	0.00044485	0
4.18583893	0.00083257	0
4.20470605	0.00119976	0
4.22357316	0.00134691	0
4.24244027	0.00121370	0
4.26130739	0.00097222	0
4.28017450	0.00088177	0
4.29904162	0.00104168	0
4.31790873	0.00129752	0
4.33677585	0.00140788	0
4.35564296	0.00130966	0
4.37451008	0.00118455	0
4.39337719	0.00123572	0
4.41224431	0.00144693	0
4.43111142	0.00163262	0
4.44997853	0.00169712	0
4.46884565	0.00172727	0
4.48771276	0.00180892	0
4.50657988	0.00189905	0
4.52544699	0.00193237	0
4.54431411	0.00188711	0
4.56318122	0.00169989	0
4.58204834	0.00130747	0
4.60091545	0.00080142	0
4.61978256	0.00037602	0
4.63864968	0.00013549	0
4.65751679	0.00003818	0
4.67638391	0.00000846	0
4.69525102	0.00000148	0
4.71411814	0.00000020	0
4.73298525	0.00000002	0
4.75185237	0.00000000	0
4.77071948	0.00000000	0
4.78958660	0.00000000	0
4.80845371	0.00000000	0
4.82732082	0.00000002	0
4.84618794	0.00000021	0
4.86505505	0.00000168	0
4.88392217	0.00000988	0
4.90278928	0.00004366	0
4.92165640	0.00014627	0
4.94052351	0.00037427	0
4.95939063	0.00073690	0
4.97825774	0.00112297	0
4.99712485	0.00132878	0
5.01599197	0.00122063	0
5.03485908	0.00086731	0
5.05372620	0.00047399	0
5.07259331	0.00020028	0
5.09146043	0.00007857	0
5.11032754	0.00008064	0
5.12919466	0.00020597	0
5.14806177	0.00048621	0
5.16692888	0.00089787	0
5.18579600	0.00127563	0
5.20466311	0.00138141	0
5.22353023	0.00112948	0
5.24239734	0.00069128	0
5.26126446	0.00031505	0
5.28013157	0.00011039	0
5.29899869	0.00005000	0
5.31786580	0.00010184	0
5.33673292	0.00029181	0
5.35560003	0.00064982	0
5.37446714	0.00108249	0
5.39333426	0.00135631	0
5.41220137	0.00128987	0
5.43106849	0.00094018	0
5.44993560	0.00052988	0
5.46880272	0.00023232	0
5.48766983	0.00007942	0
5.50653695	0.00002119	0
5.52540406	0.00000498	0
5.54427117	0.00000507	0
5.56313829	0.00002289	0
5.58200540	0.00008940	0
5.60087252	0.00026456	0
5.61973963	0.00059368	0
5.63860675	0.00101474	0
5.65747386	0.00132569	0
5.67634098	0.00132510	0
5.69520809	0.00101114	0
5.71407521	0.00058596	0
5.73294232	0.00025610	0
5.75180943	0.00008380	0
5.77067655	0.00002040	0
5.78954366	0.00000367	0
5.80841078	0.00000049	0
5.82727789	0.00000006	0
5.84614501	0.00000016	0
5.86501212	0.00000124	0
5.88387924	0.00000744	0
5.90274635	0.00003447	0
5.92161346	0.00012373	0
5.94048058	0.00034304	0
5.95934769	0.00072805	0
5.97821481	0.00117038	0
5.99708192	0.00140272	0
6.01594904	0.00126092	0
6.03481615	0.00085873	0
6.05368327	0.00044806	0
6.07255038	0.00018101	0
6.09141750	0.00005704	0
6.11028461	0.00001406	0
6.12915172	0.00000270	0
6.14801884	0.00000040	0
6.16688595	0.00000005	0
6.18575307	0.00000000	0
6.20462018	0.00000000	0
6.22348730	0.00000000	0
6.24235441	0.00000000	0
6.26122153	0.00000000	0
6.28008864	0.00000000	0
6.29895575	0.00000000	0
6.31782287	0.00000000	0
6.33668998	0.00000000	0
6.35555710	0.00000000	0
6.37442421	0.00000000	0
6.39329133	0.00000000	0
6.41215844	0.00000000	0
6.43102556	0.00000000	0
6.44989267	0.00000000	0
6.46875979	0.00000000	0
6.48762690	0.00000000	0
6.50649401	0.00000000	0
6.52536113	0.00000000	0
6.54422824	0.00000000	0
6.56309536	0.00000000	0
6.58196247	0.00000000	0
6.60082959	0.00000000	0
6.61969670	0.00000000	0
6.63856382	0.00000000	0
6.65743093	0.00000000	0
6.67629804	0.00000000	0
6.69516516	0.00000000	0
6.71403227	0.00000000	0
6.73289939	0.00000000	0
6.75176650	0.00000000	0
6.77063362	0.00000000	0
6.78950073	0.00000000	0
6.80836785	0.00000000	0
6.82723496	0.00000000	0
6.84610207	0.00000000	0
6.86496919	0.00000000	0
6.88383630	0.00000000	0
6.90270342	0.00000000	0
6.92157053	0.00000000	0
6.94043765	0.00000000	0
6.95930476	0.00000000	0
6.97817188	0.00000000	0
6.99703899	0.00000000	0
7.01590611	0.00000000	0
7.03477322	0.00000000	0
7.05364033	0.00000000	0
7.07250745	0.00000000	0
7.09137456	0.00000000	0
7.11024168	0.00000000	0
7.12910879	0.00000000	0
7.14797591	0.00000000	0
7.16684302	0.00000000	0
7.18571014	0.00000000	0
7.20457725	0.00000000	0
7.22344436	0.00000000	0
7.24231148	0.00000000	0
7.26117859	0.00000000	0
7.28004571	0.00000000	0
7.29891282	0.00000000	0
7.31777994	0.00000000	0
7.33664705	0.00000000	0
7.35551417	0.00000000	0
7.37438128	0.00000000	0
7.39324840	0.00000000	0
7.41211551	0.00000000	0
7.43098262	0.00000000	0
7.44984974	0.00000000	0
7.46871685	0.00000000	0
7.48758397	0.00000000	0
7.50645108	0.00000000	0
7.52531820	0.00000000	0
7.54418531	0.00000000	0
7.56305243	0.00000000	0
7.58191954	0.00000000	0
7.60078665	0.00000000	0
7.61965377	0.00000000	0
7.63852088	0.00000000	0
7.65738800	0.00000000	0
7.67625511	0.00000000	0
7.69512223	0.00000000	0
7.71398934	0.00000000	0
7.73285646	0.00000000	0
7.75172357	0.00000000	0
7.77059069	0.00000000	0
7.78945780	0.00000000	0
7.80832491	0.00000000	0
7.82719203	0.00000000	0
7.84605914	0.00000000	0
7.86492626	0.00000000	0
7.88379337	0.00000000	0
7.90266049	0.00000000	0
7.92152760	0.00000000	0
7.94039472	0.00000000	0
7.95926183	0.00000000	0
7.97812894	0.00000000	0
7.99699606	0.00000000	0
8.01586317	0.00000000	0
8.03473029	0.00000000	0
8.05359740	0.00000000	0
8.07246452	0.00000000	0
8.09133163	0.00000000	0
8.11019875	0.00000000	0
8.12906586	0.00000000	0
8.14793298	0.00000000	0
8.16680009	0.00000000	0
8.18566720	0.00000000	0
8.20453432	0.00000000	0
8.22340143	0.00000000	0
8.24226855	0.00000000	0
8.26113566	0.00000000	0
8.28000278	0.00000000	0
8.29886989	0.00000000	0
8.31773701	0.00000000	0
8.33660412	0.00000000	0
8.35547123	0.00000000	0
8.37433835	0.00000000	0
8.39320546	0.00000000	0
8.41207258	0.00000000	0
8.43093969	0.00000000	0
8.44980681	0.00000000	0
8.46867392	0.00000000	0
8.48754104	0.00000001	0
8.50640815	0.00000012	0
8.52527527	0.00000091	0
8.54414238	0.00000554	0
8.56300949	0.00002593	0
8.58187661	0.00009436	0
8.60074372	0.00026781	0
8.61961084	0.00059153	0
8.63847795	0.00101120	0
8.65734507	0.00132855	0
8.67621218	0.00133211	0
8.69507930	0.00101005	0
8.71394641	0.00058605	0
8.73281352	0.00026235	0
8.75168064	0.00009125	0
8.77054775	0.00002476	0
}\normalone

\pgfplotstableread{
X Y
-1.02322249	0.00006598
-1.01644853	0.00013744
-1.00967458	0.00026763
-1.00290063	0.00048760
-0.99612667	0.00083184
-0.98935272	0.00132930
-0.98257877	0.00199258
-0.97580481	0.00279277
-0.96903086	0.00364507
-0.96225691	0.00443470
-0.95548295	0.00503410
-0.94870900	0.00533625
-0.94193505	0.00528551
-0.93516109	0.00489383
-0.92838714	0.00423625
-0.92161319	0.00342782
-0.91483923	0.00259166
-0.90806528	0.00182970
-0.90129133	0.00120525
-0.89451737	0.00074005
-0.88774342	0.00042318
-0.88096947	0.00022513
-0.87419551	0.00011133
-0.86742156	0.00005113
-0.86064761	0.00002228
-0.85387365	0.00000928
-0.84709970	0.00000363
-0.84032575	0.00000132
-0.83355179	0.00000045
-0.82677784	0.00000014
-0.82000389	0.00000004
-0.81322993	0.00000001
-0.80645598	0.00000001
-0.79968202	0.00000002
-0.79290807	0.00000008
-0.78613412	0.00000025
-0.77936016	0.00000076
-0.77258621	0.00000217
-0.76581226	0.00000575
-0.75903830	0.00001424
-0.75226435	0.00003304
-0.74549040	0.00007193
-0.73871644	0.00014973
-0.73194249	0.00029306
-0.72516854	0.00053361
-0.71839458	0.00090468
-0.71162063	0.00142961
-0.70484668	0.00210792
-0.69807272	0.00290340
-0.69129877	0.00374061
-0.68452482	0.00451518
-0.67775086	0.00511862
-0.67097691	0.00547227
-0.66420296	0.00555940
-0.65742900	0.00544352
-0.65065505	0.00526420
-0.64388110	0.00520904
-0.63710714	0.00546822
-0.63033319	0.00618346
-0.62355924	0.00740401
-0.61678528	0.00907135
-0.61001133	0.01095359
-0.60323738	0.01275854
-0.59646342	0.01418215
-0.58968947	0.01497506
-0.58291552	0.01500985
-0.57614156	0.01432667
-0.56936761	0.01314505
-0.56259366	0.01183889
-0.55581970	0.01088049
-0.54904575	0.01076521
-0.54227179	0.01192876
-0.53549784	0.01466723
-0.52872389	0.01906900
-0.52194993	0.02497008
-0.51517598	0.03194793
-0.50840203	0.03936885
-0.50162807	0.04649444
-0.49485412	0.05254264
-0.48808017	0.05709775
-0.48130621	0.06005957
-0.47453226	0.06163637
-0.46775831	0.06225075
-0.46098435	0.06240267
-0.45421040	0.06255118
-0.44743645	0.06306010
-0.44066249	0.06421734
-0.43388854	0.06629767
-0.42711459	0.06961528
-0.42034063	0.07451933
-0.41356668	0.08132108
-0.40679273	0.09018670
-0.40001877	0.10105883
-0.39324482	0.11366178
-0.38647087	0.12759782
-0.37969691	0.14248470
-0.37292296	0.15803922
-0.36614901	0.17395337
-0.35937505	0.18985580
-0.35260110	0.20517754
-0.34582715	0.21923460
-0.33905319	0.23148803
-0.33227924	0.24188819
-0.32550529	0.25114464
-0.31873133	0.26077330
-0.31195738	0.27285458
-0.30518343	0.28955259
-0.29840947	0.31254888
-0.29163552	0.34258949
-0.28486156	0.37931288
-0.27808761	0.42142542
-0.27131366	0.46715961
-0.26453970	0.51484078
-0.25776575	0.56335834
-0.25099180	0.61239076
-0.24421784	0.66225274
-0.23744389	0.71350043
-0.23066994	0.76652440
-0.22389598	0.82129996
-0.21712203	0.87732197
-0.21034808	0.93372987
-0.20357412	0.98957162
-0.19680017	1.04410676
-0.19002622	1.09702630
-0.18325226	1.14849065
-0.17647831	1.19895822
-0.16970436	1.24887193
-0.16293040	1.29834420
-0.15615645	1.34699498
-0.14938250	1.39404062
-0.14260854	1.43863140
-0.13583459	1.48028324
-0.12906064	1.51965854
-0.12228668	1.55891157
-0.11551273	1.60138499
-0.10873878	1.65136126
-0.10196482	1.71338713
-0.09519087	1.79135509
-0.08841692	1.88751345
-0.08164296	2.00165911
-0.07486901	2.13078641
-0.06809506	2.26937811
-0.06132110	2.41033413
-0.05454715	2.54630193
-0.04777320	2.67100100
-0.04099924	2.78011323
-0.03422529	2.87147066
-0.02745133	2.94455507
-0.02067738	2.99960369
-0.01390343	3.03609588
-0.00712947	3.05376052
-0.00035552	3.05241454
0.00641843	3.03183029
0.01319239	2.99282855
0.01996634	2.93757162
0.02674029	2.86942253
0.03351425	2.79243679
0.04028820	2.71067932
0.04706215	2.62759170
0.05383611	2.54557658
0.06061006	2.46585902
0.06738401	2.38858435
0.07415797	2.31306585
0.08093192	2.23811129
0.08770587	2.16240245
0.09447983	2.08492041
0.10125378	2.00536349
0.10802773	1.92443811
0.11480169	1.84396567
0.12157564	1.76589659
0.12834959	1.69157059
0.13512355	1.62113499
0.14189750	1.55342344
0.14867145	1.48637845
0.15544541	1.41782481
0.16221936	1.34625963
0.16899331	1.27135235
0.17576727	1.19400943
0.18254122	1.11605776
0.18931517	1.03974201
0.19608913	0.96725943
0.20286308	0.90047740
0.20963703	0.84084901
0.21641099	0.78943164
0.22318494	0.74687412
0.22995890	0.71373455
0.23673285	0.68872445
0.24350680	0.66993212
0.25028076	0.65462161
0.25705471	0.63954992
0.26382866	0.62155785
0.27060262	0.59819651
0.27737657	0.56820529
0.28415052	0.53171544
0.29092448	0.49014610
0.29769843	0.44584381
0.30447238	0.40157073
0.31124634	0.35996588
0.31802029	0.32309512
0.32479424	0.29217490
0.33156820	0.26750694
0.33834215	0.24860249
0.34511610	0.23443308
0.35189006	0.22378809
0.35866401	0.21505008
0.36543796	0.20706710
0.37221192	0.19899066
0.37898587	0.19036813
0.38575982	0.18117670
0.39253378	0.17176767
0.39930773	0.16271589
0.40608168	0.15461082
0.41285564	0.14785622
0.41962959	0.14254564
0.42640354	0.13844916
0.43317750	0.13509965
0.43995145	0.13192861
0.44672540	0.12839227
0.45349936	0.12404953
0.46027331	0.11859118
0.46704726	0.11183249
0.47382122	0.10372669
0.48059517	0.09455156
0.48736913	0.08470657
0.49414308	0.07472339
0.50091703	0.06516068
0.50769099	0.05648111
0.51446494	0.04895410
0.52123889	0.04261858
0.52801285	0.03731650
0.53478680	0.03278291
0.54156075	0.02875951
0.54833471	0.02509269
0.55510866	0.02178346
0.56188261	0.01897353
0.56865657	0.01687321
0.57543052	0.01565718
0.58220447	0.01536701
0.58897843	0.01586950
0.59575238	0.01683329
0.60252633	0.01785864
0.60930029	0.01866660
0.61607424	0.01915552
0.62284819	0.01941453
0.62962215	0.01965780
0.63639610	0.02011108
0.64317005	0.02090152
0.64994401	0.02199714
0.65671796	0.02321676
0.66349191	0.02429761
0.67026587	0.02498440
0.67703982	0.02509939
0.68381377	0.02456915
0.69058773	0.02340937
0.69736168	0.02168986
0.70413563	0.01950579
0.71090959	0.01696333
0.71768354	0.01421683
0.72445749	0.01144916
0.73123145	0.00884625
0.73800540	0.00658088
0.74477936	0.00478562
0.75155331	0.00353575
0.75832726	0.00284501
0.76510122	0.00267034
0.77187517	0.00291999
0.77864912	0.00346261
0.78542308	0.00413878
0.79219703	0.00477870
0.79897098	0.00522745
0.80574494	0.00537389
0.81251889	0.00517443
0.81929284	0.00466186
0.82606680	0.00393446
0.83284075	0.00313154
0.83961470	0.00240469
0.84638866	0.00186540
0.85316261	0.00158576
0.85993656	0.00159912
0.86671052	0.00190163
0.87348447	0.00245220
0.88025842	0.00317188
0.88703238	0.00394784
0.89380633	0.00464684
0.90058028	0.00513933
0.90735424	0.00532902
0.91412819	0.00517845
0.92090214	0.00472080
0.92767610	0.00405274
0.93445005	0.00331093
0.94122400	0.00264066
0.94799796	0.00216706
0.95477191	0.00199166
0.96154586	0.00213514
0.96831982	0.00256715
0.97509377	0.00321118
0.98186772	0.00394993
0.98864168	0.00464385
0.99541563	0.00515958
1.00218958	0.00540285
1.00896354	0.00534534
1.01573749	0.00503552
1.02251145	0.00458756
1.02928540	0.00415038
1.03605935	0.00386545
1.04283331	0.00382585
1.04960726	0.00404844
1.05638121	0.00446753
1.06315517	0.00495207
1.06992912	0.00534240
1.07670307	0.00547695
1.08347703	0.00529221
1.09025098	0.00480053
1.09702493	0.00408080
1.10379889	0.00324855
1.11057284	0.00242071
1.11734679	0.00168779
1.12412075	0.00110043
1.13089470	0.00067042
1.13766865	0.00038132
1.14444261	0.00020228
1.15121656	0.00009998
1.15799051	0.00004599
1.16476447	0.00001968
1.17153842	0.00000782
1.17831237	0.00000289
1.18508633	0.00000099
1.19186028	0.00000032
1.19863423	0.00000010
1.20540819	0.00000003
1.21218214	0.00000001
1.21895609	0.00000000
1.22573005	0.00000000
1.23250400	0.00000000
1.23927795	0.00000000
1.24605191	0.00000000
1.25282586	0.00000000
1.25959981	0.00000000
1.26637377	0.00000000
1.27314772	0.00000000
1.27992168	0.00000000
1.28669563	0.00000000
1.29346958	0.00000000
1.30024354	0.00000000
1.30701749	0.00000000
1.31379144	0.00000000
1.32056540	0.00000000
1.32733935	0.00000000
1.33411330	0.00000000
1.34088726	0.00000000
1.34766121	0.00000000
1.35443516	0.00000000
1.36120912	0.00000000
1.36798307	0.00000000
1.37475702	0.00000000
1.38153098	0.00000000
1.38830493	0.00000000
1.39507888	0.00000000
1.40185284	0.00000000
1.40862679	0.00000000
1.41540074	0.00000000
1.42217470	0.00000000
1.42894865	0.00000000
1.43572260	0.00000000
1.44249656	0.00000000
1.44927051	0.00000000
1.45604446	0.00000000
1.46281842	0.00000000
1.46959237	0.00000000
1.47636632	0.00000000
1.48314028	0.00000000
1.48991423	0.00000000
1.49668818	0.00000000
1.50346214	0.00000000
1.51023609	0.00000000
1.51701004	0.00000000
1.52378400	0.00000000
1.53055795	0.00000000
1.53733191	0.00000000
1.54410586	0.00000000
1.55087981	0.00000000
1.55765377	0.00000000
1.56442772	0.00000000
1.57120167	0.00000000
1.57797563	0.00000000
1.58474958	0.00000000
1.59152353	0.00000000
1.59829749	0.00000000
1.60507144	0.00000000
1.61184539	0.00000000
1.61861935	0.00000000
1.62539330	0.00000000
1.63216725	0.00000000
1.63894121	0.00000000
1.64571516	0.00000000
1.65248911	0.00000000
1.65926307	0.00000000
1.66603702	0.00000000
1.67281097	0.00000000
1.67958493	0.00000000
1.68635888	0.00000000
1.69313283	0.00000000
1.69990679	0.00000000
1.70668074	0.00000000
1.71345469	0.00000000
1.72022865	0.00000000
1.72700260	0.00000000
1.73377655	0.00000000
1.74055051	0.00000000
1.74732446	0.00000000
1.75409841	0.00000000
1.76087237	0.00000000
1.76764632	0.00000000
1.77442027	0.00000000
1.78119423	0.00000000
1.78796818	0.00000000
1.79474214	0.00000000
1.80151609	0.00000000
1.80829004	0.00000000
1.81506400	0.00000000
1.82183795	0.00000000
1.82861190	0.00000000
1.83538586	0.00000000
1.84215981	0.00000000
1.84893376	0.00000000
1.85570772	0.00000000
1.86248167	0.00000000
1.86925562	0.00000000
1.87602958	0.00000000
1.88280353	0.00000000
1.88957748	0.00000000
1.89635144	0.00000000
1.90312539	0.00000000
1.90989934	0.00000000
1.91667330	0.00000000
1.92344725	0.00000000
1.93022120	0.00000000
1.93699516	0.00000000
1.94376911	0.00000000
1.95054306	0.00000000
1.95731702	0.00000000
1.96409097	0.00000000
1.97086492	0.00000000
1.97763888	0.00000000
1.98441283	0.00000000
1.99118678	0.00000000
1.99796074	0.00000000
2.00473469	0.00000000
2.01150864	0.00000000
2.01828260	0.00000000
2.02505655	0.00000000
2.03183050	0.00000000
2.03860446	0.00000000
2.04537841	0.00000000
2.05215237	0.00000000
2.05892632	0.00000000
2.06570027	0.00000000
2.07247423	0.00000000
2.07924818	0.00000000
2.08602213	0.00000000
2.09279609	0.00000000
2.09957004	0.00000000
2.10634399	0.00000000
2.11311795	0.00000000
2.11989190	0.00000000
2.12666585	0.00000000
2.13343981	0.00000000
2.14021376	0.00000000
2.14698771	0.00000000
2.15376167	0.00000000
2.16053562	0.00000000
2.16730957	0.00000000
2.17408353	0.00000000
2.18085748	0.00000000
2.18763143	0.00000000
2.19440539	0.00000000
2.20117934	0.00000000
2.20795329	0.00000000
2.21472725	0.00000001
2.22150120	0.00000003
2.22827515	0.00000011
2.23504911	0.00000036
2.24182306	0.00000110
2.24859701	0.00000314
2.25537097	0.00000837
2.26214492	0.00002087
2.26891887	0.00004880
2.27569283	0.00010720
2.28246678	0.00022476
2.28924073	0.00044636
2.29601469	0.00082639
2.30278864	0.00142768
2.30956260	0.00230391
2.31633655	0.00347626
2.32311050	0.00490888
2.32988446	0.00649297
2.33665841	0.00805000
2.34343236	0.00935955
2.35020632	0.01020775
2.35698027	0.01044275
2.36375422	0.01001814
2.37052818	0.00900768
2.37730213	0.00758517
2.38407608	0.00597648
2.39085004	0.00440156
2.39762399	0.00302676
2.40439794	0.00195376
2.41117190	0.00118678
2.41794585	0.00067683
2.42471980	0.00036219
2.43149376	0.00018168
2.43826771	0.00008533
}\normalonea	

To explain why different assets earn different average returns, the empirical asset pricing literature has focused on exploiting the ability of stock characteristics to explain/predict asset returns. Two main approaches are characteristic-based models \citep{RosenbergMcKibben_ThePrediction_1973,Lewellen_Crosssection_2014,Greenetal_Characteristics_2017} and risk-based factor models \citep{FamaFrench_Commonrisk_1993,FamaFrench_FiveFactor_2015}. The latter are motivated by the arbitrage pricing theory of \citet{Ross_APT_1976} and its extensions \citep*{ChamberlainRothschild_FactorStuctures_1982,ConnorKorajczyk_Performance_1986,ConnorKorajczyk_RiskReturn_1988}. The main issue with the two approaches is their inability to untwist the two roles played by characteristics: capturing risk exposures and representing mispricing errors \citep{DanielTitman_Characteristics_1997}. To solve the issue, \citet{Kellyetal_Characteristics_2019} and \citet{Chenetal_SeimiparametricFactor_2021} introduce characteristic-based factor models, which allow characteristics to simultaneously appear in both risk exposures and pricing errors. The models are also referred to as conditional factor models, since they capture time-variation in the risk exposures and the pricing errors. The models enable us to study the cross section of a large set of individual assets utilizing characteristics without the need to pre-specify factors, while allowing us to untwist the risk and mispricing explanations.

\begin{figure}[!htbp]
\centering
\resizebox{0.9\linewidth}{!}{
\begin{tikzpicture}
\begin{axis}[
    height=6cm,
    width=8cm,
    grid = minor,
    xmax=9.5,xmin=-1,
    ymax=,ymin=-0.2,
    xtick={0,2,4,6,8},
    ytick={0.0,0.5,1.0,1.5,2.0,2.5},
    ]
\addplot[smooth,tension=0.3, color=blue, line width=0.5pt] table[x = X,y=Y] from \normalone ;
\end{axis}
\end{tikzpicture}}
\caption{The distribution of individual US stock returns in January 2001} \label{Graph: Figure1}
\end{figure}
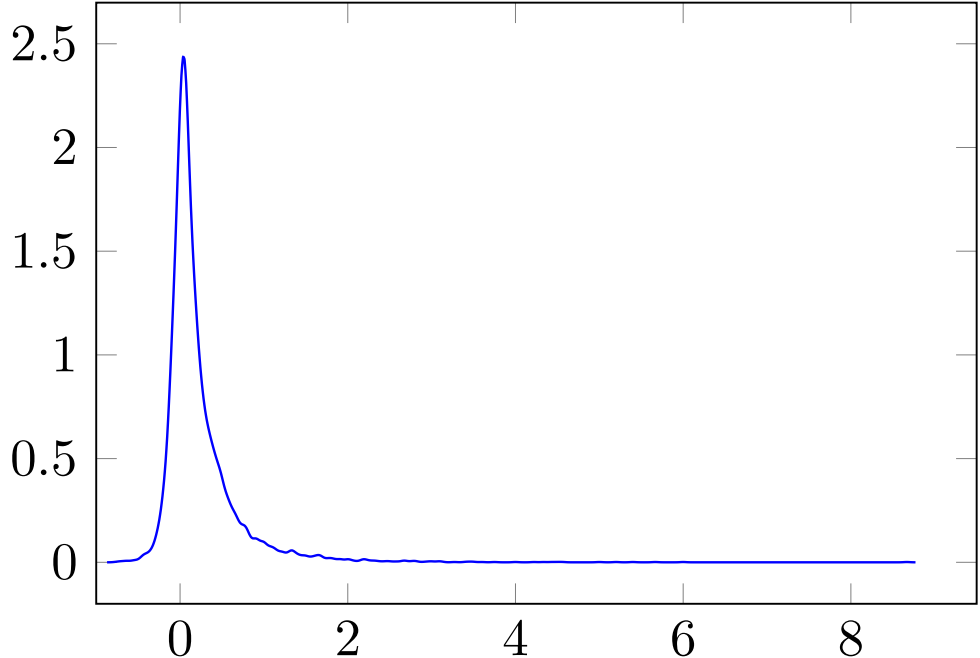

However, the models are far from being satisfactory for analyzing individual stock returns. First, the existing works that adopt least squares estimation methods may not be well suited for analyzing individual stock returns, which may exhibit heavy tails with low signal-to-noise ratios \citep{BradleyTaqqu_HeavyTails_2003,Guetal_EmpiricalAsset_2020}. Figure \ref{Graph: Figure1} shows the cross-sectional distribution of individual US stock returns in January 2021, where some observations are far away from the center of the distribution. Second, except mean factors,  the existing methods are not able to extract factors of return distributions' other information, including volatility, skewness, kurtosis and quantiles. Extracting volatility factors is an important concern in the ``idiosyncratic volatility pricing puzzle'' literature \citep{Angetal_CSVolatility_2006}. Here, we use the QR-PCA, which is robust to heavy tailed errors and allows us to estimate conditional factor structures of distributions of asset returns utilizing characteristics.

\subsection{Data and Methodology}

We use the same data set in \citet{Chenetal_SeimiparametricFactor_2021}, which contains monthly returns of $12,813$ individual stocks in the US market and $36$ characteristics with sample periods from September, 1968 to May, 2014. It is an unbalanced panel dataset, for which the QR-PCA is applicable. We implement the QR-PCA at various quantiles, and compare the results with those of the regressed-PCA. We implement both for three different specifications of the intercept function and the factor loading functions. First, we consider linear specifications by setting $\phi(z_{it})=(1,z_{it}^{\prime})^{\prime}$, denoted Model S1. Second, we consider continuous piecewise linear specifications with one or two internal knots by setting $\phi(z_{it})$ as B-splines of $z_{it}$. To fix the dimension of $\phi(z_{it})$ between the linear and nonlinear cases, we include only 18 characteristics in the case with one internal knot\textemdash denoted Model S2, and 12 characteristics in the case with two internal knots\textemdash denoted Model S3; see \citet{Chenetal_SeimiparametricFactor_2021} for the list of the 12 and 18 characteristics. We let $K_{\max}$ be the greatest integer no larger than $18$ in the implementation of $\hat{K}(\tau)$. We let the number of bootstrap draws be equal to 499 for the weighted bootstrap. In the following, we refer to the factors $\hat{F}(\tau)$ from the QR-PCA as the quantile factors, and $\hat{F}$ from the regressed-PCA as the mean factors.

\subsection{Empirical Results}\label{Sec: 72}

We first report the estimated number of factors at various quantiles, and $R^{2}$ of regressing each of the quantile factors on the mean factors in Table \ref{Tab: Empirical1}. We choose quantile $\tau = 0.01,0.05,0.1,0.25,0.5,0.75,0.9,0.95$ and $0.99$. In calculating the $R^{2}$'s, we set the number of estimated factors to $10$. The main findings are as follows. First, Model S1 has more factors at extreme quantiles (e.g.,$\tau=0.01,0.05,0.95,0.99$) than at median (i.e., $\tau=0.5$). One potential explanation for this is that $\alpha(\tau,\cdot)$ and $\beta(\tau,\cdot)$ are incorrectly specified, and misspecification error induces extra factors at extreme quantiles. Second, Models S2 and S3 have the same number of factors across quantiles, but factors are different. In particular, the quantile factors at $\tau =0.25, 0.5,0.75,0.9$ have higher correlation with the mean factors than others. Third, the median factors (that is, the quantile factors at $\tau =0.5$) are different from the mean factors, though their estimated dimensions are identical ($ \hat{K}(0.5)=\hat{K}=1$ in Model S1 and $\hat{K}(0.5)=\hat{K}=2$ in Models S1 and S2). This may be due to the heavy tails of individual stock returns. We also use our formal test to examine whether $\alpha(\tau,\cdot)=0$ for each $\tau$. The null hypothesis of $\alpha(\tau,\cdot)=0$ is rejected at $1\%$ level for all $\tau$ in all models ($p$-values which are not reported here are available upon request), regardless of the number of estimated factors used (up to 10). Hence, imposing $\alpha(\tau,\cdot)=0$ may lead to misspeciation.

\setlength{\tabcolsep}{4.5pt}
\begin{table}[!htbp]
\footnotesize
\centering
\begin{threeparttable}
\renewcommand{\arraystretch}{1.5}
\caption{Comparison between $\hat{F}(\tau)$ and $\hat{F}$}\label{Tab: Empirical1}
\begin{tabular}{ccccccccccccccccccc}
\hline\hline
&\multirow{2}{*}{Model}&\multirow{2}{*}{$\tau$}&&\multirow{2}{*}{$\hat{K}(\tau)$}&&\multicolumn{10}{c}{Column of $\hat{F}(\tau)$}&&\multirow{2}{*}{Avg}&\\
\cline{7-16}
&&&&&&$1$&$2$&$3$&$4$&$5$&$6$&$7$&$8$&$9$&$10$&&&\\
\cline{2-18}
&\multirow{9}{*}{S1}&$0.01$  &&6&&9.4	&11.1	&9.0	&8.9	&9.0	&13.7	&7.4	&8.4	&6.0	&5.4&&8.8&	\\	
&&$0.05$                     &&4&&26.5	&14.7	&19.8	&18.9	&20.5	&9.0	&13.8	&26.5	&17.9	&24.1&&19.2&	\\	
&&$0.1$                      &&1&&38.1	&20.7	&28.1	&22.7	&14.3	&16.3	&33.6	&17.4	&23.2	&34.6&&24.9&	\\	
&&$0.25$                     &&1&&52.1	&46.3	&34.2	&31.9	&38.5	&36.4	&46.1	&41.0	&46.3	&29.7&&40.3&	\\
&&$0.5$                      &&1&&61.7	&78.5	&60.3	&53.6	&48.4	&48.8	&61.4	&71.3	&43.0	&34.4&&56.1&	\\
&&$0.75$                     &&1&&68.9	&62.3	&58.2	&63.7	&49.9	&56.7	&65.9	&64.0	&46.9	&52.3&&58.9&	\\
&&$0.9$                      &&1&&67.0	&40.9	&44.9	&45.6	&47.0	&37.4	&51.1	&48.0	&40.9	&55.3&&47.8&	\\
&&$0.95$                     &&5&&60.1	&34.4	&41.9	&34.3	&43.5	&36.3	&35.4	&26.3	&34.9	&33.9&&38.1&	\\
&&$0.99$                     &&5&&44.3	&20.1	&24.1	&22.1	&22.4	&21.7	&21.5	&22.0	&14.8	&19.9&&23.3&	\\
\cline{2-18}
&\multirow{9}{*}{S2}&$0.01$  &&2&&31.9	&52.4	&11.7	&12.7	&21.4	&8.0	&16.0	&9.9	&15.3	&17.8&&19.7&	\\
&&$0.05$                     &&2&&50.2	&53.7	&39.7	&47.0	&45.4	&56.3	&22.2	&35.9	&16.8	&22.6&&39.0&	\\
&&$0.1$                      &&2&&55.4	&55.4	&58.0	&60.7	&41.4	&60.3	&51.9	&44.7	&39.6	&9.5&&47.7&	\\
&&$0.25$                     &&2&&61.8	&62.0	&71.6	&72.1	&53.2	&78.0	&56.0	&53.1	&64.2	&62.3&&63.4&	\\
&&$0.5$                      &&2&&79.4	&73.3	&77.1	&84.6	&83.4	&53.2	&77.2	&67.3	&69.1	&22.0&&68.7&	\\
&&$0.75$                     &&2&&88.2	&70.6	&80.4	&82.8	&81.9	&62.4	&70.1	&48.1	&71.3	&17.5&&67.3&	\\
&&$0.9$                      &&2&&75.2	&60.9	&70.6	&62.9	&70.2	&53.3	&50.9	&39.8	&53.5	&26.5&&56.4&	\\
&&$0.95$                     &&2&&69.3	&54.7	&60.9	&46.4	&47.4	&47.5	&45.6	&24.6	&47.8	&3.5&&44.8&	\\
&&$0.99$                     &&2&&57.8	&44.4	&32.6	&30.9	&9.5	&16.1	&26.2	&32.4	&16.2	&25.0&&29.1&	\\
\cline{2-18}
&\multirow{9}{*}{S3}&$0.01$  &&2&&34.9	&48.3	&14.6	&11.9	&11.8	&13.0	&25.1	&7.5	&41.8	&11.9&&22.1&	\\
&&$0.05$                     &&2&&50.8	&49.0	&49.1	&50.5	&36.0	&49.6	&25.5	&32.8	&42.7	&30.1&&41.6&	\\
&&$0.1$                      &&2&&52.8	&49.6	&63.8	&57.8	&51.5	&64.8	&42.6	&39.1	&55.0	&49.3&&52.6&	\\
&&$0.25$                     &&2&&61.2	&56.7	&77.2	&70.4	&64.4	&83.3	&58.0	&55.1	&58.5	&56.1&&64.1&    \\
&&$0.5$                      &&2&&80.0	&76.1	&85.9	&78.9	&77.2	&87.2	&71.2	&60.8	&74.7	&62.6&&75.5&    \\
&&$0.75$                     &&2&&89.5	&72.8	&86.8	&82.8	&83.8	&81.6	&75.4	&64.6	&57.8	&66.5&&76.2&    \\
&&$0.9$                      &&2&&78.3	&56.6	&74.7	&78.5	&75.5	&71.0	&40.6	&44.5	&44.8	&55.7&&62.0&    \\
&&$0.95$                     &&2&&72.5	&51.1	&66.1	&59.8	&67.5	&42.5	&36.0	&54.5	&29.4	&19.5&&49.9&	\\
&&$0.99$                     &&2&&60.7	&45.2	&42.4	&33.7	&42.5	&26.6	&10.7	&26.3	&21.6	&16.8&&32.7&	\\
\hline\hline
\end{tabular}
\begin{tablenotes}
      \item[\dag] {\footnotesize{The third column reports $\hat{K}(\tau)$ at each $\tau$. The fourth to the second last columns report the $R^2$ of regressing each column of $\hat{F}(\tau)$ on $\hat{F}$ ($\%$). The last column reports the average of the fourth to the second last columns. For both $\hat{F}(\tau)$ and $\hat{F}$, the number of estimated factors is set to $10$. }}
    \end{tablenotes}
\end{threeparttable}
\end{table}%

We then report $R^{2}$ of regressing each of the median factors and the mean factors on six observed factors in Table \ref{Tab: Empirical2}. The six observed factors are market excess return, ``small minus big'' factor, ``high minus low'' factor, ``momentum'' factor, ``robust minus weak'' factor, and ``conservative minus aggressive'' factor. On average, the median factors have higher correlation with the six factors than the mean factors in all models.

\setlength{\tabcolsep}{5pt}
\begin{table}[!htbp]
\footnotesize
\centering
\begin{threeparttable}
\renewcommand{\arraystretch}{1.5}
\caption{Comparison between $\hat{F}(0.5)$/$\hat{F}$ and six observed factors}\label{Tab: Empirical2}
\begin{tabular}{ccccccccccccccccc}
\hline\hline
&\multirow{2}{*}{Model}&\multirow{2}{*}{Factor}&&\multicolumn{10}{c}{Column of $\hat{F}(0.5)$/$\hat{F}$}&&\multirow{2}{*}{Avg}&\\
\cline{5-14}
&&&&$1$&$2$&$3$&$4$&$5$&$6$&$7$&$8$&$9$&$10$&&&\\
\cline{2-16}
&\multirow{2}{*}{S1}&$\hat{F}(0.5)$  &&10.1	&9.8	&59.2&19.7	&9.4	&3.2	&8.7	&3.9	&34.1	&1.8&&16.0&	\\
&&$\hat{F}$                          &&10.1	&12.6	&13.9	&12.3	&46.5	&3.2	&14.1	&13.2	&23.5	&9.4&&15.9&	\\
\cline{2-16}
&\multirow{2}{*}{S2}&$\hat{F}(0.5)$  &&10.1	&26.8	&37.9	&41.9	&51.8	&6.1	&13.9	&15.7	&12.4	&9.2&&22.6&  \\
&&$\hat{F}$                          &&10.1	&28.0	&30.4	&47.9	&31.4	&8.4	&7.9	&10.5	&9.4	&6.7&&19.1&  \\
\cline{2-16}
&\multirow{2}{*}{S3}&$\hat{F}(0.5)$  &&10.1	&26.7	&31.9	&57.0	&40.8	&7.0	&16.9	&33.2	&2.3	&11.7&&23.8&  \\
&&$\hat{F}$                          &&10.1	&25.4	&27.1	&59.1	&25.8	&6.0	&13.1	&5.2	&23.9	&20.3&&21.6&  \\
\hline\hline
\end{tabular}
\begin{tablenotes}
      \item[\dag] {\footnotesize{The sixed observed factors are market excess return, ``small minus big'' factor, ``high minus low'' factor, ``momentum'' factor, ``robust minus weak'' factor, and ``conservative minus aggressive'' factor. The third to the second last columns report the $R^2$ of regressing each column of $\hat{F}(0.5)$/$\hat{F}$ on the six observed factors ($\%$). The last column reports the average of the third to the second last columns. For both $\hat{F}(0.5)$ and $\hat{F}$, the number of estimated factors is set to $10$. }}
    \end{tablenotes}
\end{threeparttable}
\end{table}%

We further evaluate the performance of the median factors and the mean factors by comparing their ability to explain the cross section of portfolio returns. We note that as the QR-PCA and the regressed-PCA have different objective functions, it is hard to choose a measure to compare their goodness of fit for the individual stock returns.
Our analysis includes 202 portfolios available on Kenneth French's website: 25 portfolios sorted by size and book-to-market ratio, 17 industry portfolios, 25 portfolios sorted by operating profitability and investment, 25 portfolios sorted by size and variance, 35 portfolios sorted by size and net issuance, 25 portfolios sorted by size and accruals, 25 portfolios sorted by size and momentum, and 25 portfolios sorted by size and beta.

Let $r_{it}$ be the excess return of portfolio $i$ in time period $t$. For each $i$, we run the time series regression of $r_{it}$ on constant and the median factors $\hat{f}_t(0.5)$ (resp. the mean factors $\hat{f}_{t}$) using all observations to obtain $\hat{\alpha}_i(0.5)$ and $\hat{\beta}_i(0.5)$ (resp. $\hat{\alpha}_i$ and $\hat{\beta}_i$). First, we compute three types of $R^2$'s that directly speak to the ability of $\hat{f}_{t}(0.5)$/$\hat{f}_{t}$ to explain the portfolio returns. The first one is total $R^2$. The second one is the cross-sectional average of time series $R^2$ across portfolios, which reflects the ability of $\hat{f}_{t}(0.5)$/$\hat{f}_{t}$ to capture common variation in the portfolio returns. The third one is the time series average of cross-sectional $R^2$, which is most relevant for evaluating factors' ability to explain the cross section of average returns.
\begin{align}
R^2 & = 1-\frac{\sum_{i, t}[r_{it}- \{\hat{r}_{it}(0.5) \text{ or }\hat{r}_{it}\}]^2}{\sum_{i, t} r_{it}^{2}}, \label{Eqn: R21}\\
R^2_{T,N} & = 1 - \frac{1}{N} \sum_{i} \frac{\sum_{t}[r_{it}- \{\hat{r}_{it}(0.5) \text{ or }\hat{r}_{it}\}]^2}{\sum_{t}r_{i t}^{2}},\label{Eqn: R22}\\
R^2_{N,T} & = 1 - \frac{1}{T} \sum_{t} \frac{\sum_{i}[r_{it}-\{\hat{r}_{it}(0.5) \text{ or }\hat{r}_{it}\}]^2}{\sum_{i}r_{it}^{2}},\label{Eqn: R23}
\end{align}
where $\hat{r}_{it}(0.5) = \hat{\alpha}_i(0.5) + \hat{\beta}_i(0.5)^{\prime}\hat{f}_t(0.5)$ and $\hat{r}_{it} = \hat{\alpha}_i + \hat{\beta}_i^{\prime}\hat{f}_t$. Second, we consider another version of these goodness-of-fit measures by excluding $\hat{\alpha}_i(0.5)$ and $\hat{\alpha}_i$:
\begin{align}
R^2_{f}& = 1-\frac{\sum_{i, t}[r_{it}- \{\hat{\beta}_i(0.5)^{\prime}\hat{f}_t(0.5) \text{ or } \hat{\beta}_i^{\prime}\hat{f}_t\}]^2}{\sum_{i, t} r_{it}^{2}}, \label{Eqn: R24}\\
R^2_{f, T,N} & = 1 - \frac{1}{N} \sum_{i} \frac{\sum_{t}[r_{it}- \{\hat{\beta}_i(0.5)^{\prime}\hat{f}_t(0.5) \text{ or } \hat{\beta}_i^{\prime}\hat{f}_t\}]^2}{\sum_{t}r_{it}^{2}},\label{Eqn: R52}\\
R^2_{f, N,T} & = 1 - \frac{1}{T} \sum_{t} \frac{\sum_{i}[r_{it}- \{\hat{\beta}_i(0.5)^{\prime}\hat{f}_t(0.5) \text{ or } \hat{\beta}_i^{\prime}\hat{f}_t\}]^2}{\sum_{i}r_{it}^{2}}.\label{Eqn: R26}
\end{align}
Third, we assess the out-of-sample prediction. For $s\geq 240$, we run the time series regression of $r_{it}$ on constant and $\hat{f}_t(0.5)$ (resp. $\hat{f}_t$) using observations through $s$ to obtain $\hat{\beta}_{is}(0.5)$ (resp. $\hat{\beta}_{is}$) for each $i$, and compute the out-of-sample predictor of $r_{i,s+1}$ as $\hat{\beta}_{is}(0.5)^{\prime}\hat{\lambda}_{s}(0.5)$ (resp. $\hat{\beta}_{is}^{\prime}\hat{\lambda}_{s}$), where $\hat{\lambda}_{s}(0.5)$ (resp. $\hat{\lambda}_{s}$) is estimated risk premium that is obtained by regressing $\sum_{t=1}^{s}r_{it}/s$ on $\hat{\beta}_{is}(0.5)$ (resp. $\hat{\beta}_{is}$) without intercept. We can define three types of out-of-sample predictive $R^2$ analogously:
\begin{align}
R^2_O & = 1-\frac{\sum_{i, s\geq 240}[r_{i,s+1}- \{\hat{\beta}_{is}(0.5)^{\prime}\hat{\lambda}_{s}(0.5) \text{ or } \hat{\beta}_{is}^{\prime}\hat{\lambda}_{s}\}]^2}{\sum_{i, s\geq 240} r_{i, s+1}^{2}}, \label{Eqn: R21Predictive}\\
R^2_{T,N,O} & = 1 - \frac{1}{N} \sum_{i} \frac{\sum_{s\geq 240}[r_{i,s+1}-\{\hat{\beta}_{is}(0.5)^{\prime}\hat{\lambda}_{s}(0.5) \text{ or } \hat{\beta}_{is}^{\prime}\hat{\lambda}_{s}\}]^2}{\sum_{s\geq 240}r_{i, s+1}^{2}},\label{Eqn: R22Predictive}\\
R^2_{N,T,O} & = 1 - \frac{1}{T-240} \sum_{s\geq 240} \frac{\sum_{i}[r_{i,s+1}- \{\hat{\beta}_{is}(0.5)^{\prime}\hat{\lambda}_{s}(0.5) \text{ or } \hat{\beta}_{is}^{\prime}\hat{\lambda}_{s}\}]^2}{\sum_{i}r_{i, s+1}^{2}}.\label{Eqn: R23Predictive}
\end{align}

\begin{figure}[!htbp]
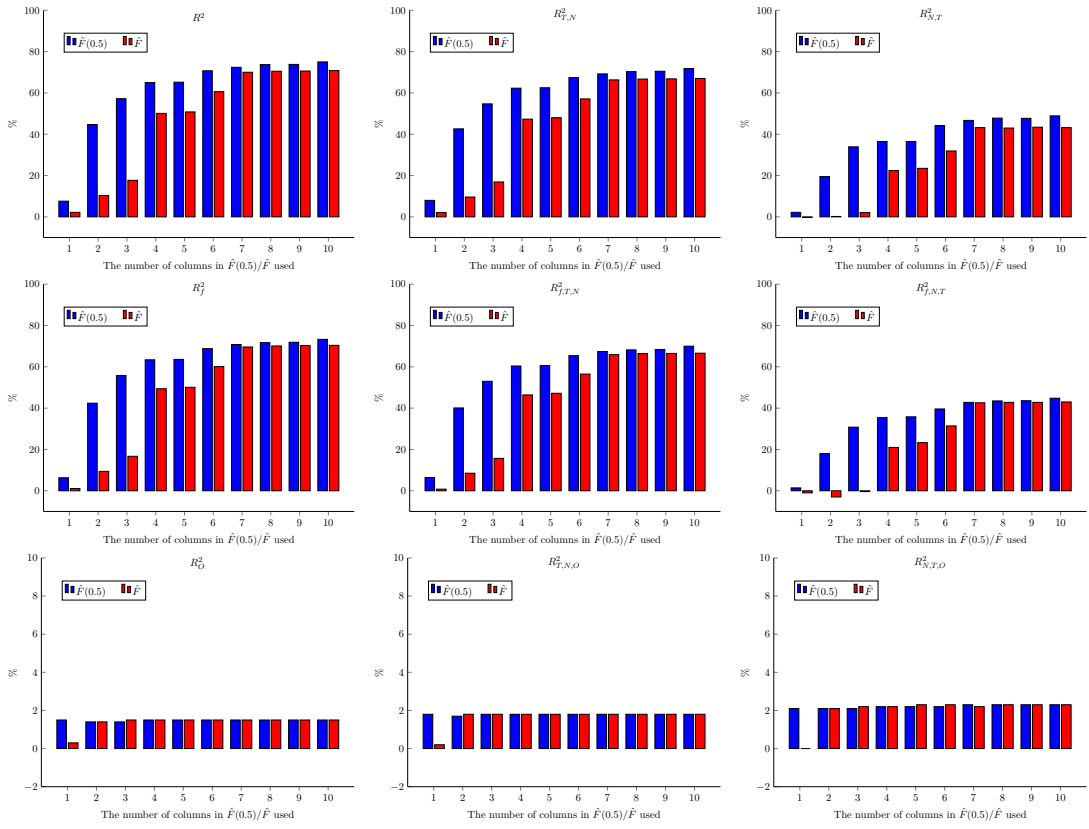

\centering
\begin{subfigure}[b]{0.32\textwidth}
\centering
\resizebox{\linewidth}{!}{
\pgfplotsset{title style={at={(0.5,0.91)}}}
}
\end{subfigure}
\caption{Evaluating $\hat{F}(0.5)$/$\hat{F}$ using portfolios: Model S1} \label{Fig: Emprical6}
\end{figure}

The results are reported in Figures \ref{Fig: Emprical6}-\ref{Fig: Emprical8}. The main findings are as follows. First, in Model S1, the median factors have better in-sample explanatory power for the portfolio returns, regardless of the number of factors used and which in-sample $R^2$ used. For example, when two factors are used, $\hat{f}_{t}(0.5)$ has much higher $R^{2}_{N,T}$ and $R^{2}_{f,N,T}$ than $\hat{f}_{t}$ ($19.5\%$ and $18.0\%$ v.s. $0.2\%$ and $-3.0\%$). Second, in Models S2 and S3, the mean factors may have better in-sample explanatory power for the portfolio returns when only the first factor is used, but the median factors do better when more than three factors are used regardless of which in-sample $R^2$ used. Third, the median factors and the mean factors have similar out-of-sample prediction power for the portfolio returns in all cases, except for Model S1 with only one factor used, where $\hat{f}_{t}(0.5)$ has much higher $R^{2}_{O}$, $R^{2}_{T,N,O}$ and $R^{2}_{N,T,O}$ than $\hat{f}_{t}$ ($1.4\%$, $1.8\%$ and $2.1\%$ v.s. $0.3\%$, $0.2\%$ and $0.0\%$).

\begin{figure}[!htbp]
\centering
\begin{subfigure}[b]{0.32\textwidth}
\centering
\resizebox{\linewidth}{!}{
\pgfplotsset{title style={at={(0.5,0.91)}}}
%
}
\end{subfigure}
\caption{Evaluating $\hat{F}(0.5)$/$\hat{F}$ using portfolios: Model S2} \label{Fig: Emprical7}
\end{figure}

\begin{figure}[!htbp]
\centering
\begin{subfigure}[b]{0.32\textwidth}
\centering
\resizebox{\linewidth}{!}{
\pgfplotsset{title style={at={(0.5,0.91)}}}
%
}
\end{subfigure}
\caption{Evaluating $\hat{F}(0.5)$/$\hat{F}$ using portfolios: Model S3} \label{Fig: Emprical8}
\end{figure}

To conclude, the QR-PCA has extracted very different quantile factors from the regressed-PCA's mean factors, and the median factors can help improve the mean factors' ability to explain the cross section of portfolio returns. We also conduct analysis for daily observations in 2005, and have similar findings; see Appendix \ref{App: F}.

\section{Conclusion}\label{Sec: 8}
In this paper, we developed a simple sieve estimation for conditional quantile factor models. We established large sample properties of the estimators under large $N$ without requiring large $T$. We also developed a weighted bootstrap for estimating the asymptotic distributions of the estimators. The results enable us to estimate conditional factor structures of distributions of asset returns utilizing characteristics. They also provide a robust, parsimonious and meaningful way of utilizing characteristics to analyze the cross-sectional differences of average individual stock returns in the presence of large return observations without the need to pre-specify factors.

\addcontentsline{toc}{section}{References}
\putbib
\end{bibunit}

\clearpage
\newpage

\begin{bibunit}
\begin{appendices} \sloppy
\bookmarksetup{open=false}
\allowdisplaybreaks 
\titleformat{\section}{\Large\center}{{\sc Appendix} \thesection}{0.25em}{- }
\setcounter{page}{1}
\renewcommand{\thetable}{\thesection.\Roman{table}}
\emptythanks
\phantomsection
\pdfbookmark[1]{Appendix Title}{title1}
\title{\vspace{-2cm}Supplement to ``Robust Estimation of Conditional Factor Models''}
\author{Qihui Chen\\ School of Management and Economics\\ The Chinese University of Hong Kong, Shenzhen\\ qihuichen@cuhk.edu.cn}
\date{}
\maketitle
\vspace{-0.3in}

\section{Proof of Theorem \ref{Thm: RateConv}}
\renewcommand{\theequation}{A.\arabic{equation}}
\setcounter{equation}{0}
\subsection{Proof of Theorem \ref{Thm: RateConv}}
\noindent{\sc Proof of Theorem \ref{Thm: RateConv}:} By the triangle inequality, it follows from the fact that $T$ is finite, Assumption \ref{Ass: Basis}(iv) and Lemma \ref{Lem: TechA1} that
\begin{align}\label{Eqn: Thm: RateConv: 1}
\sup_{\tau\in\mathcal{T}}\|\tilde{Y}(\tau) - a(\tau)1_{T}^{\prime} - B(\tau)F(\tau)^{\prime}\|^2_{F} = O_{p}\left(\frac{{J}}{{N}}+\frac{1}{J^{2\kappa}\xi_{J}^{2}}\right),
\end{align}
where recall that $1_{T}$ denote a $T\times 1$ vector of ones. Recall that $M_T= I_{T} - 1_T1_T^{\prime}/T$. Since $\|\tilde{Y}(\tau) - a1_{T}^{\prime} - B(\tau)F(\tau)^{\prime}\|^2_{F} = \|\tilde{Y}(\tau)M_T - B(\tau)[M_TF(\tau)]^{\prime}\|^{2}_{F} + \|\bar{\tilde{Y}}(\tau)1_T^{\prime} - a(\tau)1_{T}^{\prime}-B(\tau)\bar{f}(\tau)1_{T}^{\prime}\|^{2}_{F}$, it then follows from \eqref{Eqn: Thm: RateConv: 1} that
\begin{align}\label{Eqn: Thm: RateConv: 2}
\sup_{\tau\in\mathcal{T}}\|\bar{\tilde{Y}}(\tau) -  a(\tau)- B(\tau)\bar{f}(\tau)\|^{2} = O_{p}\left(\frac{{J}}{{N}}+\frac{1}{J^{2\kappa}\xi_{J}^{2}}\right)
\end{align}
and
\begin{align}\label{Eqn: Thm: RateConv: 3}
\sup_{\tau\in\mathcal{T}}\|\tilde{Y}(\tau)M_T - B(\tau)[M_TF(\tau)]^{\prime}\|^{2}_{F} = O_{p}\left(\frac{{J}}{{N}}+\frac{1}{J^{2\kappa}\xi_{J}^{2}}\right).
\end{align}
Let $V(\tau)$ be a $K(\tau)\times K(\tau)$ diagonal matrix of the first $K(\tau)$ largest eigenvalues of $\tilde{Y}(\tau)M_T\tilde{Y}(\tau)^{\prime}/T$. By the definitions of $\hat{B}(\tau)$ and $\hat{F}(\tau)$, $\hat{F}(\tau)^{\prime}M_T\hat{F}(\tau)/T =\hat{B}(\tau)^{\prime}[\tilde{Y}(\tau)M_T\tilde{Y}(\tau)^{\prime}/T]\hat{B}(\tau)=V(\tau)$. We have $\hat{B}(\tau) = [\tilde{Y}(\tau)M_T\tilde{Y}(\tau)^{\prime}/T]\hat{B}(\tau)V(\tau)^{-1}$ and $[F(\tau)^{\prime}M_T\tilde{Y}(\tau)^{\prime}\hat{B}(\tau)/T]V(\tau)^{-1}=H(\tau)$, so
\begin{align}\label{Eqn: Thm: RateConv: 4}
\hat{B}(\tau)-B(\tau)H(\tau) = \frac{1}{T}[\tilde{Y}(\tau)M_T - B(\tau)[M_TF(\tau)]^{\prime}][\tilde{Y}(\tau)M_T]^{\prime}\hat{B}(\tau)V(\tau)^{-1}.
\end{align}
By the triangle inequality, $\sup_{\tau\in\mathcal{T}}\|\tilde{Y}(\tau)M_T\|_{2}=O_{p}(1)$ follows from \eqref{Eqn: Thm: RateConv: 3}, \eqref{Eqn: Thm: RateConv: 4}, and Assumptions \ref{Ass: Basis}(ii) and \ref{Ass: InterceptLoadingsFactors}(iii). Thus, the second result of the theorem follows from \eqref{Eqn: Thm: RateConv: 3}, \eqref{Eqn: Thm: RateConv: 4} and Lemma \ref{Lem: TechA2}(i). By the definition of $\hat{a}(\tau)$,
\begin{align}\label{Eqn: Thm: RateConv: 5}
\hat{a}(\tau) - a(\tau) &= -\hat{B}(\tau)[\hat{B}(\tau)-B(\tau)H(\tau)]^{\prime}a(\tau) \notag\\
&\hspace{0.5cm}- [I_{JM}- \hat{B}(\tau)\hat{B}(\tau)^{\prime}][\hat{B}(\tau)-B(\tau)H(\tau)]H(\tau)^{-1}\bar{f}(\tau)\notag\\
&\hspace{0.5cm} +[I_{JM} - \hat{B}(\tau)\hat{B}(\tau)^{\prime}][\bar{\tilde{Y}}(\tau) -  a(\tau)- B(\tau)\bar{f}(\tau)],
\end{align}
where $H(\tau)^{-1}$ is well defined with probability approaching one by the first result of the theorem and Lemma \ref{Lem: TechA2}(ii), and we have used $a(\tau)^{\prime}B(\tau)=0$ and $[I_{JM} - \hat{B}(\tau)\hat{B}(\tau)^{\prime}]\hat{B}(\tau)=0$. Thus, by the triangle inequality, the first result of the theorem follows from the second result, \eqref{Eqn: Thm: RateConv: 2}, \eqref{Eqn: Thm: RateConv: 5}, Assumptions \ref{Ass: Basis}(ii) and \ref{Ass: InterceptLoadingsFactors}(ii), and Lemma \ref{Lem: TechA2}(ii). By the definition of $\hat{F}(\tau)$,
\begin{align}\label{Eqn: Thm: RateConv: 6}
\hat{F}(\tau) - F(\tau)[H(\tau)^{\prime}]^{-1} &= 1_{T}a(\tau) ^{\prime}[\hat{B}(\tau)-B(\tau)H(\tau)] \notag\\
&\hspace{0.5cm} - F(\tau)[H(\tau)^{\prime}]^{-1}[\hat{B}(\tau)-B(\tau)H(\tau)]^{\prime}\hat{B}(\tau) \notag\\
&\hspace{0.5cm}+ [\tilde{Y}(\tau) - a(\tau)1_{T}^{\prime} - B(\tau)F(\tau)^{\prime}]^{\prime}\hat{B}(\tau).
\end{align}
Thus, by the triangle inequality, the third result of the theorem follows from the second result, \eqref{Eqn: Thm: RateConv: 1}, \eqref{Eqn: Thm: RateConv: 6}, Assumptions \ref{Ass: Basis}(ii) and \ref{Ass: InterceptLoadingsFactors}(ii), and Lemma \ref{Lem: TechA2}(ii). Since $\hat{\beta}(\tau,z) = \hat{B}(\tau)^{\prime}\phi(z)$ and $\beta(\tau,z)=B(\tau)^{\prime}\phi(z) + \delta(\tau,z)$,
\begin{align}\label{Eqn: Thm: RateConv: 7}
\hat{\beta}(\tau,z)-H(\tau)^{\prime}\beta(\tau,z)= [\hat{B}(\tau)-B(\tau)H(\tau)]^{\prime}\phi(z) + H(\tau)^{\prime}\delta(\tau,z).
\end{align}
Thus, by the triangle inequality, the fifth result of the theorem follows from the second result, Assumption \ref{Ass: Basis}(iii), and Lemma \ref{Lem: TechA2}(i). The proof of the fourth result is similar. To prove the last result, we use Lemma \ref{Lem: TechA3} instead of Lemma \ref{Lem: TechA1}. By a similar argument for \eqref{Eqn: Thm: RateConv: 1}, for each $\tau\in\mathcal{T}$, $\|\lambda^{\prime}[\tilde{Y}(\tau) - a(\tau)1_{T}^{\prime} - B(\tau)F(\tau)^{\prime}]\|=O_{p}(1/N+J^{-2\kappa}\xi_{J}^{-2})$ for any $\lambda\in\mathbf{R}^{JM}$ such that $\|\lambda\|$ is bounded. This implies that for each $\tau\in\mathcal{T}$,
\begin{align}\label{Eqn: Thm: RateConv: 8}
\|B(\tau)^{\prime}[\tilde{Y}(\tau) - a(\tau)1_{T}^{\prime} - B(\tau)F(\tau)^{\prime}]\|^2_{F} = O_{p}\left(\frac{{1}}{{N}}+\frac{1}{J^{2\kappa}\xi_{J}^{2}}\right)
\end{align}
for each $\tau\in\mathcal{T}$. By a similar argument for the second result of the theorem, for each $\tau\in\mathcal{T}$, $\|\lambda^{\prime}[\hat{B}(\tau)-B(\tau)H(\tau)]\| = O_{p}(1/N+J^{-2\kappa}\xi_{J}^{-2})$ for any $\lambda\in\mathbf{R}^{JM}$ such that $\|\lambda\|$ is bounded. This implies that for each $\tau\in\mathcal{T}$,
\begin{align}\label{Eqn: Thm: RateConv: 9}
\|a(\tau)^{\prime}[\hat{B}(\tau)-B(\tau)H(\tau)]\| =  O_{p}\left(\frac{{1}}{{N}}+\frac{1}{J^{2\kappa}\xi_{J}^{2}}\right)
\end{align}
and
\begin{align}\label{Eqn: Thm: RateConv: 10}
\|B(\tau)^{\prime}[\hat{B}(\tau)-B(\tau)H(\tau)]\|_{F} =  O_{p}\left(\frac{{1}}{{N}}+\frac{1}{J^{2\kappa}\xi_{J}^{2}}\right).
\end{align}
Thus, by the triangle inequality, the last result of the theorem follows from the second result, \eqref{Eqn: Thm: RateConv: 1}, \eqref{Eqn: Thm: RateConv: 6}, \eqref{Eqn: Thm: RateConv: 8}-\eqref{Eqn: Thm: RateConv: 10}, Assumptions \ref{Ass: Basis}(ii), \ref{Ass: InterceptLoadingsFactors}(ii) and (iii), and Lemma \ref{Lem: TechA2}, noting that $J=o(\sqrt{N})$. This completes the proof of the theorem. \qed

\subsection{Auxiliary Lemmas}
\begin{lem}\label{Lem: TechA1}
Suppose Assumptions \ref{Ass: DGP} and \ref{Ass: Basis} hold. Assume (i) $N\to\infty$; (ii) $T$ is finite; (iii) $J\to\infty$ with $J\xi^{2}_{J}\log^{2} N=o(N)$ and $J^{-\kappa}\log N=o(1)$. Then $\sup_{t\leq T}\sup_{\tau\in\mathcal{T}}\|\tilde{Y}_t(\tau) -\beta_{t}(\tau)\|=O_{p}(\sqrt{{J}/{N}})$.
\end{lem}
\noindent{\sc Proof:} It is noted that $Q_{y_{t}|z_{t}}(\tau,z)=\alpha(\tau,z)+ \beta(\tau,z)^{\prime}f_t(\tau)$ for each $t\leq T$. By the triangle inequality, for each $t\leq T$,
\begin{align}\label{Eqn: Lem: TechA1: 1}
&\hspace{0.5cm}\sup_{\tau\in\mathcal{T},z\in\mathcal{Z}_t}|Q_{y_{t}|z_{t}}(\tau,z)-\phi(z)^{\prime}\beta_{t}(\tau)|\notag\\
&\leq\sup_{\tau\in\mathcal{T},z\in\mathcal{Z}_t}|\alpha(\tau,z)+ \beta(\tau,z)^{\prime}f_t(\tau)-\phi(z)^{\prime}[a(\tau)+B(\tau)f_t(\tau)]|\notag\\
&+\sup_{\tau\in\mathcal{T},z\in\mathcal{Z}_t}|\phi(z)^{\prime}[a(\tau)+B(\tau)f_t(\tau) - \beta_{t}(\tau)]|\notag\\
&\leq M\max_{m\leq M}\sup_{\tau\in\mathcal{T},z\in\mathcal{Z}_{t,m}}\hspace{-0.1cm}|r_{m,J}(\tau,z)|+M\sqrt{K(\tau)}\max_{k\leq K,m\leq M}\sup_{\tau\in\mathcal{T},z\in\mathcal{Z}_{t,m}}\hspace{-0.1cm}|\delta_{km,J}(\tau,z)|\sup_{\tau\in\mathcal{T}}\|f_{t}(\tau)\|\notag\\
&+ \sup_{z\in\mathcal{Z}_t}\|\phi(z)\|\sup_{\tau\in\mathcal{T}}\|a(\tau)+B(\tau)f_t(\tau) - \beta_{t}(\tau)\|\notag\\
&= O\left(J^{-\kappa}\right)+O\left(J^{-\kappa}\right)+O\left(\xi_{J}J^{-\kappa}\xi_{J}^{-1}\right)=O\left(J^{-\kappa}\right),
\end{align}
where the last equality follows by Assumptions \ref{Ass: Basis}(iii) and (iv). Since $T$ is finite, the result of the lemma thus follows by Theorem 1 of \citet{Bellonietal_ConditionalQuantile_2019}. \qed

\begin{lem}\label{Lem: TechA2}
Suppose Assumptions \ref{Ass: DGP}-\ref{Ass: InterceptLoadingsFactors} hold. Let $V(\tau)$ be given in the proof of Theorem \ref{Thm: RateConv}. Assume (i) $N\to\infty$; (ii) $T\geq K(\tau)+1$ is finite; (iii) $J\to\infty$ with $J\xi^{2}_{J}\log^{2} J=o(N)$ and $J^{-\kappa}\log N=o(1)$. Then (i) $\sup_{\tau\in\mathcal{T}}\|V(\tau)\|_{2} = O_{p}(1)$, $\sup_{\tau\in\mathcal{T}}\|V(\tau)^{-1}\|_{2}=O_{p}(1)$, and $\sup_{\tau\in\mathcal{T}}\|H(\tau)\|_{2}=O_{p}(1)$; (ii) $\sup_{\tau\in\mathcal{T}}\|H(\tau)^{-1}\|_{2}=O_{p}(1)$, if $\sup_{\tau\in\mathcal{T}}\|\hat{B}(\tau) - B(\tau) H(\tau)\|_{F}=o_{p}(1)$.
\end{lem}
\noindent{\sc Proof:} (i) Let $\mathcal{V}(\tau)$ be a $K(\tau)\times K(\tau)$ diagonal matrix of the eigenvalues of $[F(\tau)^{\prime}M_TF(\tau)/T]B(\tau)^{\prime}B(\tau)$, which are equal to the first $K(\tau)$ largest eigenvalues of $B(\tau)F(\tau)^{\prime}M_TF(\tau)B(\tau)^{\prime}/T$. Using the Weyl's inequality,
\begin{align}\label{Eqn: Lem: TechA2: 1}
\|V(\tau)-\mathcal{V}(\tau)\|_2 \leq \|{\tilde{Y}(\tau)M_T\tilde{Y}(\tau)^{\prime}}/{T}-{B(\tau)F(\tau)^{\prime}M_TF(\tau)B(\tau)^{\prime}}/{T}\|_{F}.
\end{align}
Thus, combining \eqref{Eqn: Thm: RateConv: 3} and \eqref{Eqn: Lem: TechA2: 1} implies $\sup_{\tau\in\mathcal{T}}\|V(\tau)-\mathcal{V}(\tau)\|_2 = o_{p}(1)$, therefore $\sup_{\tau\in\mathcal{T}}\|V(\tau)\|_{2} = O_{p}(1)$ and $\sup_{\tau\in\mathcal{T}}\|V(\tau)^{-1}\|_{2} = \sup_{\tau\in\mathcal{T}}\lambda^{-1}_{\min}(V(\tau)) =O_{p}(1)$ follow by Assumptions \ref{Ass: Basis}(ii), \ref{Ass: InterceptLoadingsFactors}(iii) and (iv). Let $H^{\diamond}(\tau)\equiv[F(\tau)^{\prime}M_TF(\tau)/T]B(\tau)^{\prime}\hat{B}(\tau)V(\tau)^{-1}$. Recall that $H(\tau) = [F(\tau)^{\prime}M_T\tilde{Y}(\tau)^{\prime}\hat{B}(\tau)/T]V(\tau)^{-1}$. Then,
\begin{align}\label{Eqn: Lem: TechA2: 2}
\hspace{-0.2cm}\|H(\tau) \hspace{-0.05cm}-\hspace{-0.05cm} H^{\diamond}(\tau)\|_{2}\leq\frac{1}{T}\|F(\tau)\|_{2}\|\tilde{Y}(\tau)M_T\hspace{-0.05cm}-\hspace{-0.05cm}B(\tau)[M_TF(\tau)]^{\prime}\|_{F}\|\hat{B}(\tau)\|_2\|V(\tau)^{-1}\hspace{-0.05cm}\|_2.
\end{align}
Thus, combining \eqref{Eqn: Thm: RateConv: 3} and \eqref{Eqn: Lem: TechA2: 2} implies $\sup_{\tau\in\mathcal{T}}\|H(\tau)-H^{\diamond}(\tau)\|_2 = o_{p}(1)$ by Assumptions \ref{Ass: Basis}(ii) and the second result of the lemma. Since $\|H^{\diamond}(\tau)\|_{2} \leq \|F(\tau)^{\prime}M_{T}F(\tau)/T\|_{2}$ $\|B(\tau)\|_{2}\|\hat{B}(\tau)\|_2\|V(\tau)^{-1}\|_2$, the third result of the lemma follows by the second result and Assumptions \ref{Ass: Basis}(ii) and \ref{Ass: InterceptLoadingsFactors}(iii).

(ii) Since $\|C+D\|_{F}\leq \|C\|_{F}+\|D\|_{F}$ and $\|CD\|_{F}\leq \|C\|_{2}\|D\|_{F}$,
\begin{align}\label{Eqn: Lem: TechA2: 3}
&\|\hat{B}(\tau)^{\prime}\hat{B}(\tau)-H(\tau)^{\prime}B(\tau)^{\prime}B(\tau)H(\tau)\|_{F}\notag\\
&\hspace{-0.5cm}\leq\|\hat{B}(\tau)\|_{2}\|\hat{B}(\tau) - B(\tau) H(\tau)\|_{F} + \|\hat{B}(\tau) - B(\tau)H(\tau)\|_{F}\|B(\tau)\|_{2}\|H(\tau)\|_{2}.
\end{align}
Thus, $\sup_{\tau\in\mathcal{T}}\|I_{K}-H(\tau)^{\prime}B(\tau)^{\prime}B(\tau)H(\tau)\|_{F}=o_{p}(1)$ by Assumption \ref{Ass: InterceptLoadingsFactors}(iii), since $\sup_{\tau\in\mathcal{T}}\|\hat{B}(\tau) - B(\tau) H(\tau)\|_{F}=o_{p}(1)$ and $\sup_{\tau\in\mathcal{T}}\|H(\tau)\|_2=O_{p}(1)$. It then follows that $\lambda_{\max}(B(\tau)^{\prime}B(\tau))H(\tau)^{\prime}H(\tau)-I_{K}$ is positive semidefinite with probability approaching one uniformly over $\tau\in\mathcal{T}$, since $H(\tau)^{\prime}B(\tau)^{\prime}B(\tau)H(\tau)-\lambda_{\max}(B(\tau)^{\prime}B(\tau))H(\tau)^{\prime}H(\tau)$ is negative semidefinite. Therefore, the eigenvalues of $H(\tau)^{\prime}H(\tau)$ are larger than or equal to $\lambda^{-1}_{\max}(B(\tau)^{\prime}B(\tau))$ with probability approaching one uniformly over $\tau\in\mathcal{T}$. Thus, the result of the lemma follows from Assumption \ref{Ass: InterceptLoadingsFactors}(iii).\qed

\begin{lem}\label{Lem: TechA3}
Suppose Assumptions \ref{Ass: DGP} and \ref{Ass: Basis} hold. Assume (i) $N\to\infty$; (ii) $T$ is finite; (iii) $J\to\infty$ with $J^{3}\xi^{2}_{J}\log^{2} N=o(N)$ and $J^{-\kappa+1}\log N=o(1)$. Then $\sup_{t\leq T}|\lambda^{\prime}[\tilde{Y}_t(\tau) -\beta_{t}(\tau)]|=O_{p}(1/\sqrt{N})$ for any $\lambda\in\mathbf{R}^{JM}$ such that $\|\lambda\|$ is bounded.
\end{lem}
\noindent{\sc Proof:} Since $\|\lambda\|$ is bounded, Lemma \ref{Lem: TechA5} implies that
\begin{align}\label{Eqn: Lem: TechA3: 1}
\sup_{t\leq T}|\lambda^{\prime}\sqrt{N}[\tilde{Y}_t(\tau) -\beta_{t}(\tau)]-\lambda^{\prime}J_{t}^{-1}(\tau)\mathbb{U}_t(\tau)|=o_{p}(1).
\end{align}
All eigenvalues of $J_{t}(\tau)$ are bounded from above and away from zero uniformly over $\tau\in\mathcal{T}$ by Assumptions \ref{Ass: DGP}(ii), (iii) and \ref{Ass: Basis}(i). Using the Markov inequality, $\lambda^{\prime}J_{t}^{-1}(\tau)\mathbb{U}_t(\tau)=O_{p}(1)$ by Assumption \ref{Ass: Basis}(i). Thus, by the triangle inequality, the result of the lemma follows from \eqref{Eqn: Lem: TechA3: 1}.\qed

\begin{lem}\label{Lem: TechA4}
Suppose Assumptions \ref{Ass: DGP}-\ref{Ass: InterceptLoadingsFactors} hold, $\sup_{\tau\in\mathcal{T}}|a(\tau)^{\prime}J_{t}^{-1}(\tau)\mathbb{U}_t(\tau)|=O_{p}(1)$ and $\sup_{\tau\in\mathcal{T}}\|B(\tau)^{\prime}J_{t}^{-1}(\tau)\mathbb{U}_t(\tau)\|=O_{p}(1)$ for each $t\leq T$. Assume (i) $N\to\infty$; (ii) $T\geq K(\tau)+1$ is finite; (iii) $J\to\infty$ with $J^{3}\xi^{2}_{J}\log^{2} N=o(N)$ and $J^{-\kappa+1}\log N=o(1)$. Then
\[\sup_{\tau\in\mathcal{T}}\frac{1}{T}\|\hat{F}(\tau)-F(\tau)[H(\tau)^{\prime}]^{-1}\|_{F}^{2}=O_{p}\left(\frac{{1}}{{N}}+\frac{1}{J^{2\kappa}\xi_{J}^{2}}\right).\]
\end{lem}
\noindent{\sc Proof:} Since $\sup_{\tau\in\mathcal{T}}\|a(\tau)\|$ is bounded, Lemma \ref{Lem: TechA5} implies that
\begin{align}\label{Eqn: Lem: TechA4: 1}
\sup_{t\leq T}\sup_{\tau\in\mathcal{T}}|a(\tau)^{\prime}\sqrt{N}[\tilde{Y}_t(\tau) -\beta_{t}(\tau)]-a(\tau)^{\prime}J_{t}^{-1}(\tau)\mathbb{U}_t(\tau)|=o_{p}(1).
\end{align}
Since $\sup_{\tau\in\mathcal{T}}|a(\tau)^{\prime}J_{t}^{-1}(\tau)\mathbb{U}_t(\tau)|=O_{p}(1)$ and $T$ is finite, by the triangle inequality, \eqref{Eqn: Lem: TechA4: 1} implies that $\sup_{t\leq T}\sup_{\tau\in\mathcal{T}}|a(\tau)^{\prime}[\tilde{Y}_t(\tau) -\beta_{t}(\tau)]|=O_{p}(1/\sqrt{N})$. Similarly, we have $\sup_{t\leq T}\sup_{\tau\in\mathcal{T}}\|B(\tau)^{\prime}[\tilde{Y}_t(\tau) -\beta_{t}(\tau)]\|=O_{p}(1/\sqrt{N})$. Given these, the result of the lemma follows by a similar argument for the last result of Theorem \ref{Thm: RateConv}. \qed

\begin{lem}\label{Lem: TechA5}
Suppose Assumptions \ref{Ass: DGP} and \ref{Ass: Basis} hold. Assume (i) $N\to\infty$; (ii) $T$ is finite; (iii) $J\to\infty$ with $J^{3}\xi^{2}_{J}\log^{2} N=o(N)$ and $J^{-\kappa+1}\log N=o(1)$. Then
\[\sup_{t\leq T}\sup_{\tau\in\mathcal{T}}\|\sqrt{N}[\tilde{Y}_t(\tau) -\beta_{t}(\tau)]-J_{t}^{-1}(\tau)\mathbb{U}_t(\tau)\|=O_{p}\left(\frac{J^{3/4}\xi_{J}^{1/2}\log^{1/2}N}{N^{1/4}}+\frac{\log^{1/2}N}{J^{(\kappa-1)/2}}\right),\]
where $J_{t}(\tau)= E[f_{y_t|z_t}(Q_{t}(\tau,z_{t})|z_t)\phi(z_{t})\phi(z_{t})^{\prime}]$ and $\mathbb{U}_t(\tau) = \sum_{i=1}^{N}\phi(z_{it})(\tau-1\{u_{it}\leq \tau\})/\sqrt{N}$, where $\{u_{it}=F_{y_{t}|z_{t}}(y_{it}|z_{it})\}_{i=1}^{N}$ is a sequence of i.i.d. $U(0,1)$ random variables for each $t$ and $F_{y_t|z_{t}}(y|z)$ denotes the conditional distribution function.
\end{lem}
\noindent{\sc Proof:} Since $T$ is finite, in view of \eqref{Eqn: Lem: TechA1: 1}, the result of the lemma follows by Theorem 2 of \citet{Bellonietal_ConditionalQuantile_2019}. \qed

\section{Proof of Theorem \ref{Thm: AsympDis}}
\renewcommand{\theequation}{B.\arabic{equation}}
\setcounter{equation}{0}
\subsection{Proof of Theorem \ref{Thm: AsympDis}}
\noindent{\sc Proof of Theorem \ref{Thm: AsympDis}}: Since the eigenvalues of $J_{t}(\tau)$ are bounded from above and away from zero uniformly over $\tau\in\mathcal{T}$, by the triangle inequality, it follows from Assumptions \ref{Ass: Basis}(iv) and \ref{Ass: AsymDis}(ii) and Lemma \ref{Lem: TechA5} that
\begin{align}\label{Eqn: Thm: AsympDis: 1}
\sup_{\tau\in\mathcal{T}}\|\sqrt{N}[\tilde{Y}(\tau) - a(\tau)1_{T}^{\prime} - B(\tau)F(\tau)^{\prime}]-\mathbb{D}(\tau)\|_{F} = O_{p}\left(\eta_{N}+\delta_{N} + \frac{\sqrt{N}}{J^{\kappa}\xi_{J}}\right),
\end{align}
where $\eta_{N}= [{J^{3/4}\xi_{J}^{1/2}\log^{1/2}N}]/{N^{1/4}}+{J^{-(\kappa-1)/2}\log^{1/2}N}$. Since $\|M_{T}\|_2=1$,
\begin{align}\label{Eqn: Thm: AsympDis: 2}
\sup_{\tau\in\mathcal{T}}\|\sqrt{N}[\bar{\tilde{Y}}(\tau) -  a(\tau)- B(\tau)\bar{f}(\tau)]-\mathbb{D}(\tau)1_{T}/T\| = O_{p}\left(\eta_{N}+\delta_{N} + \frac{\sqrt{N}}{J^{\kappa}\xi_{J}}\right)
\end{align}
and
\begin{align}\label{Eqn: Thm: AsympDis: 3}
\sup_{\tau\in\mathcal{T}}\|\sqrt{N}[\tilde{Y}(\tau)M_T - B(\tau)[M_TF(\tau)]^{\prime}]-\mathbb{D}(\tau)M_T\|_{F} = O_{p}\left(\eta_{N}+\delta_{N} + \frac{\sqrt{N}}{J^{\kappa}\xi_{J}}\right).
\end{align}
Thus, by the triangle inequality, the second result of the theorem follows from \eqref{Eqn: Thm: RateConv: 3}, \eqref{Eqn: Thm: RateConv: 4}, \eqref{Eqn: Thm: AsympDis: 3}, Lemma \ref{Lem: TechB1} and Theorem \ref{Thm: RateConv}.
The proofs of the other two results are similar. This completes the proof of the theorem.\qed

\subsection{Auxiliary Lemmas}
\begin{lem}\label{Lem: TechB1}
Suppose Assumptions \ref{Ass: DGP}-\ref{Ass: InterceptLoadingsFactors} and \ref{Ass: AsymDis}(i) hold. Let $V(\tau)$ be given in the proof of Theorem \ref{Thm: RateConv}. Assume (i) $N\to\infty$; (ii) $T\geq K(\tau) +1$ is finite; (iii) $J\to\infty$ with $J\xi^{2}_{J}\log^{2} N=o(N)$ and $J^{-\kappa}\log N=o(1)$. Then
\begin{align*}
\sup_{\tau\in\mathcal{T}}\|H(\tau)-\mathcal{H}(\tau)\|_{F} &= O_{p}\left(\frac{\sqrt{J}}{\sqrt{N}}+\frac{1}{J^{\kappa}\xi_{J}}\right),\notag\\
\sup_{\tau\in\mathcal{T}}\|H(\tau)V(\tau)^{-1}-\mathcal{M}(\tau)\|_{F} &=O_{p}\left(\frac{\sqrt{J}}{\sqrt{N}}+\frac{1}{J^{\kappa}\xi_{J}}\right),
\end{align*}
where $\mathcal{H}(\tau) = [F(\tau)^{\prime}M_TF(\tau)/T]^{1/2}\Upsilon(\tau)\mathcal{V}(\tau)^{-1/2}$, $\mathcal{M}(\tau) = \mathcal{H}(\tau)\mathcal{V}(\tau)^{-1}$, $\mathcal{V}(\tau)$ is a diagonal matrix of the eigenvalues of $[F(\tau)^{\prime}M_TF(\tau)/T]^{1/2}B(\tau)^{\prime}B(\tau)[F(\tau)^{\prime}M_TF(\tau)/T]^{1/2}$ and $\Upsilon(\tau)$ is the corresponding eigenvector matrix such that $\Upsilon(\tau)^{\prime}\Upsilon(\tau) = I_{K(\tau)}$.
\end{lem}
\noindent{\sc Proof:} By the definition of $\hat{B}(\tau)$, $[\tilde{Y}(\tau)M_T\tilde{Y}(\tau)^{\prime}/T]\hat{B}(\tau) = \hat{B}(\tau)V(\tau)$. Pre-multiply both sides by $[F(\tau)^{\prime}M_TF(\tau)/T]^{1/2}B(\tau)^{\prime}$ to obtain
\begin{align}\label{Eqn: Lemma: TechB1: 1}
&[{F(\tau)^{\prime}M_TF(\tau)}/{T}]^{1/2}B(\tau)^{\prime}[{\tilde{Y}(\tau)M_T\tilde{Y}(\tau)^{\prime}}/{T}]\hat{B}(\tau)\notag\\
&\hspace{-0.5cm}= [{F(\tau)^{\prime}M_TF(\tau)}/{T}]^{1/2}B(\tau)^{\prime}\hat{B}(\tau)V(\tau).
\end{align}
To simplify the notation, let $\epsilon_{N}(\tau) \equiv [F(\tau)^{\prime}M_TF(\tau)/T]^{1/2}B(\tau)^{\prime}\{\tilde{Y}(\tau)M_T\tilde{Y}(\tau)^{\prime}/T-B(\tau)[F(\tau)^{\prime}M_TF(\tau)/T]B(\tau)^{\prime}\}\hat{B}(\tau)$ and $R_{N}(\tau)\equiv[{F(\tau)^{\prime}M_TF(\tau)}/{T}]^{1/2}B(\tau)^{\prime}\hat{B}(\tau)$. Then we can rewrite \eqref{Eqn: Lemma: TechB1: 1} as
\begin{align}\label{Eqn: Lemma: TechB1: 2}
&\{[{F(\tau)^{\prime}M_TF(\tau)}/{T}]^{1/2}B(\tau)^{\prime}B(\tau)[{F(\tau)^{\prime}M_TF(\tau)}/{T}]^{1/2} + \epsilon_{N}(\tau)R_{N}(\tau)^{-1}\} R_{N}(\tau)\notag\\
&\hspace{-0.5cm}= R_{N}(\tau)V(\tau).
\end{align}
Let $D_{N}(\tau)$ be a diagonal matrix consisting of the diagonal elements of $R_{N}(\tau)^{\prime}R_{N}(\tau)$. Denote $\Upsilon_{N}(\tau)\equiv R_{N}(\tau)D_{N}(\tau)^{-1/2}$, which has a unit length. Then we can further rewrite \eqref{Eqn: Lemma: TechB1: 2} as
\begin{align}\label{Eqn: Lemma: TechB1: 3}
&\{[{F(\tau)^{\prime}M_TF(\tau)}/{T}]^{1/2}B(\tau)^{\prime}B(\tau)[{F(\tau)^{\prime}M_TF(\tau)}/{T}]^{1/2} + \epsilon_{N}(\tau)R_{N}(\tau)^{-1}\} \Upsilon_{N}(\tau)\notag\\
&\hspace{-0.5cm}= \Upsilon_{N}(\tau)V(\tau),
\end{align}
so we have that $\Upsilon_{N}(\tau)$ is the eigenvector matrix of $\{[{F(\tau)^{\prime}M_TF(\tau)}/{T}]^{1/2}B(\tau)^{\prime}B(\tau)$ $[{F(\tau)^{\prime}M_TF(\tau)}/{T}]^{1/2} + \epsilon_{N}(\tau)R_{N}(\tau)^{-1}\}$ and $V(\tau)$ is the diagonal eigenvalue matrix. Since $R_{N}(\tau)= [{F(\tau)^{\prime}M_TF(\tau)}/{T}]^{1/2}B(\tau)^{\prime}B(\tau)H(\tau)+o_{p}(1)$ uniformly over $\tau\in\mathcal{T}$ by simple algebra and Theorem \ref{Thm: RateConv}, $R_{N}(\tau)^{-1}=O_{p}(1)$ uniformly over $\tau\in\mathcal{T}$ by Assumptions \ref{Ass: InterceptLoadingsFactors}(iii) and (iv) and Lemma \ref{Lem: TechA2}(ii). This together with \eqref{Eqn: Thm: RateConv: 3} implies that
\begin{align}\label{Eqn: Lemma: TechB1: 4}
\sup_{\tau\in\mathcal{T}}\|\epsilon_{N}(\tau)R_{N}(\tau)^{-1}\|_{F}=O_{p}\left(\frac{\sqrt{J}}{\sqrt{N}}+\frac{1}{J^{\kappa}\xi_{J}}\right).
\end{align}
Because the eigenvalues of $[{F(\tau)^{\prime}M_TF(\tau)}/{T}]B(\tau)^{\prime}B(\tau)$ equal the eigenvalues of $A(\tau)\equiv [{F(\tau)^{\prime}M_TF(\tau)}/{T}]^{1/2}B(\tau)^{\prime}B(\tau)[{F(\tau)^{\prime}M_TF(\tau)}/{T}]^{1/2}$, the eigenvalues of $A(\tau)$ are distinct by Assumption \ref{Ass: AsymDis}(i). Following the eigenvector perturbation theory, there exists a unique eigenvector matrix $\Upsilon(\tau)$ of $A(\tau)$ such that
\begin{align}\label{Eqn: Lemma: TechB1: 5}
\sup_{\tau\in\mathcal{T}}\|\Upsilon_{N}(\tau)-\Upsilon(\tau)\|_{F} = O_{p}\left(\frac{\sqrt{J}}{\sqrt{N}}+\frac{1}{J^{\kappa}\xi_{J}}\right).
\end{align}
By \eqref{Eqn: Thm: RateConv: 3} and simple algebra, $R_{N}(\tau)^{\prime}R_{N}(\tau) = \hat{B}(\tau)^{\prime}B(\tau)[{F(\tau)^{\prime}M_TF(\tau)}/{T}]B(\tau)^{\prime}\hat{B}(\tau)= \hat{B}(\tau)^{\prime}[\tilde{Y}(\tau)M_T\tilde{Y}(\tau)^{\prime}/T]\hat{B}(\tau)+O_{p}(J^{-\kappa}\xi_{J}^{-1}+{\sqrt{J}}/{\sqrt{N}}) = V(\tau)+O_{p}(J^{-\kappa}\xi_{J}^{-1}+{\sqrt{J}}/{\sqrt{N}})$ uniformly over $\tau\in\mathcal{T}$. This implies that
\begin{align}\label{Eqn: Lemma: TechB1: 6}
\sup_{\tau\in\mathcal{T}}\|D_{N}(\tau)- V(\tau)\|_{F} = O_{p}\left(\frac{\sqrt{J}}{\sqrt{N}}+\frac{1}{J^{\kappa}\xi_{J}}\right).
\end{align}
Recall that $H^{\diamond}(\tau)=[F(\tau)^{\prime}M_TF(\tau)/T]B(\tau)^{\prime}\hat{B}(\tau)V(\tau)^{-1}$ in the proof of Lemma \ref{Lem: TechA2}(i). Thus, by \eqref{Eqn: Lem: TechA2: 1}, \eqref{Eqn: Lemma: TechB1: 5} and \eqref{Eqn: Lemma: TechB1: 6}, we have $H^{\diamond}(\tau)= [{F(\tau)^{\prime}M_TF(\tau)}/{T}]^{1/2}\Upsilon_{N}(\tau) D_{N}(\tau)^{1/2}$ $V(\tau)^{-1}= \mathcal{H}(\tau) + O_{p}(J^{-\kappa}\xi_{J}^{-1}+{\sqrt{J}}/{\sqrt{N}})$ uniformly over $\tau\in\mathcal{T}$, which together with \eqref{Eqn: Lem: TechA2: 2} leads to the first result of the lemma. The second result of the lemma holds by the first result of the lemma, \eqref{Eqn: Lem: TechA2: 1} and Lemma \ref{Lem: TechA2}(i). \qed

\section{Proof of Theorem \ref{Thm: BootDis}}
\renewcommand{\theequation}{C.\arabic{equation}}
\setcounter{equation}{0}
\subsection{Proof of Theorem \ref{Thm: BootDis}}
\noindent{\sc Proof of Theorem \ref{Thm: BootDis}}: Recall that in the proof of Theorem \ref{Thm: AsympDis}, we define $\eta_{N}= [{J^{3/4}\xi_{J}^{1/2}\log^{1/2}N}]/{N^{1/4}}+{J^{-(\kappa-1)/2}\log^{1/2}N}$. Since the eigenvalues of $J_{t}(\tau)$ are bounded from above and away from zero uniformly over $\tau\in\mathcal{T}$, by the triangle inequality, it follows from Assumption \ref{Ass: BootWeight}(ii) and Lemma \ref{Lem: TechC1} that
\begin{align}\label{Eqn: Thm: BootDis: 1}
\sup_{\tau\in\mathcal{T}}\|\sqrt{N}[\tilde{Y}^{\ast}(\tau) - \tilde{Y}(\tau)]-\mathbb{D}^{\ast}(\tau)\|_{F} = O_{p}\left(\eta_{N}+\delta_{N}\right).
\end{align}
Recall that $\hat{F}(\tau)^{\prime}M_T\hat{F}(\tau)/T =V(\tau)$ and $\hat{F}(\tau)=\tilde{Y}(\tau)^{\prime}\hat{B}(\tau)$. By the definition of $\hat{B}^{\ast}(\tau)$,
\begin{align}\label{Eqn: Thm: BootDis: 2}
\hat{B}^{\ast}(\tau)-B(\tau)H(\tau) = \frac{1}{T}[\tilde{Y}^{\ast}(\tau)M_T - B(\tau)[M_TF(\tau)]^{\prime}][\tilde{Y}(\tau)M_T]^{\prime}\hat{B}(\tau)V(\tau)^{-1}.
\end{align}
Combining \eqref{Eqn: Thm: RateConv: 4} and \eqref{Eqn: Thm: BootDis: 2}, we obtain
\begin{align}\label{Eqn: Thm: BootDis: 3}
\hat{B}^{\ast}(\tau)-\hat{B}(\tau) = \frac{1}{T}[\tilde{Y}^{\ast}(\tau) - \tilde{Y}(\tau)][\tilde{Y}(\tau)M_T]^{\prime}\hat{B}(\tau)V(\tau)^{-1}.
\end{align}
Thus, by the triangle inequality, the second result of the theorem follows from \eqref{Eqn: Thm: RateConv: 3}, \eqref{Eqn: Thm: BootDis: 1}, \eqref{Eqn: Thm: BootDis: 3}, Lemma \ref{Lem: TechB1} and Theorem \ref{Thm: RateConv} with $\mathbb{G}^{\ast}_{B}(\tau) = \mathbb{D}^{\ast}(\tau)M_TF(\tau)B(\tau)^{\prime}B(\tau)$ $\mathcal{M}(\tau)/T$. By the definition of $\hat{a}^{\ast}(\tau)$,
\begin{align}\label{Eqn: Thm: BootDis: 4}
\hat{a}^{\ast}(\tau) - a(\tau) &= -\hat{B}^{\ast}(\tau)[\hat{B}^{\ast}(\tau)^{\prime}\hat{B}^{\ast}(\tau)]^{-1}[\hat{B}^{\ast}(\tau)-B(\tau)H(\tau)]^{\prime}a(\tau) \notag\\
&\hspace{-1.5cm}- [I_{JM}- \hat{B}^{\ast}(\tau)[\hat{B}^{\ast}(\tau)^{\prime}\hat{B}^{\ast}(\tau)]^{-1}\hat{B}^{\ast}(\tau)^{\prime}][\hat{B}^{\ast}(\tau)-B(\tau)H(\tau)]H(\tau)^{-1}\bar{f}(\tau)\notag\\
&\hspace{-1.5cm} +[I_{JM} - \hat{B}^{\ast}(\tau)[\hat{B}^{\ast}(\tau)^{\prime}\hat{B}^{\ast}(\tau)]^{-1}\hat{B}^{\ast}(\tau)^{\prime}][\bar{\tilde{Y}}^{\ast}(\tau) -  a(\tau)- B(\tau)\bar{f}(\tau)],
\end{align}
where we have used $a(\tau)^{\prime}B(\tau)=0$ and $[I_{JM} - \hat{B}^{\ast}(\tau)[\hat{B}^{\ast}(\tau)^{\prime}\hat{B}^{\ast}(\tau)]^{-1}\hat{B}^{\ast}(\tau)^{\prime}]\hat{B}^{\ast}(\tau)=0$. By the triangle inequality, we can replace $\hat{B}^{\ast}(\tau)$ in ``$\hat{B}^{\ast}(\tau)[\hat{B}^{\ast}(\tau)^{\prime}\hat{B}^{\ast}(\tau)]^{-1}$'' and ``$\hat{B}^{\ast}(\tau)[\hat{B}^{\ast}(\tau)^{\prime}\hat{B}^{\ast}(\tau)]^{-1}\hat{B}^{\ast}(\tau)^{\prime}$'' with $\hat{B}(\tau)$. Specifically, we have
\begin{align}\label{Eqn: Thm: BootDis: 5}
\hat{a}^{\ast}(\tau) - a(\tau) &= -\hat{B}(\tau)[\hat{B}^{\ast}(\tau)-B(\tau)H(\tau)]^{\prime}a(\tau) \notag\\
&\hspace{0.5cm}- [I_{JM}- \hat{B}(\tau)\hat{B}(\tau)^{\prime}][\hat{B}^{\ast}(\tau)-B(\tau)H(\tau)]H(\tau)^{-1}\bar{f}(\tau)\notag\\
&\hspace{0.5cm} +[I_{JM} - \hat{B}(\tau)\hat{B}(\tau)^{\prime}][\bar{\tilde{Y}}^{\ast}(\tau) -  a(\tau)- B(\tau)\bar{f}(\tau)]+\Delta_{a,N}(\tau),
\end{align}
where $\sup_{\tau\in\mathcal{T}}\|\Delta_{a,N}(\tau)\|=O_{p}(\sqrt{J/N}[\sqrt{J/N}+J^{-\kappa}\xi_{J}^{-1}])$. Combining \eqref{Eqn: Thm: RateConv: 5} and \eqref{Eqn: Thm: BootDis: 5}, we thus obtain
\begin{align}\label{Eqn: Thm: BootDis: 6}
\hat{a}^{\ast}(\tau) - \hat{a}(\tau) &= -\hat{B}(\tau)[\hat{B}^{\ast}(\tau)-\hat{B}(\tau)]^{\prime}a(\tau) \notag\\
&\hspace{0.5cm}- [I_{JM}- \hat{B}(\tau)\hat{B}(\tau)^{\prime}][\hat{B}^{\ast}(\tau)-\hat{B}(\tau)]H(\tau)^{-1}\bar{f}(\tau)\notag\\
&\hspace{0.5cm} +[I_{JM} - \hat{B}(\tau)\hat{B}(\tau)^{\prime}][\bar{\tilde{Y}}^{\ast}(\tau) - \bar{\tilde{Y}}(\tau)]+\Delta_{a,N}(\tau).
\end{align}
Thus, by the triangle inequality, the first result of the theorem follows from \eqref{Eqn: Thm: BootDis: 1}, \eqref{Eqn: Thm: BootDis: 6}, the second result of the theorem, Lemma \ref{Lem: TechB1} and Theorem \ref{Thm: RateConv} with $\mathbb{G}^{\ast}_{a}(\tau) = -B(\tau)\mathcal{H}(\tau)\mathbb{G}^{\ast}_{B}(\tau)^{\prime} a(\tau)-[I_{JM}-B(\tau)\mathcal{H}(\tau)\mathcal{H}(\tau)^{\prime}B(\tau)^{\prime}][\mathbb{G}^{\ast}_{B}(\tau)\mathcal{H}(\tau)^{-1}\bar{f}(\tau)-\mathbb{D}^{\ast}(\tau)1_{T}/T]$. By the definition of $\hat{F}^{\ast}(\tau)$,
\begin{align}\label{Eqn: Thm: BootDis: 7}
\hat{F}^{\ast}(\tau) - F(\tau)[H(\tau)^{\prime}]^{-1} &= 1_{T}a(\tau) ^{\prime}[\hat{B}^{\ast}(\tau)-B(\tau)H(\tau)][\hat{B}^{\ast}(\tau)^{\prime}\hat{B}^{\ast}(\tau)]^{-1} \notag\\
&\hspace{-1.5cm} - F(\tau)[H(\tau)^{\prime}]^{-1}[\hat{B}^{\ast}(\tau)-B(\tau)H(\tau)]^{\prime}\hat{B}^{\ast}(\tau)[\hat{B}^{\ast}(\tau)^{\prime}\hat{B}^{\ast}(\tau)]^{-1} \notag\\
&\hspace{-1.5cm}+ [\tilde{Y}^{\ast}(\tau) - a(\tau)1_{T}^{\prime} - B(\tau)F(\tau)^{\prime}]^{\prime}\hat{B}^{\ast}(\tau)[\hat{B}^{\ast}(\tau)^{\prime}\hat{B}^{\ast}(\tau)]^{-1}.
\end{align}
where we have used $\hat{B}^{\ast}(\tau)^{\prime}\hat{B}^{\ast}(\tau)[\hat{B}^{\ast}(\tau)^{\prime}\hat{B}^{\ast}(\tau)]^{-1}=I_{K}$. By the triangle inequality, we can replace $\hat{B}^{\ast}(\tau)$ in ``$[\hat{B}^{\ast}(\tau)^{\prime}\hat{B}^{\ast}(\tau)]^{-1}$'' and ``$\hat{B}^{\ast}(\tau)[\hat{B}^{\ast}(\tau)^{\prime}\hat{B}^{\ast}(\tau)]^{-1}$'' with $\hat{B}(\tau)$. Specifically, we have
\begin{align}\label{Eqn: Thm: BootDis: 8}
\hat{F}^{\ast}(\tau) - F(\tau)[H(\tau)^{\prime}]^{-1} &= 1_{T}a(\tau) ^{\prime}[\hat{B}^{\ast}(\tau)-B(\tau)H(\tau)] \notag\\
&\hspace{0.5cm} - F(\tau)[H(\tau)^{\prime}]^{-1}[\hat{B}^{\ast}(\tau)-B(\tau)H(\tau)]^{\prime}\hat{B}(\tau) \notag\\
&\hspace{0.5cm}+ [\tilde{Y}^{\ast}(\tau) - a(\tau)1_{T}^{\prime} - B(\tau)F(\tau)^{\prime}]^{\prime}\hat{B}(\tau).
\end{align}
where $\sup_{\tau\in\mathcal{T}}\|\Delta_{F,N}(\tau)\|=O_{p}(\sqrt{J/N}[\sqrt{J/N}+J^{-\kappa}\xi_{J}^{-1}])$. Combining \eqref{Eqn: Thm: RateConv: 6} and \eqref{Eqn: Thm: BootDis: 8}, we thus obtain
\begin{align}\label{Eqn: Thm: BootDis: 9}
\hat{F}^{\ast}(\tau) - \hat{F}(\tau) &= 1_{T}a(\tau) ^{\prime}[\hat{B}^{\ast}(\tau)-\hat{B}(\tau)] \notag\\
&\hspace{0.5cm} - F(\tau)[H(\tau)^{\prime}]^{-1}[\hat{B}^{\ast}(\tau)-\hat{B}(\tau)]^{\prime}\hat{B}(\tau) \notag\\
&\hspace{0.5cm}+ [\tilde{Y}^{\ast}(\tau) - \tilde{Y}(\tau)]^{\prime}\hat{B}(\tau).
\end{align}
Thus, by the triangle inequality, the third result of the theorem follows from \eqref{Eqn: Thm: BootDis: 1}, \eqref{Eqn: Thm: BootDis: 9}, the second result of the theorem, Lemma \ref{Lem: TechB1} and Theorem \ref{Thm: RateConv} with $\mathbb{G}^{\ast}_{F}(\tau) = 1_{T}a(\tau)^{\prime}\mathbb{G}^{\ast}_{B}(\tau)-F(\tau)[\mathcal{H}(\tau)^{\prime}]^{-1}\mathbb{G}^{\ast}_{B}(\tau)^{\prime}B(\tau)\mathcal{H}(\tau)+\mathbb{D}^{\ast}(\tau)B(\tau)\mathcal{H}(\tau)$. \qed

\subsection{Auxiliary Lemmas}
\begin{lem}\label{Lem: TechC1}
Suppose Assumptions \ref{Ass: DGP}, \ref{Ass: Basis} and \ref{Ass: BootWeight}(i) hold. Assume (i) $N\to\infty$; (ii) $T$ is finite; (iii) $J\to\infty$ with $J^{3}\xi^{2}_{J}\log^{2} N=o(N)$ and $J^{-\kappa+1}\log N=o(1)$. Then
\[\sup_{t\leq T}\sup_{\tau\in\mathcal{T}}\|\sqrt{N}[\tilde{Y}^{\ast}_t(\tau) -\tilde{Y}_t(\tau)]-J_{t}^{-1}(\tau)\mathbb{U}^{\ast}_t(\tau)\|=O_{p}\left(\frac{J^{3/4}\xi_{J}^{1/2}\log^{1/2}N}{N^{1/4}}+\frac{\log^{1/2}N}{J^{(\kappa-1)/2}}\right).\]
\end{lem}
\noindent{\sc Proof:} Since $T$ is finite, in view of \eqref{Eqn: Lem: TechA1: 1}, the result of the lemma follows by of lemma 7 in \citet{Bellonietal_ConditionalQuantile_2019}. \qed

\section{Proof of Theorem \ref{Thm: NumFactors}}
\renewcommand{\theequation}{D.\arabic{equation}}
\setcounter{equation}{0}
\subsection{Proof of Theorem \ref{Thm: NumFactors}}
\noindent{\sc Proof of Theorem \ref{Thm: NumFactors}}: (A) To simplify the notation, for each $k$, we let $\theta_{k}(\tau)\equiv{\lambda_{k}(\tilde{Y}(\tau)M_T\tilde{Y}(\tau)^{\prime}/T)}/$ $\lambda_{k+1}(\tilde{Y}(\tau)M_T\tilde{Y}(\tau)^{\prime}/T)$. Suppose that there is $\tau^{\ast}\in\mathcal{T}$ such that $\hat{K}(\tau^{\ast})\neq K(\tau^{\ast})$. It follows that
\begin{align}\label{Eqn: Thm: NumFactors: 1}
\frac{\lambda_{1}(\tilde{Y}(\tau^{\ast})M_T\tilde{Y}(\tau^{\ast})^{\prime}/T)}{\lambda_ {K(\tau^{\ast})}(\tilde{Y}(\tau^{\ast})M_T\tilde{Y}(\tau^{\ast})^{\prime}/T)}\geq \theta_{K(\tau^{\ast})}
\end{align}
or
\begin{align}\label{Eqn: Thm: NumFactors: 2}
\frac{\lambda_{K(\tau^{\ast})+1}(\tilde{Y}(\tau^{\ast})M_T\tilde{Y}(\tau^{\ast})^{\prime}/T)}{\lambda_{T-K(\tau^{\ast})-1}(\tilde{Y}(\tau^{\ast})M_T\tilde{Y}(\tau^{\ast})^{\prime}/T)}\geq \theta_{K(\tau^{\ast})}.
\end{align}
Thus,
\begin{align}\label{Eqn: Thm: NumFactors: 3}
&P(\hat{K}(\tau)\neq {K}(\tau) \text{ for some } \tau\in\mathcal{T})\leq P\left(\frac{\lambda_{1}(\tilde{Y}(\tau^{\ast})M_T\tilde{Y}(\tau^{\ast})^{\prime}/T)}{\lambda_ {K(\tau^{\ast})}(\tilde{Y}(\tau^{\ast})M_T\tilde{Y}(\tau^{\ast})^{\prime}/T)}\geq \theta_{K(\tau^{\ast})}\right)\notag\\
&\hspace{1.5cm}+P\left(\frac{\lambda_{K(\tau^{\ast})+1}(\tilde{Y}(\tau^{\ast})M_T\tilde{Y}(\tau^{\ast})^{\prime}/T)}{\lambda_{T-K(\tau^{\ast})-1}(\tilde{Y}(\tau^{\ast})M_T\tilde{Y}(\tau^{\ast})^{\prime}/T)}\geq \theta_{K(\tau^{\ast})}\right).
\end{align}
By Lemmas \ref{Lem: TechD1} and \ref{Lem: TechD2}, ${\lambda_{1}(\tilde{Y}(\tau^{\ast})M_T\tilde{Y}(\tau^{\ast})^{\prime}/T)}/{\lambda_{K(\tau^{\ast})}(\tilde{Y}(\tau^{\ast})M_T\tilde{Y}(\tau^{\ast})^{\prime}/T)}=O_{p}(1)$ and $NJ^{-1}\theta^{-1}_{K(\tau^{\ast})}=O_{p}(1)$. Thus, the first term on the right hand side of \eqref{Eqn: Thm: NumFactors: 3} tends to zero as $N\to\infty$. Similarly, by Assumption \ref{Ass: NumFactors} and Lemmas \ref{Lem: TechD1} and \ref{Lem: TechD2}, the second term on the right hand side of \eqref{Eqn: Thm: NumFactors: 3} tends to zero as $N\to\infty$. Therefore, the result of the theorem follows from \eqref{Eqn: Thm: NumFactors: 3}.

(B) Suppose that there is $\tau^{\ast}\in\mathcal{T}$ such that $\tilde{K}(\tau^{\ast})\neq K(\tau^{\ast})$. It follows that $\lambda_{K(\tau^{\ast})-1}(\tilde{Y}(\tau^{\ast})M_T\tilde{Y}(\tau^{\ast})^{\prime}/T)$ $<\lambda_{N}$ or $\lambda_{K(\tau^{\ast})+1}(\tilde{Y}(\tau^{\ast})M_T\tilde{Y}(\tau^{\ast})^{\prime}/T)\geq\lambda_{N}$. Thus,
\begin{align}\label{Eqn: Thm: NumFactors: 4}
P(\tilde{K}(\tau)\neq {K}(\tau) \text{ for some } \tau\in\mathcal{T})&\leq P(\lambda_{K(\tau^{\ast})-1}(\tilde{Y}(\tau^{\ast})M_T\tilde{Y}(\tau^{\ast})^{\prime}/T)<\lambda_{N})\notag\\
&\hspace{0.5cm}+ P(\lambda_{K(\tau^{\ast})+1}(\tilde{Y}(\tau^{\ast})M_T\tilde{Y}(\tau^{\ast})^{\prime}/T)\geq\lambda_{N}).
\end{align}
Since $\lambda_{N}\to\infty$, $P(\lambda_{K(\tau^{\ast})-1}(\tilde{Y}(\tau^{\ast})M_T\tilde{Y}(\tau^{\ast})^{\prime}/T)<\lambda_{N})\to 0$ by Lemma \ref{Lem: TechD1}. Denote the $k$th largest singular value of a matrix $A$ by $\sigma_{k}(A)$. Noting that $\lambda_{k}(AA^{\prime})=\sigma^{2}_{k}(A)$, it follows that
\begin{align}\label{Eqn: Thm: NumFactors: 5}
&\sup_{\tau\in\mathcal{T}}\lambda_{K(\tau)+1}(\tilde{Y}(\tau)M_T\tilde{Y}(\tau)^{\prime}/T) = \sigma^{2}_{K(\tau)+1}(\tilde{Y}(\tau)M_T/\sqrt{T})\notag\\
&\hspace{-0.5cm}= \sup_{\tau\in\mathcal{T}}| \sigma_{K(\tau)+1}(\tilde{Y}(\tau)M_T/\sqrt{T})-\sigma_{K(\tau)+1}(B(\tau)[M_{T}F(\tau)]^{\prime}/\sqrt{T})|^{2}\notag\\
&\hspace{-0.5cm}\leq \sup_{\tau\in\mathcal{T}}\frac{1}{T}\|\tilde{Y}(\tau)M_T - B(\tau)[M_TF(\tau)]^{\prime}\|^{2}_{F}=O_{p}\left(\frac{{J}}{{N}}+\frac{1}{J^{2\kappa}\xi_{J}^{2}}\right),
\end{align}
where the second equality follows since $\sigma_{K(\tau)+1}(B(\tau)[M_{T}F(\tau)]^{\prime}/\sqrt{T})=0$, the first inequality follows by the Weyl's inequality, and the last equality follows by \eqref{Eqn: Thm: RateConv: 3}. Since $\lambda_N\min\{N/J,J^{2\kappa}\xi_{J}^{2}\}\to\infty$, \eqref{Eqn: Thm: NumFactors: 5} implies that $P(\lambda_{K(\tau^{\ast})+1}(\tilde{Y}(\tau^{\ast})M_T\tilde{Y}(\tau^{\ast})^{\prime}/T)\geq\lambda_{N})\to 0 $. This completes the proof of the theorem.\qed

\subsection{Auxiliary Lemmas}
\begin{lem}\label{Lem: TechD1}
Suppose Assumptions \ref{Ass: DGP}-\ref{Ass: InterceptLoadingsFactors} hold. Assume (i) $N\to\infty$; (ii) $T\geq K(\tau) +1$ is finite; (iii) $J\to\infty$ with $J\xi^{2}_{J}\log^{2} N=o(N)$ and $J^{-\kappa}\log N=o(1)$. Then there exist positive constants $d_1$ and $d_2$ such that
\[d_1+o_{p}(1)\leq \lambda_{K(\tau)}(\tilde{Y}(\tau)M_T\tilde{Y}(\tau)^{\prime}/T)\leq \lambda_{1}(\tilde{Y}(\tau)M_T\tilde{Y}(\tau)^{\prime}/T)\leq d_2+o_{p}(1),\]
where both ``$o_{p}(1)$'' hold uniformly over $\tau\in\mathcal{T}$, and $d_1$ and $d_2$ do not depend on $\tau\in\mathcal{T}$.
\end{lem}
\noindent{\sc Proof:} By \eqref{Eqn: Lem: TechA2: 1},  $\sup_{\tau\in\mathcal{T}}\sup_{1\leq k\leq K(\tau)}|\lambda_{k}(\tilde{Y}(\tau)M_T\tilde{Y}(\tau)^{\prime}/T) - \lambda_{k}([F(\tau)^{\prime}M_TF(\tau)/T]$ $B(\tau)^{\prime}B(\tau))|=o_{p}(1)$. The result of the lemma immediately holds by Assumptions \ref{Ass: Basis}(ii), \ref{Ass: InterceptLoadingsFactors}(iii) and (iv). \qed

\begin{lem}\label{Lem: TechD2}
Suppose Assumptions \ref{Ass: DGP}-\ref{Ass: InterceptLoadingsFactors} hold. Assume (i) $N\to\infty$; (ii) $T\geq 2[K(\tau) +1]$ is finite; (iii) $J\to\infty$ with $J^{3}\xi^{2}_{J}\log^{2} N=o(N)$ and $J^{-\kappa+1}\log N=o(1)$. Then
\begin{align*}
\lambda_{T}(\mathbb{E}(\tau)\mathbb{E}(\tau)^{\prime}/T)+o_{p}(1)&\leq N\lambda_{T-K(\tau)-1}(\tilde{Y}(\tau)M_T\tilde{Y}(\tau)^{\prime}/T)\notag\\
&\hspace{-2.5cm}\leq N\lambda_{K(\tau)+1}(\tilde{Y}(\tau)M_T\tilde{Y}(\tau)^{\prime}/T)\leq \lambda_{1}(\mathbb{E}(\tau)\mathbb{E}(\tau)^{\prime}/T)+o_{p}(1),
\end{align*}
where both ``$o_{p}(1)$'' hold uniformly over $\tau\in\mathcal{T}$.
\end{lem}
\noindent{\sc Proof:} Let $E(\tau)\equiv \mathbb{E}(\tau)/\sqrt{N}$ and $\Delta(\tau)\equiv \tilde{Y}(\tau)M_T-[B(\tau)F(\tau)^{\prime}+{E}(\tau)]M_T$. Denote the $k$th largest singular value of a matrix $A$ by $\sigma_{k}(A)$. Noting that $\lambda_{k}(AA^{\prime})=\sigma^{2}_{k}(A)$, we have that for $k=1,\ldots,T-K(\tau)$,
\begin{align}\label{Eqn: TechD2: 1}
&\hspace{0.5cm}|\lambda_{K(\tau)+k}(\tilde{Y}(\tau)M_T\tilde{Y}(\tau)^{\prime})-\lambda_{K(\tau)+k}([B(\tau)F(\tau)^{\prime}+{E}(\tau)]M_T[B(\tau)F(\tau)^{\prime}+{E}(\tau)]^{\prime})|\notag\\
&\leq|\sigma_{K(\tau)+k}(\tilde{Y}(\tau)M_T)-\sigma_{K(\tau)+k}([B(\tau)F(\tau)^{\prime}+{E}(\tau)]M_T)|^2+2|\sigma_{K(\tau)+k}(\tilde{Y}(\tau)M_T)\notag\\
&\hspace{0.5cm}-\sigma_{K(\tau)+k}([B(\tau)F(\tau)^{\prime}+{E}(\tau)]M_T)|\sigma_{K(\tau)+k}([B(\tau)F(\tau)^{\prime}+{E}(\tau)]M_T)\notag\\
&\leq\|\tilde{Y}(\tau)M_T-[B(\tau)F(\tau)^{\prime}+{E}(\tau)]M_T\|^{2}_{F}+2\|\tilde{Y}(\tau)M_T-[B(\tau)F(\tau)^{\prime}+{E}(\tau)]M_T\|_{F}\notag\\
&\hspace{0.5cm} \times \lambda^{1/2}_{K(\tau)+k}([B(\tau)F(\tau)^{\prime}+{E}(\tau)]M_T[B(\tau)F(\tau)^{\prime}+{E}(\tau)]^{\prime})\notag\\
& \leq\|\Delta(\tau)\|^{2}_{F}+2\|\Delta(\tau)\|_{F}\lambda^{1/2}_{K(\tau)+1}([B(\tau)F(\tau)^{\prime}+{E}(\tau)]M_{T}[B(\tau)F(\tau)^{\prime}+{E}(\tau)]^{\prime}),
\end{align}
where the first inequality holds the triangle inequality, the second one holds by the Weyl's inequality, and the third one holds by the fact that $\lambda_{K(\tau)+k}([B(\tau)F(\tau)^{\prime}+{E}(\tau)]M_T[B(\tau)F(\tau)^{\prime}+{E}(\tau)]^{\prime})\leq \lambda_{K(\tau)+1}([B(\tau)F(\tau)^{\prime}+{E}(\tau)]M_T[B(\tau)F(\tau)^{\prime}+{E}(\tau)]^{\prime})$ for $k\geq 1$. Next, we prove that the right-hand side of \eqref{Eqn: TechD2: 1} is negligible and then examine the behavior of $\lambda_{K(\tau)+k}([B(\tau)F(\tau)^{\prime}+{E}(\tau)]M_T[B(\tau)F(\tau)^{\prime}+{E}(\tau)]^{\prime})$. Let $\tilde{B}(\tau)\equiv B(\tau)+{E}(\tau)M_TF(\tau)[F(\tau)^{\prime}M_TF(\tau)]^{-1}$ and $M_{F(\tau)} \equiv I_{T}-M_TF(\tau)[F(\tau)^{\prime}M_TF(\tau)]^{-1}[M_TF(\tau)]^{\prime}$. We may decompose $[B(\tau)F(\tau)^{\prime}+{E}(\tau)]M_T[B(\tau)F(\tau)^{\prime}+{E}(\tau)]^{\prime}$ by
\begin{align}\label{Eqn: TechD2: 2}
&[B(\tau)F(\tau)^{\prime}+{E}(\tau)]M_T[B(\tau)F(\tau)^{\prime}+{E}(\tau)]^{\prime}\notag\\
&\hspace{-0.5cm}= \tilde{B}(\tau)F(\tau)^{\prime}M_TF(\tau)\tilde{B}(\tau)^{\prime}+{E}(\tau)M_TM_{F(\tau)}M_T{E}(\tau)^{\prime}.
\end{align}
Then, \eqref{Eqn: TechD2: 2} implies that for $k=1,\ldots,T-K(\tau)$,
\begin{align}\label{Eqn: TechD2: 3}
&\lambda_{K(\tau)+k}([B(\tau)F(\tau)^{\prime}+{E}(\tau)]M_T[B(\tau)F(\tau)^{\prime}+{E}(\tau)]^{\prime})\notag\\
&\hspace{-0.5cm}\leq \lambda_{K(\tau)+1}(\tilde{B}(\tau)F(\tau)^{\prime}M_TF(\tau)\tilde{B}(\tau)^{\prime})+\lambda_{k}({E}(\tau)M_TM_{F(\tau)}M_T{E}(\tau)^{\prime})\notag\\
&\hspace{-0.5cm}= \lambda_{k}({E}(\tau)M_TM_{F(\tau)}M_T{E}(\tau)^{\prime})\leq \lambda_{k}({E}(\tau){E}(\tau)^{\prime}),
\end{align}
where the first inequality follows by Lemma \ref{Lem: TechD3}(i), the equality follows by the fact that $\lambda_{K(\tau)+1}(\tilde{B}(\tau)F(\tau)^{\prime}M_TF(\tau)\tilde{B}(\tau)^{\prime})=0$, and the second inequality follows since $I-M_T$ and $I-M_{F(\tau)}$ are positive semi-definite. Moreover, \eqref{Eqn: TechD2: 2} also implies that for $k=1,\ldots,T-2K(\tau)-1$,
\begin{align}\label{Eqn: TechD2: 4}
&\lambda_{K(\tau)+k}([B(\tau)F(\tau)^{\prime}+{E}(\tau)]M_T[B(\tau)F(\tau)^{\prime}+{E}(\tau)]^{\prime})\notag\\
&\hspace{-0.5cm}\geq\lambda_{K(\tau)+k}({E}(\tau)M_TM_{F(\tau)}M_T{E}(\tau)^{\prime})\notag\\
&\hspace{-0.5cm}=\lambda_{K(\tau)+k}({E}(\tau)M_TM_{F(\tau)}M_T{E}(\tau)^{\prime})+\lambda_{K(\tau)+1}({E}(\tau)M_T[I-M_{F(\tau)}]M_T{E}(\tau)^{\prime})\notag\\
&\hspace{-0.5cm}\geq\lambda_{2K(\tau)+k}({E}(\tau)M_T{E}(\tau)^{\prime})\notag\\
&\hspace{-0.5cm}=\lambda_{2K(\tau)+k}({E}(\tau)M_T{E}(\tau)^{\prime})+\lambda_{2}({E}(\tau)[I_T-M_T]{E}(\tau)^{\prime})\notag\\
&\hspace{-0.5cm}\geq \lambda_{2K(\tau)+k+1}({E}(\tau){E}(\tau)^{\prime}),
\end{align}
where the first inequality holds by Lemma \ref{Lem: TechD3}(ii), the first equality holds since the rank of $\lambda_{K(\tau)+1}({E}(\tau)M_T[I-M_{F(\tau)}]M_T{E}(\tau)^{\prime})=0$, the second inequality holds by Lemma \ref{Lem: TechD3}(i), and the last two lines hold similarly. Putting \eqref{Eqn: TechD2: 3} and \eqref{Eqn: TechD2: 4} together implies that eigenvalues of $[B(\tau)F(\tau)^{\prime}+{E}(\tau)]M_T[B(\tau)F(\tau)^{\prime}+{E}(\tau)]^{\prime}$ are bounded by those of ${E}(\tau){E}(\tau)^{\prime}$. Thus, we may just study the behavior of the eigenvalues of ${E}(\tau){E}(\tau)^{\prime}$. Then, it follows from \eqref{Eqn: TechD2: 3} that
\begin{align}\label{Eqn: TechD2: 5}
&\sup_{\tau\in\mathcal{T}}\lambda_{K(\tau)+1}(N[B(\tau)F(\tau)^{\prime}+{E}(\tau)]M_T[B(\tau)F(\tau)^{\prime}+{E}(\tau)]^{\prime}/T)\notag\\
&\hspace{-0.5cm}\leq \sup_{\tau\in\mathcal{T}}\lambda_{1}({N}{E}(\tau){E}(\tau)^{\prime}/T)  =O_{p}(J),
\end{align}
where the equality follows by lemma 32 of \citet{Bellonietal_ConditionalQuantile_2019}. By Assumption \ref{Ass: Basis}(iv) and Lemma \ref{Lem: TechA5}, $\sup_{\tau\in\mathcal{T}}\|\sqrt{N}\Delta(\tau)\|=O_{p}(r_{N})$. Thus, combining \eqref{Eqn: TechD2: 1} and \eqref{Eqn: TechD2: 5} yields
\begin{align}\label{Eqn: TechD2: 6}
&\sup_{\tau\in\mathcal{T}}\sup_{k\leq T-K(\tau)}\left|N\lambda_{K(\tau)+k}(\tilde{Y}(\tau)M_T\tilde{Y}(\tau)^{\prime}/T)\right.\notag\\
&\hspace{1cm}\left.-N\lambda_{K(\tau)+k}([B(\tau)F(\tau)^{\prime}+{E}(\tau)]M_T[B(\tau)F(\tau)^{\prime}+{E}(\tau)]^{\prime}/T)\right|=o_{p}(1).
\end{align}
This means that $N\lambda_{K(\tau)+k}(\tilde{Y}(\tau)M_T\tilde{Y}(\tau)^{\prime}/T)$ and $N\lambda_{K(\tau)+k}([B(\tau)F(\tau)^{\prime}+{E}(\tau)]M_T$ $\times [B(\tau)F(\tau)^{\prime}+{E}(\tau)]^{\prime}/T)$ are asymptotically equivalent, which holds uniformly over $\tau\in\mathcal{T}$. By the triangle inequality, it follows from \eqref{Eqn: TechD2: 3}, \eqref{Eqn: TechD2: 4} and \eqref{Eqn: TechD2: 6} that
\begin{align}\label{Eqn: TechD2: 7}
&\lambda_{T}({N}{E}(\tau){E}(\tau)^{\prime}/T) + o_{p}(1)\leq N\lambda_{T-K(\tau)-1}({\tilde{Y}(\tau)M_T\tilde{Y}(\tau)^{\prime}}/{T})\notag\\
&\hspace{1cm}\leq N\lambda_{K(\tau)+1}({\tilde{Y}(\tau)M_T\tilde{Y}(\tau)^{\prime}}/{T})\leq \lambda_{1}({N}{E}(\tau){E}(\tau)^{\prime}/T) + o_{p}(1),
\end{align}
where both ``$o_{p}(1)$'' hold uniformly over $\tau\in\mathcal{T}$. This completes the proof of the lemma.
\qed

\begin{lem}[Weyl's inequalities]\label{Lem: TechD3}
Let $C$ and $D$ be $L\times L$ symmetric matrices.\\
(i) For every $s,t\geq 1$ and $s+t-1\leq L$,
\[\lambda_{s+t-1}(C+D)\leq \lambda_{s}(C)+\lambda_{t}(D).\]
(ii) If $D$ is positive semi-definite, for all $1\leq s\leq L$,
\[\lambda_{s}(C+D)\geq \lambda_{s}(C).\]
\end{lem}
\noindent{\sc Proof:} The results can be found in \citet{Bhatia_MatrixAnalysis_1997}, \citet{AhnHorenstein_EigenvalueRatio_2013} and \citet*{Fanetal_ProjectedPCASupp_2016}.\qed

\section{Auxiliary Propositions}
\renewcommand{\theequation}{E.\arabic{equation}}
\setcounter{equation}{0}

\begin{pro}\label{Pro: Strong}
Let $\mathbb{V}(\cdot) = \mathrm{vec}(\mathbb{U}(\cdot))$, where $\mathbb{U}(\cdot)$ is given in Assumption \ref{Ass: AsymDis}(ii). Suppose Assumption \ref{Ass: DGP} and \ref{Ass: Basis}(i) hold. Assume (i) $N\to\infty$; (ii) $T\geq 1$ is finite; (iii) $J\to\infty$ with $J^{7}\xi^{6}_{J}\log^{6} N=o(N)$. Then there exists a zero-mean Gaussian process $\mathbb{G}(\cdot)$ on $\mathcal{T}$ with a.s. continuous path such that the covariance function of $\mathbb{G}(\cdot)$ coincides with that of $\mathbb{V}(\cdot)$ and
\[\sup_{\tau\in\mathcal{T}}\|\mathbb{V}(\tau)-\mathbb{G}(\tau)\|= O_{p}\left(\frac{J^{7/6}\xi^{3/8}\log^{3/8} N}{N^{1/16}}\right).\]
\end{pro}
\noindent{\sc Proof:} Let $\{\pi_{j}: (0,1]\to(0,1]\}_{j=0}^{\infty}$ be a sequence of projection operators given by $\pi_j(\tau)=k/2^j$ if $\tau\in((k-1)/2^j,k/2^j],k=1,\ldots,2^j$. In what follows, for a given process $G(\cdot)\in[\ell^{\infty}(\mathcal{T})]^{JMT}$, we approximate $G(\cdot)$ by its projection $G\circ \pi_{j}(\cdot)$. Since the path of $G\circ \pi_{j}(\cdot)$ is a step function with at most $2^j$ steps, we can identify $G\circ \pi_{j}(\cdot)$ with a random vector $G\circ \pi_{j}$ in $\mathbf{R}^{2^jJMT}$. The proof consists of four steps.
\begin{enumerate}
  \item[(1)] Finite-dimensional approximation: under some appropriate choice of $j$,
  \begin{align*}
    \sup_{\tau\in\mathcal{T}}\|\mathbb{V}(\tau)-\mathbb{V}\circ \pi_{j}(\tau)\|=O_{p}(\delta_{1,N}) \text{ for some }\delta_{1,N}\downarrow 0.
  \end{align*}
  \item[(2)] Gaussian vector approximation: under some appropriate choice of $j$, there exists a Gaussian random vector $\mathbb{N}_{j}\sim N(0,\mathrm{var}(\mathbb{V}\circ \pi_{j}))$ such that
  \begin{align*}
    \|\mathbb{N}_{j}-\mathbb{V}\circ \pi_{j}\|=O_{p}(\delta_{2,N}) \text{ for some }\delta_{2,N}\downarrow 0.
  \end{align*}
    \item[(3)] Gaussian process embedding: there exists a Gaussian process $\mathbb{G}(\cdot)$ on $\mathcal{T}$ with properties stated in the proposition such that $\mathbb{N}_{j}=\mathbb{G}\circ\pi_{j}$ a.s..
  \item[(4)] Infinite-dimensional approximation: under some appropriate choice of $j$,
  \begin{align*}
    \sup_{\tau\in\mathcal{T}}\|\mathbb{G}(\tau)-\mathbb{G}\circ \pi_{j}(\tau)\|=O_{p}(\delta_{3,N}) \text{ for some }\delta_{3,N}\downarrow 0.
  \end{align*}
\end{enumerate}

We now prove the four steps by extending the proof of Lemma 14 of \citet{Bellonietal_ConditionalQuantile_2019}. The main difference here is that the covariance structure of $\mathbb{V}(\cdot)=(\mathbb{U}_1(\cdot)^{\prime}, \mathbb{U}_2(\cdot)^{\prime},\ldots, \mathbb{U}_T(\cdot)^{\prime})^{\prime}$ involves the temporal dependence of $\{y_{it},z_{it}\}_{i\leq N,t\leq T}$, that is, the covariance function matrix of $\mathbb{V}(\cdot)$ does not have a diagonal block structure. However, since $T$ is finite, steps (1)-(4) can be proved by similar arguments. To prove step (1), we obtain
\begin{align}\label{Eqn: Pro: Strong: 1}
  \sup_{\tau\in\mathcal{T}}\|\mathbb{V}(\tau)-\mathbb{V}\circ \pi_{j}(\tau)\|^{2} \leq \sum_{t=1}^{T}\sup_{\tau\in\mathcal{T}}\|\mathbb{U}_t(\tau)-\mathbb{U}_t\circ \pi_{j}(\tau)\|^{2}.
\end{align}
A rate for each term in the summation has been established in the proof of Lemma 14 of \citet{Bellonietal_ConditionalQuantile_2019}, so step (1) follows since $T$ is finite. Specifically, we have
\begin{align}\label{Eqn: Pro: Strong: 2}
  \sup_{\tau\in\mathcal{T}}\|\mathbb{V}(\tau)-\mathbb{V}\circ \pi_{j}(\tau)\|=O_{p}\left(\sqrt{\frac{J\log N}{2^{j}}}+\sqrt{\frac{J^2\xi_{J}^{2}\log^{4}N}{N}}\right).
\end{align}
Since Yurinskii's coupling \citep{Pollard_Probability_2002} only requires existence of third moment, step (2) follows by similar arguments as in the proof of Lemma 14 of \citet{Bellonietal_ConditionalQuantile_2019}.  Specifically, we have
  \begin{align}\label{Eqn: Pro: Strong: 3}
    \|\mathbb{N}_{j}-\mathbb{V}\circ \pi_{j}\|=O_{p}\left(\left(\frac{(2^{j})^{5}J^2\xi_{J}^6\log N}{N}\right)^{1/6}\right).
  \end{align}
Step (3) follows by the exactly same argument as in the proof of Lemma 14 of \citet{Bellonietal_ConditionalQuantile_2019} or their Lemma 17. Write $\mathbb{G}(\cdot)=(\mathbb{G}_1(\cdot)^{\prime}, \mathbb{G}_2(\cdot)^{\prime}, \ldots, \mathbb{G}_T(\cdot)^{\prime})^{\prime}$, where $\mathbb{G}_t(\cdot)\in[\ell^{\infty}(\mathcal{T})]^{JM}$.  It follows that
\begin{align}\label{Eqn: Pro: Strong: 4}
  \sup_{\tau\in\mathcal{T}}\|\mathbb{G}(\tau)-\mathbb{G}\circ \pi_{j}(\tau)\|^{2} \leq \sum_{t=1}^{T}\sup_{\tau\in\mathcal{T}}\|\mathbb{G}_t(\tau)-\mathbb{G}_t\circ \pi_{j}(\tau)\|^{2}.
\end{align}
A rate for each term in the summation has been established in the proof of Lemma 14 of \citet{Bellonietal_ConditionalQuantile_2019}, so step (4) follows since $T$ is finite. Specifically, we have
\begin{align}\label{Eqn: Pro: Strong: 5}
  \sup_{\tau\in\mathcal{T}}\|\mathbb{G}(\tau)-\mathbb{G}\circ \pi_{j}(\tau)\|=O_{p}\left(\sqrt{\frac{J\log N}{2^{j}}}\right).
\end{align}

Thus, by setting $2^{j} = N^{1/8}J^{1/8} \log^{1/4}N/\xi_{J}^{3/4}$, we may combine \eqref{Eqn: Pro: Strong: 2}, \eqref{Eqn: Pro: Strong: 3}, step (3) and \eqref{Eqn: Pro: Strong: 5} to yield the desired result, since $J^{7}\xi^{6}_{J}\log^{6} N=o(N)$.\qed

\begin{pro}\label{Pro: StrongBoot}
Let $\mathbb{V}^{\ast}(\cdot) = \mathrm{vec}(\mathbb{U}^{\ast}(\cdot))$, where $\mathbb{U}^{\ast}(\cdot)$ is given in Assumption \ref{Ass: BootWeight}(ii). Suppose Assumption \ref{Ass: DGP}, \ref{Ass: Basis}(i) and \ref{Ass: BootWeight}(i) hold. Assume (i) $N\to\infty$; (ii) $T\geq 1$ is finite; (iii) $J\to\infty$ with $J^{7}\xi^{6}_{J}\log^{6} N=o(N)$. Then there exists a zero-mean Gaussian process $\mathbb{G}^{\ast}(\cdot)$ on $\mathcal{T}$ conditional on $\{y_{it},z_{it}\}_{i\leq N,t\leq T}$ with a.s. continuous path such that the covariance function of $\mathbb{G}^{\ast}(\cdot)$ coincides with that of $\mathbb{V}(\cdot)$ (in Proposition \ref{Pro: Strong}) and
\[\sup_{\tau\in\mathcal{T}}\|\mathbb{V}^{\ast}(\tau)-\mathbb{G}^{\ast}(\tau)\|= O_{p}\left(\frac{J^{7/6}\xi^{3/8}\log^{3/8} N}{N^{1/16}}\right).\]
\end{pro}
\noindent{\sc Proof:} Let $\{\pi_{j}: (0,1]\to(0,1]\}_{j=0}^{\infty}$ be a sequence of projection operators defined in the proof of Proposition \ref{Pro: Strong}. The proof also consists of four steps.
\begin{enumerate}
  \item[(1)] Finite-dimensional approximation: under some appropriate choice of $j$,
  \begin{align*}
    \sup_{\tau\in\mathcal{T}}\|\mathbb{V}^{\ast}(\tau)-\mathbb{V}^{\ast}\circ \pi_{j}(\tau)\|=O_{p}\left(\sqrt{\frac{J\log N}{2^{j}}}+\sqrt{\frac{J^2\xi_{J}^{2}\log^{4}N}{N}}\right).
  \end{align*}
  \item[(2)] Gaussian vector approximation: under some appropriate choice of $j$, there exists a Gaussian random vector $\mathbb{N}^{\ast}_{j}\sim N(0,\mathrm{var}(\mathbb{V}\circ \pi_{j}))$ conditional on $\{y_{it},z_{it}\}_{i\leq N,t\leq T}$ such that
  \begin{align*}
    \|\mathbb{N}^{\ast}_{j}-\mathbb{V}^{\ast}\circ \pi_{j}\|=O_{p}\left(\left(\frac{(2^{j})^{5}J^2\xi_{J}^6\log N}{N}\right)^{1/6}\right).
  \end{align*}
    \item[(3)] Gaussian process embedding: there exists a Gaussian process $\mathbb{G}^{\ast}(\cdot)$ on $\mathcal{T}$ with properties stated in the proposition such that $\mathbb{N}^{\ast}_{j}=\mathbb{G}^{\ast}\circ\pi_{j}$ a.s..
  \item[(4)] Infinite-dimensional approximation: under some appropriate choice of $j$,
  \begin{align*}
    \sup_{\tau\in\mathcal{T}}\|\mathbb{G}^{\ast}(\tau)-\mathbb{G}^{\ast}\circ \pi_{j}(\tau)\|=O_{p}\left(\sqrt{\frac{J\log N}{2^{j}}}\right).
  \end{align*}
\end{enumerate}
Steps (1), (3) and (4) follow by the same arguments for steps (1), (3) and (4) in the proof of Proposition \ref{Pro: Strong}. For step (2), we may apply the Yurinskii’s coupling conditional on $\{y_{it},z_{it}\}_{i\leq N,t\leq T}$ to obtain that there exists a Gaussian random vector $\mathbb{N}^{\ast\ast}_{j}\sim N(0,\mathrm{var}^{\ast}(\mathbb{V}^{\ast}\circ \pi_{j}))$ conditional on $\{y_{it},z_{it}\}_{i\leq N,t\leq T}$
\begin{align}\label{Eqn: Pro: StrongBoot: 1}
    \|\mathbb{N}^{\ast\ast}_{j}-\mathbb{V}^{\ast}\circ \pi_{j}\|=O_{p}\left(\left(\frac{(2^{j})^{5}J^2\xi_{J}^6\log N}{N}\right)^{1/6}\right),
\end{align}
where $\mathrm{var}^{\ast}(\mathbb{V}^{\ast}\circ \pi_{j})$ is the variance of $\mathbb{V}^{\ast}\circ \pi_{j}$ conditional on $\{y_{it},z_{it}\}_{i\leq N,t\leq T}$. It is easy to show that $\|[\mathrm{var}^{\ast}(\mathbb{V}^{\ast}\circ \pi_{j})]^{1/2} -[\mathrm{var}(\mathbb{V}\circ \pi_{j})]^{1/2}\|_{2}=O_{p}(2^{j}J/\sqrt{N})$; see Lemma D.8 of \citet{Chenetal_SeimiparametricFactor_2021}. Thus, $\|\mathbb{N}^{\ast}_{j}-\mathbb{N}^{\ast\ast}_{j}\|=O_{p}((2^{j}J)^{3/2}/\sqrt{N})$. Step (3) thus follows by the triangle inequality. This completes the proof of the proposition. \qed

\section{Additional Empirical Results}\label{App: F}
\renewcommand{\theequation}{F.\arabic{equation}}
\renewcommand{\thefigure}{F.\arabic{figure}}
\setcounter{equation}{0}
\setcounter{table}{0}
\setcounter{figure}{0}

In this section, we conduct analysis for daily returns of individual stocks in the US market. We obtain daily gross returns from CRSP for all firms listed in NYSE, AMEX and NASDAQ, and daily risk free interest rate from Kenneth French's website. Our analysis includes four characteristic variables: size (of monthly frequency), value (of yearly frequency), momentum (of daily frequency) and volatility (of daily frequency). For each day, the values of the characteristics are transformed into relative ranking values with range $[-0.5,0.5]$. The data set is an unbalanced panel one, for which the QR-PCA is applicable. There is a total number of 4417 stocks, and each day has at least 3926 stocks that have observations on both returns and the four characteristics.

We implement the QR-PCA at various quantiles, and compare the results with those of the regressed-PCA. We implement both for three different specifications of the intercept function and the factor
loading functions. First, we consider linear specifications by setting $\phi(z_{it})=(1,z_{it}^{\prime})^{\prime}$, denoted Model M1. Second, we consider cubic spline specifications with two or five internal knots by setting $\phi(z_{it})$ as B-splines of $z_{it}$, denoted Models M2 and M3. We let $K_{\max}$ be the greatest integer no larger than $18$ in the implementation of $\hat{K}(\tau)$. We implement 499 bootstrap draws for the weighted bootstrap. We refer to the factors $\hat{F}(\tau)$ from the QR-PCA as the quantile factors, and $\hat{F}$ from the regressed-PCA as the mean factors.

Table \ref{Tab: EmpiricalF1} reports the estimated number of quantile factors at $\tau = 0.01, 0.05, 0.1,$ $0.25, 0.5, 0.75, 0.9, 0.95$ and $0.99$, and $R^{2}$ of regressing each of the first 10 quantile factors on the first 10 mean factors. We have similar findings as in Table \ref{Tab: Empirical1}. The quantile factors vary across quantiles in all specifications, and the median factors are very different from the mean factors. The null hypothesis of $\alpha(\tau,\cdot)$ is rejected at $1\%$ level for all $\tau$ in all models ($p-$values which are not reported here are available upon request), regardless of the number of estimated factors used (up to 10).

\setlength{\tabcolsep}{4.5pt}
\begin{table}[!htbp]
\footnotesize
\centering
\begin{threeparttable}
\renewcommand{\arraystretch}{1.5}
\caption{Comparison between $\hat{F}(\tau)$ and $\hat{F}$: daily returns}\label{Tab: EmpiricalF1}
\begin{tabular}{ccccccccccccccccccc}
\hline\hline
&\multirow{2}{*}{Model}&\multirow{2}{*}{$\tau$}&&\multirow{2}{*}{$\hat{K}(\tau)$}&&\multicolumn{10}{c}{Column of $\hat{F}(\tau)$}&&\multirow{2}{*}{Avg}&\\
\cline{7-16}
&&&&&&$1$&$2$&$3$&$4$&$5$&$6$&$7$&$8$&$9$&$10$&&&\\
\cline{2-18}
&\multirow{9}{*}{M1}&$0.01$  &&1&&26.9  &9.4    &6.7	&6.1	&-	&-	&-	&-	&-	&-&&12.3&	\\	
&&$0.05$                     &&1&&42.4	&8.2    &29.7	&12.8	&-	&-	&-	&-	&-	&-&&23.3&	\\	
&&$0.1$                      &&1&&47.9	&8.2	&21.5	&35.2	&-	&-	&-	&-	&-	&-&&28.2&	\\	
&&$0.25$                     &&1&&70.8	&13.5	&16.9	&47.6	&-	&-	&-	&-	&-	&-&&37.2&	\\
&&$0.5$                      &&1&&98.3	&77.9	&59.2	&75.3	&-	&-	&-	&-	&-	&-&&77.7&	\\
&&$0.75$                     &&1&&76.5	&27.2	&36.4	&65.7	&-	&-	&-	&-	&-	&-&&51.5&	\\
&&$0.9$                      &&1&&47.7	&18.2	&30.7	&49.5	&-	&-	&-	&-	&-	&-&&36.5&	\\
&&$0.95$                     &&1&&36.8	&16.0	&43.4	&30.4	&-	&-	&-	&-	&-	&-&&31.6&	\\
&&$0.99$                     &&1&&34.2	&13.4	&19.7	&16.5	&-	&-	&-	&-	&-	&-&&21.0&	\\
\cline{2-18}
&\multirow{9}{*}{M2}&$0.01$  &&1&&4.3 &9.4 &9.4 &9.6 &16.5&18.9&10.7&10.2&9.5 &5.0 &&11.1&\\
&&$0.05$                     &&1&&12.4&13.3&13.3&13.3&30.8&17.8&15.8&11.3&18.8&11.1&&15.6&\\ &&$0.1$                      &&1&&22.3&9.3 &9.3 &12.9&40.3&36.4&14.6&16.5&31.8&13.6&&21.2&\\
&&$0.25$                     &&3&&61.8&9.0 &9.0 &14.4&60.7&24.0&29.3&29.3&48.1&12.8&&34.5&\\ &&$0.5$                      &&1&&94.6&72.6&72.6&63.2&59.7&42.5&51.0&46.6&24.5&10.7&&51.6&\\	
&&$0.75$                     &&3&&84.4&32.4&32.4&30.0&55.7&33.2&43.1&41.9&8.5 &40.0&&41.8&\\	
&&$0.9$                      &&3&&34.8&23.8&23.8&14.3&52.1&32.1&23.0&29.1&23.2&11.8&&27.8&\\	
&&$0.95$                     &&2&&28.0&30.3&30.3&12.4&37.1&21.0&13.1&13.3&25.4&16.6&&22.3&\\	
&&$0.99$                     &&2&&34.6&27.2&27.2&10.1&22.2&15.1&17.3&11.4&12.8&15.3&&18.1&\\

\cline{2-18}
&\multirow{9}{*}{M3}&$0.01$  &&1&&5.5 &11.3&13.5&11.9&15.0&7.8 &6.5 &4.8 &14.0&7.2 &&9.8 &\\
&&$0.05$                     &&1&&6.1 &15.0&16.4&18.9&17.7&11.1&14.0&20.5&10.6&16.6&&14.7&\\
&&$0.1$                      &&1&&12.1&14.8&17.0&20.1&21.9&12.6&14.1&20.6&11.8&25.9&&17.1&\\
&&$0.25$                     &&3&&34.7&20.7&17.1&39.1&25.1&19.6&23.7&19.5&17.6&17.6&&23.5&\\
&&$0.5$                      &&1&&84.9&45.4&53.1&69.1&53.7&39.2&42.2&39.7&29.6&13.5&&47.0&\\
&&$0.75$                     &&3&&56.9&41.0&49.3&47.0&29.7&65.6&32.8&25.8&13.2&16.9&&37.8&\\
&&$0.9$                      &&2&&39.1&31.0&29.0&19.4&24.8&48.6&22.0&31.8&25.3&18.6&&29.0&\\
&&$0.95$                     &&2&&35.1&30.1&18.2&54.7&24.5&33.2&17.3&39.7&21.2&15.3&&28.9&\\
&&$0.99$                     &&1&&33.9&31.4&13.2&16.1&15.0&21.8&16.4&14.1&10.0&14.7&&18.7&\\
\hline\hline
\end{tabular}
\begin{tablenotes}
      \item[\dag] {\footnotesize{The third column reports $\hat{K}(\tau)$ at each $\tau$. The fourth to the second last columns report the $R^2$ of regressing each column of $\hat{F}(\tau)$ on $\hat{F}$ ($\%$). The last column reports the average of the fourth to the second last columns. For both $\hat{F}(\tau)$ and $\hat{F}$, the number of estimated factors is set to $10$. }}
    \end{tablenotes}
\end{threeparttable}
\end{table}%

Table \ref{Tab: EmpiricalF2} reports $R^{2}$ of regressing each of the first 10 median factors and the first 10 mean factors on six observed factors (see Section \ref{Sec: 72}). We also find that the median factors have higher correlation with the six factors than the mean factors in all models.

\setlength{\tabcolsep}{5pt}
\begin{table}[!htbp]
\footnotesize
\centering
\begin{threeparttable}
\renewcommand{\arraystretch}{1.5}
\caption{Comparison between $\hat{F}(0.5)$/$\hat{F}$ and six observed factors: daily returns}\label{Tab: EmpiricalF2}
\begin{tabular}{ccccccccccccccccc}
\hline\hline
&\multirow{2}{*}{Model}&\multirow{2}{*}{Factor}&&\multicolumn{10}{c}{Column of $\hat{F}(0.5)$/$\hat{F}$}&&\multirow{2}{*}{Avg}&\\
\cline{5-14}
&&&&$1$&$2$&$3$&$4$&$5$&$6$&$7$&$8$&$9$&$10$&&&\\
\cline{2-16}
&\multirow{2}{*}{M1}&$\hat{F}(0.5)$  &&3.3	&95.4	&69.8	&44.8	&-	&-	&-	&-	&-	&-&&53.3&	\\
&&$\hat{F}$                          &&3.3	&92.3	&58.6   &36.7	&-	&-	&-	&-	&-	&-&&47.7&	\\
\cline{2-16}
&\multirow{2}{*}{M2}&$\hat{F}(0.5)$  &&3.3	&88.1	&42.6	&15.7	&24.8	&26.5	&16.9	&16.4	&2.9	&4.1 &&24.1&  \\
&&$\hat{F}$                          &&3.3	&44.8	&29.2	&32.4	&26.6	&15.2	&28.6	&9.1	&7.1	&11.8&&20.8&  \\
\cline{2-16}
&\multirow{2}{*}{M3}&$\hat{F}(0.5)$  &&3.3	&73.6	&14.6	&30.5	&48.4	&16.8	&1.4	&16.5	&24.3	&8.2 &&23.8&  \\
&&$\hat{F}$                          &&3.3	&23.6	&17.5	&64.1	&14.1	&2.3	&29.0	&19.8	&10.3	&7.9 &&19.2&  \\
\hline\hline
\end{tabular}
\begin{tablenotes}
      \item[\dag] {\footnotesize{The sixed observed factors are market excess return, ``small minus big'' factor, ``high minus low'' factor, ``momentum'' factor, ``robust minus weak'' factor, and ``conservative minus aggressive'' factor. The third to the second last columns report the $R^2$ of regressing each column of $\hat{F}(0.5)$/$\hat{F}$ on the six observed factors ($\%$). The last column reports the average of the third to the second last columns. For both $\hat{F}(0.5)$ and $\hat{F}$, the number of estimated factors is set to $10$. }}
    \end{tablenotes}
\end{threeparttable}
\end{table}%

\begin{figure}[!htbp]
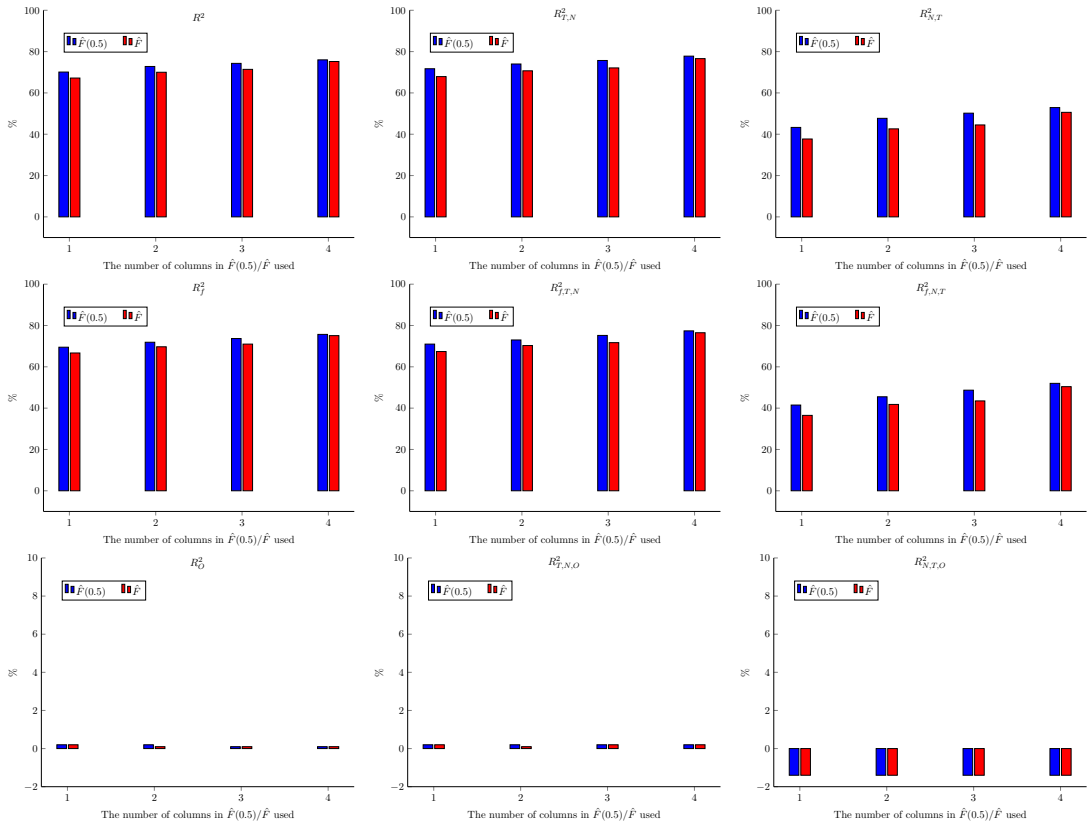

\centering
\begin{subfigure}[b]{0.32\textwidth}
\centering
\resizebox{\linewidth}{!}{
\pgfplotsset{title style={at={(0.5,0.91)}}}
}
\end{subfigure}
\caption{Evaluating $\hat{F}(0.5)$/$\hat{F}$ using portfolios: Model M1 (daily returns)} \label{Fig: EmpricalF6}
\end{figure}

We evaluate the median factors and the mean factors by comparing their ability to explain the cross section of portfolio returns. Our analysis includes 124 portfolios available on Kenneth French’s website: 25 portfolios sorted by size and book-to-market ratio, 49 industry portfolios, 25 portfolios sorted by operating profitability and investment, and 25 portfolios sorted by size and momentum. The results are reported in Figures \ref{Fig: EmpricalF6}-\ref{Fig: EmpricalF8}. We have similar findings as in Figures \ref{Fig: Emprical6}-\ref{Fig: Emprical8}. The median factors have better in-sample explanatory power for the portfolio returns than the mean factors in all cases, while they share similar out-of-sample prediction power. For example, in Model M2 with only one factor used, $\hat{F}(0.5)$ has much higher $R^{2}$, $R^{2}_{T,N}$ and $R^{2}_{N,T}$ than $\hat{F}$ ($64.9\%$, $65.9\%$ and $34.6\%$ v.s. $34.9\%$, $35.2\%$ and $13.4\%$).

\begin{figure}[!htbp]
\centering
\begin{subfigure}[b]{0.32\textwidth}
\centering
\resizebox{\linewidth}{!}{
\pgfplotsset{title style={at={(0.5,0.91)}}}
%
}
\end{subfigure}
\caption{Evaluating $\hat{F}(0.5)$/$\hat{F}$ using portfolios: Model M2 (daily returns)} \label{Fig: EmpricalF7}
\end{figure}

\begin{figure}[!htbp]
\centering
\begin{subfigure}[b]{0.32\textwidth}
\centering
\resizebox{\linewidth}{!}{
\pgfplotsset{title style={at={(0.5,0.91)}}}
%
}
\end{subfigure}
\caption{Evaluating $\hat{F}(0.5)$/$\hat{F}$ using portfolios: Model M3 (daily returns)} \label{Fig: EmpricalF8}
\end{figure}

\end{appendices}
\newpage
\addcontentsline{toc}{section}{References}
\putbib
\end{bibunit}
\bibliography{C:/Dropbox/Research_Related/BibMaker/Mybibliography_Infinity}

\end{document}